\documentclass[12pt]{article}
\usepackage{amsmath}
\usepackage{epsfig}
\newcommand{\ep}{\varepsilon}

\newcommand{\Li}[2]{{\mbox{Li}}_{#1}\left(#2\right)}

\newcommand{\LS}[3]{{\mbox{Ls}}_{#1}^{(#2)}\left(#3\right)}

\newcommand{\Snp}[2]{{\mbox{S}}_{#1\!}\left(#2\right)}

\newcommand{\myfrac}[2]{{\textstyle{\frac{#1}{#2}}}}

\newcommand{\re}{\mathop{\mathrm{Re}}\nolimits}

\catcode`@=11
\newcount\@tempcntc
\def\@citex[#1]#2{\if@filesw\immediate\write\@auxout{\string\citation{#2}}\fi
  \@tempcnta\z@\@tempcntb\m@ne\def\@citea{}\@cite{\@for\@citeb:=#2\do
    {\@ifundefined
       {b@\@citeb}{\@citeo\@tempcntb\m@ne\@citea\def\@citea{,}{\bf ?}\@warning
       {Citation `\@citeb' on page \thepage \space undefined}}%
    {\setbox\z@\hbox{\global\@tempcntc0\csname b@\@citeb\endcsname\relax}%
     \ifnum\@tempcntc=\z@ \@citeo\@tempcntb\m@ne
       \@citea\def\@citea{,}\hbox{\csname b@\@citeb\endcsname}%
     \else
      \advance\@tempcntb\@ne
      \ifnum\@tempcntb=\@tempcntc
      \else\advance\@tempcntb\m@ne\@citeo
      \@tempcnta\@tempcntc\@tempcntb\@tempcntc\fi\fi}}\@citeo}{#1}}
\def\@citeo{\ifnum\@tempcnta>\@tempcntb\else\@citea\def\@citea{,}%
  \ifnum\@tempcnta=\@tempcntb\the\@tempcnta\else
   {\advance\@tempcnta\@ne\ifnum\@tempcnta=\@tempcntb \else \def\@citea{--}\fi
    \advance\@tempcnta\m@ne\the\@tempcnta\@citea\the\@tempcntb}\fi\fi}
\catcode`@=12
\begin{document}

%
%
\title{
\vskip-3cm{\baselineskip14pt
\centerline{\normalsize DESY~10--027 \hfill ISSN 0418-9833}
\centerline{\normalsize March 2010\hfill}}
\vskip1.5cm
Differential reduction of generalized hypergeometric functions from Feynman
diagrams: One-variable case
}

\author{
{\sc Vladimir~V.~Bytev}\thanks{On leave of absence from
Joint Institute for Nuclear Research,141980 Dubna (Moscow Region), Russia.},
{\sc Mikhail~Yu.~Kalmykov},$\!\!{}^*$ 
{\sc Bernd~A.~Kniehl}
\\
\\
{\normalsize II. Institut f\"ur Theoretische Physik, Universit\"at Hamburg,}\\
{\normalsize Luruper Chaussee 149, 22761 Hamburg, Germany}
}

\date{}

\maketitle
\abstract{ 
The differential-reduction algorithm, which allows one to express generalized
hypergeometric functions with parameters of arbitrary values in terms of such
functions with parameters whose values differ from the original ones by
integers, is discussed in the context of evaluating Feynman diagrams.
Where this is possible, we compare our results with those obtained using
standard techniques.
It is shown that the criterion of reducibility of multiloop Feynman integrals
can be reformulated in terms of the criterion of reducibility of
hypergeometric functions.
The relation between the numbers of master integrals obtained by differential
reduction and integration by parts is discussed.
\medskip

}

\newpage


\renewcommand{\thefootnote}{\arabic{footnote}}
\setcounter{footnote}{0}
\section{Introduction}
\setcounter{equation}{0}

It is commonly accepted that any multiloop and/or multileg Feynman diagram in
covariant gauge within dimensional regularization \cite{dimreg} may be treated
as a generalized hypergeometric function\footnote{%
This statement is also valid for phase-space integrals (see, e.g.,
Ref.~\cite{Melnikov}).}
\cite{regge}. 
Starting from its $\alpha$ representation, any Feynman diagram may be written
in the form of a Mellin-Barnes integral \cite{MB,smirnov}, 
\begin{eqnarray}
\Phi(n) \sim
\int_{-i \infty}^{+i \infty}
\prod_{a,b,c}
\frac{\Gamma(\sum_{i=1}^m A_{ai}z_i+B_a)}{\Gamma(\sum_{j=1}^r C_{bj}z_j+D_{b})}
dz_c Y^{z_c-1}
\;,
\label{MB}
\end{eqnarray}
where $Y_a$ are algebraic functions of external kinematic invariants and
$A,B,C,D$ are some matrices depending in a linear way on the dimension of
space-time $n$, which is an arbitrary complex number, and the powers of the
propagators.

By the application of Cauchy's theorem,\footnote{%
This is true if all arguments of the $\Gamma$ functions are different.
Otherwise, an additional regularization for each propagator (or the
introduction of extra masses) is necessary.
See, e.g., Ref.~\cite{example}.}
this integral can be rewritten as a linear combination of multiple
series,\footnote{%
One of the first examples of this type of representation for a Feynman diagram
was given in Ref.~\cite{first}.}
\begin{equation}
\Phi(n,\vec{x}) \sim
\sum_{k_1,\cdots, k_{r+m}=0}^\infty
\prod_{a,b}
\frac{\Gamma(\sum_{i=1}^m \tilde{A}_{ai}k_i+\tilde{B}_a)}{\Gamma(\sum_{j=1}^r
\tilde{C}_{bj}k_j+\tilde{D}_{b})}
x_1^{k_1} \cdots x_{r+m}^{k_{r+m}}
\;.
\label{Horn}
\end{equation}
For real diagrams, some of the variables $x_k$ may be complex number.
We call this type of variable a ``hidden variable'' and the corresponding
index of summation a ``hidden index of summation.''
In all existing examples, the representation of Eq.~(\ref{Horn}) belongs to a
Horn-type hypergeometric series \cite{Gelfand} if the hidden index of
summation is considered as an independent variable.

For the reader's convenience, we recall that the multiple series 
$\sum_{\vec{m}=0}^\infty C(\vec{m}) \vec{x}^{\vec{m}}$
is called {\it Horn-type hypergeometric} if, for each $i=1, \ldots, r$, the ratio 
$C(\vec{m}+\vec{e}_i)/C(\vec{m})$ is a rational function in the index of
summation $(m_1, \cdots, m_r)$ \cite{Gelfand,bateman}.
The coefficients of such a series have the general form 
\begin{equation}
C(\vec{m})
= \prod_{i=1}^r \lambda_i^{m_i}R(\vec{m})
\frac{
\prod_{j=1}^N \Gamma(\mu_j(\vec{m})+\gamma_j)
}
{
\prod_{k=1}^M \Gamma(\nu_k(\vec{m})+\delta_k)
}\;,
\label{ore}
\end{equation}
where $N, M \geq 0$, $\lambda_j,\delta_j, \gamma_j$ are arbitrary complex
numbers, $\mu_j, \nu_k: Z^r \to Z$ are arbitrary integer-valued linear maps,
and $R$ is an arbitrary rational function \cite{Gelfand,Ore:Sato}.

However, to our knowledge, the proof that any Feynman diagram can be described
by a Horn-type series does not exist.
There is another way to proof this statement.
The classical $\alpha$ representation of a Feynman diagram is a particular
case of the generalized Euler integral representation of the
Gel'fand-Kapranov-Zelevinski system \cite{Gelfand} that is related to the 
Horn-type series representation.
We call the Feynman diagram representation of Eq.~(\ref{MB}) or the equivalent
representation of Eq.~(\ref{Horn}) a hypergeometric representation
\cite{Gelfand}.

For Horn-type hypergeometric functions, there are so-called step-up (step-down)
operators $H_\lambda^{+}$ ($H_\lambda^{-}$) \cite{bateman,acat08}.
These are differential operators which, upon application to a hypergeometric
function $S_\lambda$, shift the value of one its upper (lower) parameters by
unity, as $H_\lambda^{+} S_\lambda = S_{\lambda+1}$
($H_\lambda^{-} S_\lambda = S_{\lambda-1}$).
Takayama \cite{theorem} proposed an algorithm that allows one to construct
inverse differential operators, the step-down (step-up) operators
$B_\lambda^{-}$ ($B_\lambda^{+}$), starting from direct operators and systems
of differential equations for hypergeometric functions.
These operators satisfy the relations
$B_\lambda^{-} S_{\lambda+1} \to S_\lambda$ 
($B_\lambda^{+} S_{\lambda} \to S_{\lambda+1}$).
Takayama pointed out \cite{theorem} that inverse operators are uniquely defined
for any hypergeometric function with an irreducible monodromy group, which
implies that the parameters and the differences between upper and lower
parameters are not integer. 
By the action of such differential operators on a hypergeometric function,
the value of any parameter can be shifted by an arbitrary integer.
We call this procedure of applying differential operators to shift the
parameters by integers {\it differential reduction}.\footnote{%
There are various publications on contiguous relations for hypergeometric
functions, i.e.\ algebraic relations between hypergeometric functions of
several variables with shifted values of parameters, starting from the
classical paper by Gauss \cite{Gauss}. 
To our knowledge, a closed algorithm for the algebraic reduction of
Horn-type hypergeometric functions of several variables, i.e.\ an algorithm for
solving these algebraic relations, does not exist.} 
An important step of Takayama's algorithm is the construction of a differential
Gr\"obner basis for the system of differential equations for hypergeometric
functions.\footnote{%
The idea of using the differential-Gr\"obner-basis technique directly for the
reduction of off-shell Feynman diagrams, without spitting them into linear
combinations of hypergeometric functions, was proposed by Tarasov
\cite{Tarasov:Grobner}.}

%
%


It is quite surprising that the technique for the reduction of Feynman
diagrams advocated here, namely to split a given Feynman diagram into a
linear combination of Horn-type hypergeometric functions with rational
coefficients and to subsequently apply differential reduction, has never been
elaborated and that its interrelation with the well-known integration-by-parts
(IBP) technique \cite{ibp} has never been discussed.

The aim of this paper is to demonstrate how the differential-reduction
algorithm may be successfully applied to evaluate Feynman diagrams.
For simplicity, we consider here only the particular case of Horn-type multiple
hypergeometric functions, i.e.\ the functions ${}_{p+1}F_{p}$, and some Feynman
diagrams with arbitrary powers of propagators, which are expressible in terms
of these functions.

The structure of this paper is as follows.
In Section~{\ref{reduction}}, we discuss the differential-reduction algorithm
for generalized hypergeometric functions ${}_{p+1}F_{p}$.
This is a crucial step towards proving theorems on the construction of
all-order $\ep$ expansions presented in
Refs.~\cite{KWY07a,KWY07b,KWY07c,KK08a}.
Section~\ref{application} illustrates the application of differential
reduction to several Feynman diagrams of phenomenological interest.
In Section ~\ref{camparison}, we discuss how the counting of master integrals
in the differential-reduction approach is related to that in the IBP technique.
The results of our analysis are briefly summarized in
Section~\ref{conclusion}.
In Appendix~\ref{appendix}, we describe interrelations between a set of basis
functions generated by the differential-reduction algorithm and a set of
hypergeometric functions whose higher-order $\ep$ expansions were constructed
in Ref.~\cite{DK04}.

\boldmath
\section{Differential-reduction algorithm for the generalized hypergeometric function ${}_{p+1}F_{p}$}
\unboldmath
\label{reduction}

\subsection{Notation}
\label{definition}

Let us consider the generalized hypergeometric function $_pF_q(a;b;z)$,
defined by a series about $z=0$ as
\begin{equation}
{}_pF_q(\vec{a};\vec{b};z) 
\equiv
{}_{p}F_q \left( \begin{array}{c|}
\vec{a} \\
\vec{b}
\end{array}~ z \right)
= \sum_{k=0}^\infty \frac{z^k}{k!} \frac{\prod_{i=1}^p (a_i)_k}{\prod_{j=1}^q (b_j)_k} \;,
\label{hypergeometric}
\end{equation}
where $(a)_k = \Gamma(a+k)/\Gamma(a)$ is the Pochhammer symbol.
The sets $\vec{a}=(a_1,\cdots, a_p)$ and $\vec{b}=(b_1,\cdots, b_q)$ are
called the upper and lower parameters of the hypergeometric function,
respectively.
In terms of the differential operator $\theta$, 
\begin{equation}
\theta = z \frac{d}{d z} \;, 
\label{theta}
\end{equation}
the differential equation for the hypergeometric function ${}_pF_{q}$ can be
written as 
\begin{equation}
\left[ 
z \prod_{i=1}^p (\theta + a_i)
- \theta \prod_{i=1}^q (\theta + b_i-1)
\right] 
{}_pF_q(\vec{a}; \vec{b}; z) = 0.
\label{diff}
\end{equation}
Hypergeometric functions which differ by $\pm 1$ in the value of one of
their parameters are called contiguous, and linear relations between
contiguous functions and their derivatives are called contiguous
relations.\footnote{%
A full set of contiguous relations for an arbitrary hypergeometric function
${}_{p}F_{q}$ was considered by Rainville \cite{rainville}.
For ${}_{p}F_{q}$, there are $2(p+q)$ contiguous functions.}
In particular, the following differential identities between contiguous
functions are universal \cite{acat08}: 
\begin{eqnarray}
{}_pF_q(a_1+1, \vec{a};\vec{b};z) & = &  
B_{a_1}^{+} {}_pF_q(a_1, \vec{a};\vec{b};z) 
 = 
\frac{1}{a_1}
\left(\theta  +  a_1 \right) 
{}_pF_q(a_1, \vec{a};\vec{b};z) \;, 
\nonumber
\\
{}_pF_q(\vec{a};b_1-1,\vec{b};z) & = &  
H_{b_1}^{-} {}_pF_q(\vec{a}; b_1,\vec{b};z) 
= 
\frac{1}{b_1  -  1}
\left(\theta   +  b_1  -  1 \right) 
{}_pF_q(\vec{a};b_1,\vec{b};z) \;,
\label{universal:b}
\end{eqnarray}
which directly follow from the series representation of
Eq.~(\ref{hypergeometric}).
The operators $B_{a_1}^{+}$ ($H_{b_1}^{-}$) are the step-up (step-down)
operators for upper (lower) parameters of hypergeometric functions.

\subsection{Non-exceptional values of parameters}
\label{reduction:d}

In Ref.~\cite{theorem}, it was shown that, for given step-up (step-down)
operators, inverse step-down (step-up) operators, which are uniquely defined
modulo Eq.~(\ref{diff}), can be constructed.
This type of operators were explicitly constructed for the hypergeometric
function ${}_{p+1}F_{p}$ by Takayama in Ref.~\cite{takayama}.
For completeness, we reproduce his result here:
\begin{eqnarray}
{}_{p+1}F_p(a_i-1, \vec{a}; \vec{b}; z) &=& B_{a_i}^{-} {}_{p+1}F_p(a_i, \vec{a}; \vec{b}; z) \;,
\nonumber \\
B_{a_i}^{-} &=& - \frac{a_i}{c_i} 
\left[
t_i (\theta) - z \prod_{j\neq i} (\theta + a_j )
\right]_{a_i \to a_i - 1} \;, 
\nonumber \\ 
c_i &=& -a_i \prod_{j=1}^{p} (b_j-1-a_i) \; ,
\nonumber \\ 
t_i (x)  &=& \frac{x \prod_{j=1}^p (x + b_j - 1) - c_i }{x+a_i} \;,
\label{diff:oper:1}\\
{}_{p+1}F_p(\vec{a}; b_i+1, \vec{b}; z)  &=& H_{b_i}^{+} {}_{p+1}F_p(\vec{a}; b_1, \vec{b}; z) \;,
\nonumber \\
H_{a_i}^{+} &=& \frac{b_i-1}{d_i} 
\left[
\frac{\theta}{z} \prod_{j\neq i} (\theta + b_j - 1 )
- s_i (\theta) 
\right]_{b_i \to b_i + 1} \;, 
\nonumber \\ 
d_i &=& \prod_{j=1}^{p+1} (1+a_j-b_i) \; ,
\nonumber \\ 
s_i (x)  &=& \frac{\prod_{j=1}^{p+1} (x + a_j) - d_i }{x+b_i-1} \;,
\label{diff:oper:2}
\end{eqnarray}
where $|_{a \to a+1} $ means substitution of $a$ by $a+1$.
The functions $t_i(x)$ and $s_i(x)$ are polynomials in $x$, so that
Eqs.~(\ref{diff:oper:1}) and (\ref{diff:oper:2}) are polynomials in
the derivative $\theta$.
Let us introduce the symmetric polynomial $P^{(p)}_j(\{r_k\})$ as follows: 
\begin{equation}
\prod_{k=1}^p(z+r_k) = \sum_{j=0}^p P^{(p)}_{p-j}(\{r_k\}) z^j = \sum_{j=0}^p P_j^{(p)}(\{r_k\}) z^{p-j}\;,
\label{P}
\end{equation}
so that 
\begin{eqnarray}
P^{(p)}_0(\{r_k\}) & = & 1 \;,  
\nonumber \\ 
P^{(p)}_j(\{r_k\}) & = & \sum_{i_1,\cdots,i_r=1}^p \prod_{i_1 < \cdots < i_j } r_{i_1} \cdots r_{i_j} \;, \qquad j=1, \cdots, p \;.
\end{eqnarray}
For example, we have 
$P^{(p)}_1(\{r_k\}) =  \sum_{j=1}^p r_j$ and
$P^{(p)}_p(\{r_k\}) =  \prod_{j=1}^p r_j$.
Then, we may write
\begin{eqnarray}
t_i(x) & = &  
\sum_{j=0}^p P^{(p)}_{p-j}(\{b_r-1\}) \frac{x^{j+1} - (-a_i)^{j+1}}
{x+a_i}
= 
\sum_{j=0}^p P^{(p)}_{p-j}(\{b_r-1\}) \sum_{k=0}^j x^{j-k}(-a_i)^k \;.
\qquad
\end{eqnarray}
A similar consideration is valid also for the last relation in
Eq.~(\ref{diff:oper:2}):
\begin{eqnarray}
s_i(x) & = &  
\sum_{j=0}^{p+1} P^{(p+1)}_{p+1-j}(\{a_r\}) \frac{x^{j} - (1-b_i)^{j} }
{x-(1-b_i)}
= 
\sum_{j=0}^p P^{(p+1)}_{p-j}(\{a_r\}) \sum_{k=0}^j x^{j-k}(1 - b_i)^k 
\;.
\end{eqnarray}
The differential reduction has the form of a product of several differential
step-up and step-down operators, $H^{\pm}_{b_k}$ and $B^{\pm}_{a_k}$,
respectively:\footnote{%
Due to the relation 
$\theta\, {}_pF_q(\vec{a}; \vec{b}; z)
 = z\prod_{i=1}^p a_i/(\prod_{j=1}^q b_j) {}_pF_q(\vec{a}+1; \vec{b}+1; z)$,
not all step-up operators are independent.
In fact, 
$( \prod_{j=1}^p H^{+}_{b_j}) 
( \prod_{k=1}^{p+1} B^{+}_{a_k} ) 
F(\vec{a}; \vec{b}; z) 
= 
( \prod_{k=1}^{p+1} B^{+}_{a_k} ) 
( \prod_{j=1}^p H^{+}_{b_j} ) 
F(\vec{a}; \vec{b}; z) 
= \prod_{j=1}^{p} b_j/(\prod_{k=1}^{p+1} a_k) (d/dz) F(\vec{a}; \vec{b}; z)$.}
\begin{equation}
F(\vec{a}+\vec{m}; \vec{b}+\vec{n}; z)  = 
\left( H_{\{ a\}}^{\pm} \right)^{\sum_i m_i} \left( B_{\{ b\}}^{\pm} \right)^{\sum_j n_j} F(\vec{a}; \vec{b}; z) \;,
\label{general}
\end{equation}
so that the maximal power of $\theta$ in this expression is equal to
$r \equiv \sum_i m_i+\sum_j n_j$. 
In a symbolic form, this may be written as
\begin{eqnarray}
R(a_i,b_j,z)
F(\vec{a}+\vec{m}; \vec{b}+\vec{n}; z)  = 
\left[ 
S_1(a_i,b_j,z) \theta^r + \cdots + S_{r+1}(a_i,b_j,z)
\right] F(\vec{a}; \vec{b}; z) \;,
\quad
\label{general:2}
\end{eqnarray}
where $R$ and $\{S_j\}$ are some polynomials. 
Since the hypergeometric function  ${}_{p+1}F_{p}(\vec{a};\vec{b}; z)$
satisfies the following differential equation of order $p+1$ [see
Eq.~(\ref{diff})]: 
\begin{eqnarray} 
\lefteqn{(1-z) \theta^{p+1} 
{}_{p+1}F_{p}(\vec{a};\vec{b}; z)}
\nonumber\\
&=& 
\left\{ 
\sum_{r=1}^{p}
\left[ 
z P^{(p+1)}_{p+1-r}(\{a_j\})
- 
P^{(p)}_{p+1-r}(\{b_j-1\})
\right] \theta^r
+ 
z \prod_{k=1}^{p+1} a_k 
\right\}
{}_{p+1}F_{p}(\vec{a};\vec{b}; z) \;,
\end{eqnarray}
it is possible to express all terms containing higher powers of the operator
$\theta$, $\theta^k$ with $k \geq p+1$, as a linear combination of rational
functions of $z$ depending parametrically on $a$ and $b$ multiplied by lower
powers of $\theta$, $\theta^j$ with $j \leq p$. 
In this way, any function ${}_{p+1}F_{p}(\vec{a}+\vec{m};\vec{b}+\vec{k}; z)$,
where $\vec{m}$ and $\vec{k}$ are sets of integers, is expressible in terms of
the basic function and its first $p$ derivatives as 
\begin{eqnarray}
\label{decomposition}
\lefteqn{S(a_i,b_j,z)
{}_{p+1}F_{p}(\vec{a}+\vec{m};\vec{b}+\vec{k}; z)
= \left \{
  R_1(a_i,b_j,z) \theta^p
+ R_2(a_i,b_j,z) \theta^{p-1}
\right.}
\nonumber\\
&&{}+\left. 
\cdots 
+ R_p(a_i,b_j,z) \theta
+ R_{p+1}(a_i,b_j,z) 
\right\}{}_{p+1}F_{p}(\vec{a};\vec{b}; z) \;,
\end{eqnarray}
where $S$ and $R_i$ are polynomials in the parameters $\{a_i\}$ and $\{b_j\}$
and the argument $z$.

From Eq.~(\ref{diff:oper:1}) it follows that, if one of the upper parameters 
$a_j$ is equal to unity, then the application of the step-down operator 
$B^{-}_{a_j}$ to the hypergeometric function ${}_{p+1}F_{p}$ produces unity, 
$B^{-}_{1} {}_{p+1}F_{p}(1, \vec{a};\vec{b}; z) =1$.
Taking into account the explicit form of the step-down operator $B^{-}_{1}$, 
\begin{equation}
B^{-}_{1} =   
\frac{1}{\prod_{k=1}^p(b_k-1)}
\left[ \prod_{j=1}^p (b_j-1) 
+ \sum_{j=1}^p P^{(p)}_{p-j}(\{b_k - 1\}) \theta^j
- z \prod_{j=1}^p \left(\theta+a_j\right)
\right] \;,
\end{equation}
we obtain the differential identity
\begin{eqnarray}
\lefteqn{\left\{ 
\prod_{j=1}^p (b_j-1) 
 -  
z \prod_{j=1}^p a_j 
 +  
(1-z) \theta^p
\right\} 
{}_{p+1}F_{p}(1, \vec{a};\vec{b}; z)} 
\nonumber \\
&&{}+ 
\left\{ 
\sum_{j=1}^{p-1} \left[ P^{(p)}_{p-j}(\{b_k-1\}) - z P^{(p)}_{p-j}(\{a_k\}) \right] \theta^j 
\right\}
{}_{p+1}F_{p}(1, \vec{a};\vec{b}; z) 
= 
\prod_{k=1}^p(b_k-1) \;.\quad
\label{relation}
\end{eqnarray}
The case where two or more upper parameters are equal to unity, e.g.\
$a_1=a_2=1$, does not generate any new identities. 
As a consequence, if there is a subset $\vec{l}$ of positive integers in the
set of upper parameters, the reduction procedure has the modified form:
\begin{eqnarray}
\lefteqn{\tilde{P}(a_i,b_j,z)
{}_{p+1}F_{p}(\vec{l}, \vec{a}+\vec{m};\vec{b}+\vec{k}; z)
= \tilde{R}_{1}(a_i,b_j,z)+ 
\left\{\tilde{R_2}(a_i,b_j,z) \theta^{p-1}
\right.}
\nonumber\\
&&{}+\left.
\cdots 
+ \tilde{R}_{p}(a_i,b_j,z) \theta
+\tilde{R}_{p+1}(a_i,b_j,z) \right\}
{}_{p+1}F_{p}(\vec{1}, \vec{a};\vec{b}; z) \;.
\label{decomposition:1}
\end{eqnarray}

Let us write explicit expressions for the inverse operators for several
hypergeometric functions. 
For the Gauss hypergeometric function ${}_{2}F_{1}$, we have:
\begin{eqnarray}
{}_{2}F_{1} \left( \begin{array}{c|}
a_1-1,a_2 \\
b_1
\end{array}\, z \right) 
&=& \frac{1}{b_1-a_1}
\left[
(1-z) \theta + b_1 - a_1 - a_2 z 
\right]
{}_{2}F_{1} \left( \begin{array}{c|}
a_1,a_2 \\
b_1
\end{array}\, z \right)  \;,
\\
{}_{2}F_{1} \left( \begin{array}{c|}
a_1,a_2 \\
b_1+1
\end{array} \, z \right) 
&=& \frac{b_1}{(b_1-a_1)(b_1-a_2)}
\left[
(1-z) \frac{d}{d z} + b_1 - a_1 - a_2 
\right]
{}_{2}F_{1} \left( \begin{array}{c|}
a_1,a_2 \\
b_1
\end{array} \, z \right)  \;.
\nonumber
\end{eqnarray}

For the hypergeometric function ${}_3F_2$, the inverse differential operators
read:
\begin{eqnarray}
\lefteqn{
{}_{3}F_{2} \left( \begin{array}{c|}
a_1 - 1,a_2,a_3 \\
b_1,b_2
\end{array}\, z \right) 
(b_1 - a_1)(b_2 - a_1)}
\nonumber\\
&=& \left\{
(1 - z) \theta^2 
+ 
\left[ 
(b_1 + b_2 - 1 - a_1)  - z (a_2 + a_3)
\right] \theta 
+ (b_1 - a_1)(b_2 - a_1) - z a_2 a_3 
\right\}
\nonumber\\
&&{}\times
{}_{3}F_{2} \left( \begin{array}{c|}
a_1,a_2,a_3 \\
b_1,b_2
\end{array} \, z \right)  \;,
\nonumber\\
\lefteqn{
{}_{3}F_{2} \left( \begin{array}{c|}
a_1,a_2,a_3 \\
b_1+1,b_2
\end{array}\, z \right) 
(a_1 - b_1)(a_2 - b_1) (a_3 - b_1)}
\nonumber \\
&=& 
b_1 \left\{
\frac{1-z}{z} \theta^2 
+ 
\left[ 
\frac{b_2 - 1}{z}  -  (a_1 + a_2 + a_3 - b_1)
\right] \theta \right.
\nonumber \\ 
&&{}
- \left.\frac{1}{b_1} \left[ a_1 a_2 a_3 - (a_1 - b_1)(a_2 - b_1) (a_3 - b_1)
\right]
\right\}
{}_{3}F_{2} \left( \begin{array}{c|}
a_1,a_2,a_3 \\
b_1,b_2
\end{array} \, z \right)  \;.
\end{eqnarray}

For the hypergeometric function ${}_4F_3$, the differential operators read:
\begin{eqnarray}
\lefteqn{
{}_{4}F_{3} \left( \begin{array}{c|}
a_1-1,a_2,a_3,a_4 \\
b_1,b_2,b_3
\end{array}\, z \right) 
(b_1 - a_1)
(b_2 - a_1)
(b_3 - a_1)}
\nonumber\\
&=& \left\{
(1 - z) \theta^3
+ 
\left[ 
(b_1 + b_2 + b_3 - 2 - a_1)  
- z (a_2 + a_3 + a_4)
\right] \theta^2 \right.
\nonumber \\
&&{}+ 
\left[ 
b_1 b_2 + b_1 b_3 + b_2 b_3 + a_1^2
+ 
(a_1+1) \left(1 - \sum_{j=1}^3 b_j\right)
- z (a_2 a_3 + a_2 a_4 + a_3 a_4 )
\right] \theta 
\nonumber \\ 
&&{}\left.+ (b_1 - a_1)(b_2 - a_1) (b_3 - a_1) 
- z a_2 a_3 a_4
\right\}
{}_{4}F_{3} \left( \begin{array}{c|}
a_1,a_2,a_3,a_4 \\
b_1,b_2,b_3
\end{array} \, z \right)  \;,
\nonumber
\\
\lefteqn{
{}_{4}F_{3} \left( \begin{array}{c|}
a_1,a_2,a_3,a_4 \\
b_1+1,b_2,b_3
\end{array}\, z \right) 
(a_1 - b_1)(a_2 - b_1) (a_3 - b_1) (a_4 - b_1)}
\nonumber \\
&=&
b_1 \left\{
\frac{1-z}{z} \theta^3 
+ 
\left[ 
\frac{b_2 + b_3 - 2}{z}  - (a_1 + a_2 + a_3 + a_4 - b_1)
\right] \theta^2 \right.
\nonumber \\
&&{}+
\left[ 
\frac{(b_2 - 1 )  (b_3 - 1 )}{z}  
- (a_1 a_2 + a_1 a_3 + a_1 a_4 + a_2 a_3 + a_2 a_4 + a_3 a_4)
+ b_1 \left(\sum_{j=1}^4 a_j - b_1\right)
\right] \theta
\nonumber \\
&&{}-\left.
\frac{1}{b_1} \left[ a_1 a_2 a_3 a_4 - (a_1 - b_1)(a_2 - b_1) (a_3 - b_1) (a_4 - b_1) 
\right]
\right\}
{}_{4}F_{3} \left( \begin{array}{c|}
a_1,a_2,a_3,a_4 \\
b_1,b_2,b_3
\end{array} \, z \right)  \;.
\end{eqnarray}

\subsection{Differential-reduction algorithms for special values of parameters}
\label{exceptional}

A special consideration is necessary when $b_i = a_i+1$.
In this case, the expressions in Eqs.~(\ref{diff:oper:1}) and
(\ref{diff:oper:2}) are equal to zero.
Let us start with the case when all $a_i$ are different.
In this case, the following relation should be applied 
(see Eq.~(15) in Chapter~5 of Ref.~\cite{rainville},
Chapter~7.2 in Ref.~\cite{PBM3}, and Ref.~\cite{reduction}):
\begin{eqnarray}
(a-b)\,
{}_{p}F_q\left(\begin{array}{c|}
a, b, \cdots \\
a+1, b+1, \cdots \end{array} \,z \right) 
&=& 
a\,
{}_{p-1}F_{q-1}\left(\begin{array}{c|}
b, \cdots \\
b+1, \cdots \end{array} \,z \right) 
\nonumber\\
&&{}- 
b\,{}_{p-1}F_{q-1}\left(\begin{array}{l|}
a, \cdots \\
a+1, \cdots \end{array} \,z \right) \;.
\label{special:1}
\end{eqnarray}
Repeating this procedure several times, we are able to split any original
hypergeometric function with several parameters having unit difference into a
set of hypergeometric functions with only one kind of parameters having unit
difference (for the particular cases, see Eqs.~(7.2.3.21)--(7.2.3.23) in
Ref.~\cite{PBM3}):
\begin{eqnarray}
\lefteqn{{}_{p}F_q\left(\begin{array}{c|}
\{a_1 \}_{r_1}, \{a_2 \}_{r_2}, \cdots, \{a_m \}_{r_m}, c_{1}, \cdots, c_{k} \\
\{1 + a_1 \}_{r_1}, \{1 + a_2 \}_{r_2}, \cdots, \{1 + a_m \}_{r_m}, b_{1}, \cdots , b_{l} \end{array} \,z \right)}
\nonumber \\
&\to& 
\sum_{i=1}^m 
{}_{p-R+r_i}F_{q-R+r_i}\left(\begin{array}{c|}
\{a_j \}_{r_i},  c_{1}, \cdots, c_{k} \\
\{1 + a_j \}_{r_i}, b_{1}, \cdots , b_{l}  \end{array} \,z \right) \;,
\label{special:2}
\end{eqnarray}
where $\{a_j \}_{r_i}$ denotes $r_j$ repetitions of $a_j$ in the argument list,
\begin{equation}
R = \sum_{j=1}^m r_j \;,\qquad
c_i \neq a_i \;, \qquad b_i \neq 1+a_i \;, \qquad c_j \neq 1+b_j \;.
\end{equation}
For a special set of parameters, Eqs.~(7.2.3.21) and (7.2.3.23) in
Ref.~\cite{PBM3} are useful:
\begin{eqnarray}
&& 
{}_{p}F_q\left(\begin{array}{c|}
\{a \}_{p-2}, \rho, \sigma \\
\{b \}_{q-2}, \rho+n, \sigma+1 \end{array} ~z \right) 
= 
\frac{(\rho)_n}{(\rho-\sigma)_n} 
{}_{p-1}F_{q-1}\left(\begin{array}{c|}
\{a \}_{p-2}, \sigma \\
\{b \}_{q-2}, \sigma+1 \end{array} ~z \right) 
\nonumber \\ && 
- 
\frac{(\rho)_n \sigma}{(\rho-\sigma)_n} 
\Sigma_{k=1}^n 
\frac{(\rho-\sigma-1)_k}{(\rho)_k}
{}_{p-1}F_{q-1}\left(\begin{array}{c|}
\{a \}_{p-2}, \rho \\
\{b \}_{q-2}, \rho+k \end{array} ~z \right) \;,
\\
&& 
{}_{p}F_q\left(\begin{array}{c|}
\{a \}_{p-n}, \sigma_1, \cdots, \sigma_n \\
\{b \}_{q-n}, \sigma_1  +  m_1, \cdots, \sigma_n+m_n \end{array} ~z \right) 
= 
\prod_{j=1}^n \frac{(\sigma_j)_{m_j}}{(m_j-1)!} 
\Sigma_{k=1}^n 
\Sigma_{j_1=0}^{m_1-1}
\cdots 
\Sigma_{j_n=0}^{m_n-1}
\frac{1}{\sigma_k+j_k}
\nonumber \\ && 
\times
\prod_{l=1}^n 
\frac{(1  -  m_l)_{j_l}}{j_l!}
\prod_{i=1; i \neq k}^n 
\frac{1}{\sigma_i  +  j_i  -  \sigma_k  -  j_k}
{}_{p-n+1}F_{q-n+1}\left(\begin{array}{c|}
\{a \}_{p-n}, \sigma_k+j_k \\
\{b \}_{q-n}, \sigma_k+j_k+1 \end{array} ~z \right) \;.
\label{7.2.23}
\end{eqnarray}
In Eq.~(\ref{7.2.23}), $m_n$ are integers, all $\sigma_i$ are different,
and, if $\sigma_i-\sigma_k=N$ with $N=1,2,\cdots$, then $m_k<N$.

Let us return to the last expression in Eq.~(\ref{special:2}) and rewrite it as
\begin{eqnarray}
{}_{p+1}F_{p}\left(\begin{array}{c|}
\{a  +  m\}_r,  a_{1} + k_1, \cdots, a_{p+1-r} + k_{p+1-r} \\
\{1  +  a  +  m\}_q, b_{1} + l_1, \cdots , b_{p-q} + l_{p-q}  \end{array} ~z \right) \;,
\label{basis}
\end{eqnarray}
where $m$, $\{k_r\}$, and $\{ l_j \}$ are integers and $a$, $\{a_k\}$, and
$\{b_j\}$ are parameters of the basis function. 
Using the reduction procedure described in Sec.~\ref{reduction:d}, we may
convert this function as 
\begin{eqnarray}
{}_{p+1}F_{p}\left(\begin{array}{c|}
\{a  +  m\}_r,  \{a_{j} + k_j\}_{p+1-r} \\
\{1  +  a  +  m\}_q, \{ b_{k} + l_k \}_{p-q}  \end{array} ~z \right) 
\to 
{}_{p+1}F_{p}\left(\begin{array}{c|}
\{a  +  m\}_r,  \{a_{j} + m\}_{p+1-r} \\
\{1  +  a  +  m\}_q, \{ b_{k} + m \}_{p-q}  \end{array} ~z \right) \;,\qquad
\end{eqnarray}
and then apply the differential relation (see Eq.~(7.2.3.47) in
Ref.~\cite{PBM3})\footnote{%
Another useful relation is Eq.~(7.2.3.50) in Ref.~\cite{PBM3}:
\begin{equation}
\left( 
\frac{d}{dz}
\right)^n 
\left[ 
z^{\sigma+n-1}
{}_{p}F_{q}\left(\begin{array}{c|}
\{a_i \}_{p-1},  \sigma \\
\{b_k \}_{q-1}  \end{array} ~z \right) 
\right]
= 
(\sigma)_n
z^{\sigma-1}
{}_{p}F_{q}\left(\begin{array}{c|}
\{a_i \}_{p-1},  \sigma+n\\
\{b_k \}_{q}  \end{array} ~z \right) \;.
\nonumber
\end{equation}
}
\begin{equation}
{}_{p}F_{q}\left(\begin{array}{c|}
\{a_i  +  m\}_p \\
\{b_k  +  m\}_q  \end{array} ~z \right) 
= 
\frac{\prod_{k=1}^q\{ (b_k)_m \} }{\prod_{j=1}^p \{(a_j)_m \}}
\left( 
\frac{d}{dz}
\right)^m 
{}_{p}F_{q}\left(\begin{array}{c|}
\{a_i \}_p\\
\{b_k \}_q \end{array} ~z \right) \;,
\label{32}
\end{equation}
where the derivative $d/(dz)$ could be rewritten in terms of $\theta$, with
the help of 
\begin{equation}
\left( 
\frac{d}{dz}
\right)^m 
= 
\left( 
\frac{1}{z}
\right)^m 
\theta
(\theta-1)
\cdots 
(\theta-m+1) \;. 
\end{equation}
If $m \geq p$, Eq.~(\ref{32}) can be converted to a differential identity of
order $p-1$.
Putting everything together, we obtain 
\begin{eqnarray}
&& 
P(a_k,b_j,z)
{}_{p+1}F_{p}\left(\begin{array}{c|}
\{a  +  m\}_r,  \{a_{j} + k_j\}_{p+1-r} \\
\{1  +  a  +  m\}_q, \{ b_{k} + l_k \}_{p-q}  \end{array} ~z \right) 
\nonumber \\ && 
= 
\left[
Q_1(a_k,b_j,z) \theta^p
+ 
\cdots
+ 
Q_{p+1}(a_k,b_j,z) 
\right]
{}_{p+1}F_{p}\left(\begin{array}{c|}
\{a\}_r,  a_{1}, \cdots, a_{p-r} \\
\{1 + a \}_q, b_{1}, \cdots , b_{p-q-1}  \end{array} ~z \right) \;.\qquad
\label{result}
\end{eqnarray}
We remark that Eq.~(\ref{result}) can be further simplified by using
Eq.~(\ref{universal:b}), which can be written as  
\begin{eqnarray}
\theta^k
{}_{p+1}F_{p}\left(\begin{array}{c|}
A,         \vec{a} \\
1  +  A, \vec{b}  \end{array} ~z \right) 
= 
(-A)^k 
{}_{p+1}F_{p}\left(\begin{array}{c|}
A,         \vec{a} \\
1  +  A, \vec{b}  \end{array} ~z \right) 
-
\sum_{j=0}^{k-1}
(-A)^{k-j} \theta^j 
{}_{p}F_{p-1}\left(\begin{array}{c|}
\vec{a} \\
\vec{b}  \end{array} ~z \right) 
\;.\qquad
\label{cr3a}
\end{eqnarray}
Recursive application of this expression allows us to reduce higher powers of
derivatives in Eq.~(\ref{result}) as
\begin{eqnarray}
\left( 
\frac{\theta}{A}
\right)^q
{}_{p+1}F_{p}\left(\begin{array}{c|}
\{A \}_r,        \vec{a} \\
\{1  +  A \}_r, \vec{b}  \end{array} ~z \right) 
= 
\sum_{j=0}^q (-1)^{(j+q)}
\left( q \atop j\right)
{}_{p+1-j}F_{p-j}\left(\begin{array}{c|}
\{A \}_{r-j},        \vec{a} \\
\{1  +  A \}_{r-j}, \vec{b}  \end{array} ~z \right) \;,\qquad
\label{cr3b}
\end{eqnarray}
where $q \leq r$. 

For a particular set of parameters, a further simplification of
Eq.~(\ref{result}) can be achieved. 
For example, for $a=1$, we have
\begin{equation}
{}_{p}F_q\left(\begin{array}{c|}
1, \{a_i \}_{p-1} \\
2, \{b_k \}_{q-1}
\end{array} ~z \right) 
= 
\frac{1}{z}
\frac{\prod_{l=1}^{q-1} (b_l - 1)}
     {\prod_{j=1}^{p-1} (a_j - 1)}
\left[
{}_{p-1}F_{q-1}\left(\begin{array}{c|}
\{a_i-1 \}_{p-1} \\
\{b_k-1 \}_{q-1}
\end{array} ~z \right) 
- 1
\right] \;,
\end{equation}
where $a_j,b_k \neq 1$.

Further useful relations for particular values of parameters of the basis
function were derived in Ref.~\cite{reduction}, namely
\begin{equation}
{}_{p+1}F_p\left(\begin{array}{c|}
\{a\}_p, b  \\
\{1+a\}_p \end{array} ~z \right) 
= 
-\frac{(-a)^p}{\Gamma(p)}
\int_0^1 dt \frac{t^{a-1}}{(1-zt)^b} \ln^{p-1} t \;,
\end{equation}
and, for $b=1$,
\begin{equation}
{}_{p+1}F_p\left(\begin{array}{c|}
1,a, \cdots, a \\
a+1, \cdots, a+1 \end{array} ~z \right) 
 = a^p \Phi(z,p,a) \;,
\end{equation}
where $\Phi(z,p,a)$ is the Lerch function 
defined as~\cite{bateman}
\begin{equation}
\Phi(z,p,a) = \sum_{k=0}^\infty \frac{z^k}{(a+k)^p} 
=  \frac{1}{\Gamma(p)} \int_0^\infty 
\frac{e^{-(a-1)t}}{e^t-z} t^{p-1}dt \;,
\end{equation}
so that 
\begin{equation}
\Li{n}{z} = z \Phi(z,n,1)\;.
\end{equation}

\subsection{Criteria of reducibility of hypergeometric functions}
\label{criteria} 

In this section, we formulate the criteria of reducibility
of the hypergeometric function ${}_{p}F_{q}(\vec{a};\vec{b}; z)$, i.e.\ we
state under which conditions the hypergeometric function 
${}_{p}F_{q}(\vec{a}; \vec{b}; z)$ and its derivatives are expressible in
terms of hypergeometric functions of lower order and/or with lower
derivatives.

We call the result derived by Karlsson \cite{karlsson} the first criterion
of reducibility of the hypergeometric function
${}_{p}F_{q}(\vec{a};\vec{b}; z)$ to hypergeometric functions of lower
order,
\begin{eqnarray}
&& 
{}_{p}F_q\left(\begin{array}{c|}
b_1+m_1, \cdots, b_n+m_n, a_{n+1}, \cdots , a_p \\
b_1, \cdots, b_n, b_{n+1}, \cdots , b_q \end{array} ~z \right) 
\nonumber \\ && 
= 
\sum_{j_1=0}^{m_1}
\cdots 
\sum_{j_n=0}^{m_n}
A(j_1, \cdots j_n) z^{J_n}
{}_{p-n}F_{q-n}\left(\begin{array}{c|}
a_{n+1}+J_n, \cdots a_p+J_n \\
b_{n+1}+J_n, \cdots , b_q+J_n \end{array} ~z \right) \; , 
\label{criterion1}
\end{eqnarray}
where $m_j$ are positive integers, $J_n = j_1 + \cdots +j_n$, and 
\begin{eqnarray}
A(j_1, \cdots j_n) 
&=& 
\left( m_1 \atop j_1 \right)
\cdots 
\left( m_n \atop j_n \right)
\frac{(b_2 + m_2)_{J_1} (b_3 + m_3)_{J_2} \cdots (b_n + m_n)_{J_{n - 1}} (a_{n + 1})_{J_n} \cdots (a_{p})_{J_n} }
     {(b_1)_{J_1} (b_2)_{J_2} \cdots (b_n)_{J_n} (b_{n + 1})_{J_n} \cdots (b_{q})_{J_n} } \;.
\nonumber \\
\end{eqnarray}
In particular, we have
\begin{eqnarray}
{}_{p}F_q \left( \begin{array}{c|}
b_1 + m_1,  a_2, \cdots,  a_{p} \\
b_1, b_2, \cdots b_{q} \\
\end{array}~z\right)
 =  
\sum_{j=0}^{m_1} z^j \left( m_1 \atop j \right)
\frac{(a_2)_j \cdots (a_p)_j}{(b_1)_j \cdots (b_q)_j} 
{}_{p-1}F_{q-1} \left( \begin{array}{c|}
a_2 + j, \cdots a_{p} + j \\
b_2 + j, \cdots b_{q} + j \\
\end{array}~z\right).
\nonumber
\\ 
\end{eqnarray}
In words, the first criterion of reducibility of the hypergeometric function
reads: \\
{\it Criterion I} \\
\boldmath
{\bf 
The hypergeometric function ${}_{p}F_{q}(\vec{a};\vec{b}; z)$ which has pairs
of parameters satisfying $a_i=b_i+m_i$ with $m_i$ being positive integers is
expressible in terms of functions of lower order according to
Eq.~(\ref{criterion1}).
}
\unboldmath

Equations~(\ref{special:1}) and (\ref{special:2}) yield the
second criterion of reducibility of the hypergeometric function
${}_{p}F_{q}(\vec{a};\vec{b}; z)$ to functions of lower order.
In its explicit form, it was derived in Ref.~\cite{reduction}
(see Eqs.~(18)--(20) in Ref.~\cite{reduction}).
Assuming that 
(i) $\{a_1, \cdots a_n \}$ are different and 
(ii) if $a_r-a_i=N$ with $N=1,2,\ldots$, then $m_i<N$, we have
(see also Eqs.~(7.2.3.21) and (7.2.3.23) in Ref.~\cite{PBM3})
\begin{eqnarray}
&& 
{}_{p}F_q\left(\begin{array}{c|}
a_1, \cdots a_n, a_{n+1} \cdots a_p \\
a_1 + 1 + m_1, \cdots, a_n + 1 + m_n,b_{n+1}, \cdots , b_{q} \end{array} ~z \right) 
\prod_{r=1}^n \frac{1}{(a_r)_{m_r+1}}
\nonumber \\ && \hspace{4mm}
= 
\sum_{i=1}^n 
\sum_{j=0}^{m_i} 
\frac{(-m_i)_j}{j!(a_i+j)m_i!}
\left( 
\prod_{r=1,r\neq i}^n
\frac{1}{(a_r-a_i-j)_{m_r+1}}
\right)
\nonumber \\ && \hspace{8mm}
\times
{}_{p-n+1}F_{q-n+1}\left(\begin{array}{c|}
a_i  +  j,a_{n+1}, \cdots, a_p \\
a_i  +  1  +  j, b_{n+1}, \cdots , b_{q}  \end{array} ~z \right) \;,
\label{criterion2}
\end{eqnarray}
so that we may formulate it as:\\
{\it Criterion II} \\
\boldmath
{\bf 
The hypergeometric function ${}_{p}F_{q}(\vec{a};\vec{b}; z)$ which has
two or more pairs of parameters satisfying $b_i=a_i+m_i+1$ with $m_i$ being
positive integers is expressible in terms of functions of lower order if
additional conditions on the parameters  $a_i$ (see Eq.~(\ref{criterion2}))
are satisfied.
}
\unboldmath

Equations~(\ref{cr3a}) and (\ref{cr3b}) are considered as the third criterion
of reducibility: \\
{\it Criterion III} \\
\boldmath
{\bf 
The result of the differential reduction of a hypergeometric function of the
type 
${}_{p+1}F_{p}(\vec{A}+\vec{m}, \vec{a}+\vec{k}; \vec{1}+\vec{A}+\vec{m},
\vec{b}+\vec{l}; z)$,
where $\vec{m}$, $\vec{k}$, and $\vec{l}$ are sets of integers, is expressible
in terms of the function
${}_{p+1}F_{p}(\vec{A}, \vec{a}; \vec{1}+\vec{A}, \vec{b}; z)$
and hypergeometric functions of lower order and their derivatives.
}
\unboldmath

Equation~(\ref{relation}) constitutes the fourth criterion of reducibility: \\
{\it Criterion IV} \\
{\bf 
If one of the upper parameters of a hypergeometric function is an integer,
the result of the differential reduction of this hypergeometric
function has one less derivative and is described by
Eq.~(\ref{decomposition:1}).
}

Criteria I--IV of reducibility of hypergeometric functions are much simpler
than the ordinary criterion of reducibility of Feynman diagrams
\cite{baikov:criterion}.

\boldmath
\subsection{All-order $\ep$ expansions of hypergeometric functions}
\unboldmath

Recently, several theorems on the structure of the coefficients of all-order
$\ep$ expansions of hypergeometric functions about integer and/or rational
values of their parameters have been proven
\cite{KWY07a,KWY07b,KWY07c,KK08a,nested,D,DK01,KKWY08a}. 
For a recent review, see Ref.~\cite{KKWY08a}. 

In this paper, we mainly consider hypergeometric functions of the type of
Eq.~(\ref{dk04}). 
There still does not exist a rigorous mathematical proof regarding the
structure of the coefficients of the all-order $\ep$ expansions of
hypergeometric functions of this type beyond the Gauss hypergeometric 
functions $_{2}F_1$ \cite{KWY07a,KK08a}. 
Through functions of weight 4, the coefficients of the $\ep$ expansions were
constructed in Ref.~\cite{DK04,MKL}, as discussed in details in
Appendix~\ref{appendix}), which is sufficient for next-to-next-to-leading-order
(two-loop) calculations. 
Recently, more coefficients of the $\ep$ expansions of the Clausen functions
$_{3}F_2$ have been evaluated in Ref.~\cite{HM}.

\boldmath
\subsection{Reduction at $z=1$ and construction of $\ep$ expansion}
\unboldmath
\label{z=1}

The value $z=1$ is a particular case of a ``hidden'' variable. 
It is evident that the application of Eq.~(\ref{diff}) to the r.h.s.\ of
Eq.~(\ref{decomposition}) gives rise to the generation of factors $1/(1-z)^k$,
so that the direct limit $z \to 1$ cannot be taken. 
Let us recall that the hypergeometric series defined by
Eq.~(\ref{hypergeometric}) converges \cite{bateman} for $|z|=1$, when
$ \re \sum_{j=1}^p b_j - \re \sum_{j=1}^{p+1} a_j > 0$, so that, 
if the l.h.s.\ of Eq.~(\ref{decomposition}) converges, the r.h.s.\ of the
equation does also exist.

The main idea is to convert the original hypergeometric function
${}_{p+1}F_{p}(\vec{a};\vec{b};z)$ to a function of argument $1-z$.
However, beyond type ${}_2F_1$, the analytical continuation of the
hypergeometric function
$\left. {}_{p+1}F_{p}(\vec{A};\vec{B};z) \right|_{z \to 1-z}$ is not
expressible in terms of functions of the same type, but has a more complicated
structure \cite{analytical}.
Nevertheless, we can perform the analytical continuation $z \to 1-z$ of the
coefficients of the $\ep$ expansions of hypergeometric functions entering the
r.h.s.\ of Eq.~(\ref{decomposition}).
It was shown in Ref.~\cite{hyper} that, under the transformation $z \to 1-z$,
hyperlogarithms are expressible again in terms of hyperlogarithms. 
In this way, if the coefficients of the $\ep$ expansion are expressible in
terms of hyperlogarithms, there is an opportunity to find the limit
$z \to 1-z$ of the differential reduction, but only in fixed orders of the 
$\ep$ expansions and only for such values of parameters of the hypergeometric
functions for which the analytical structures of the coefficients are known. 

An alternative approach to evaluate hypergeometric functions at $z=1$ was
discussed in Refs.~\cite{takayama,nested}.

\section{Application to Feynman diagrams}
\label{application}

\begin{figure}[th]
\centering{\vbox{\epsfysize=90mm \epsfbox{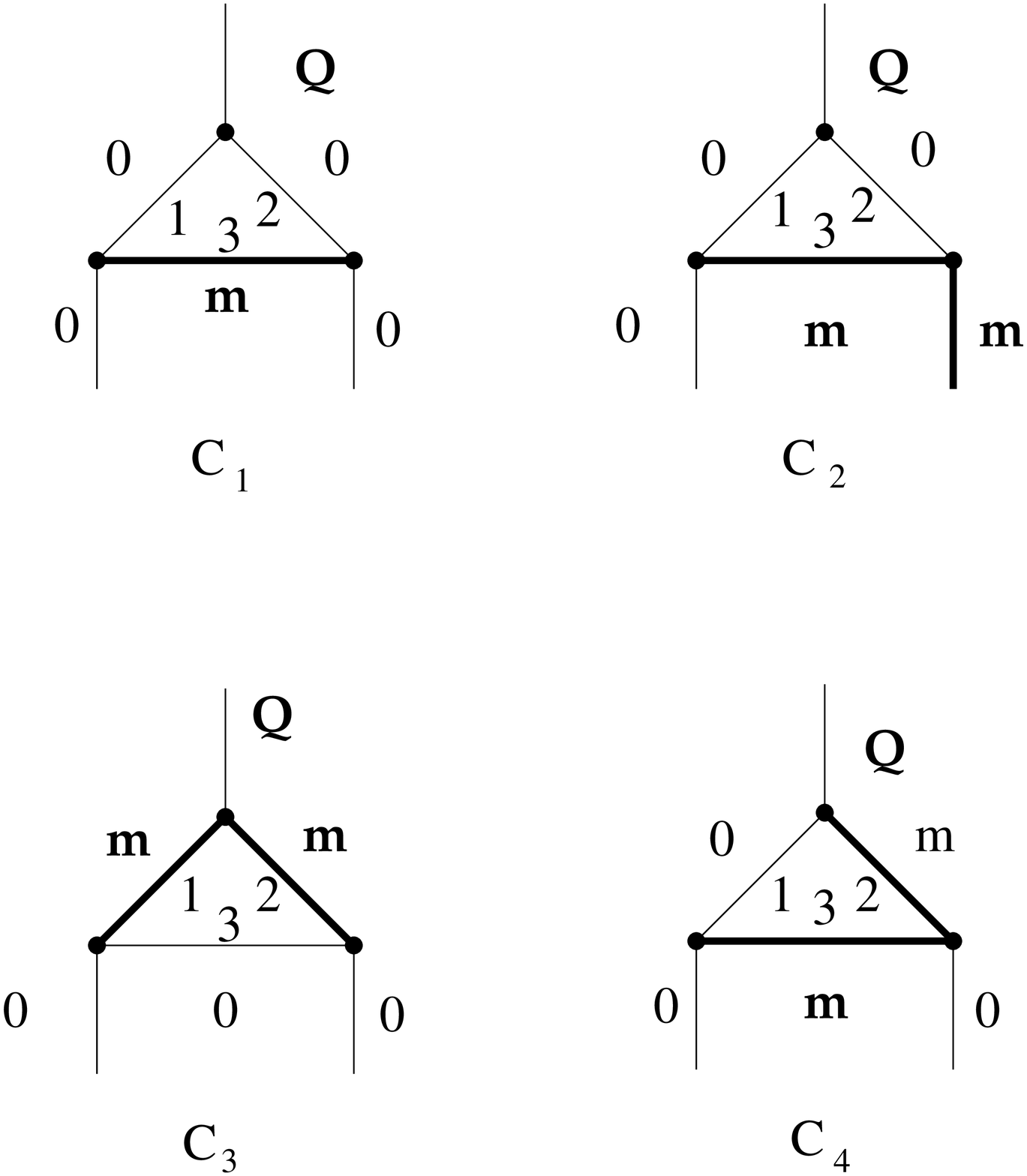}}}
\caption{\label{one-loop} 
One-loop vertex diagrams expressible in terms of generalized
hypergeometric functions.
Bold and thin lines correspond to massive and massless propagators,
respectively.}
\end{figure}

\begin{figure}[th]
\centering
{\vbox{\epsfysize=130mm \epsfbox{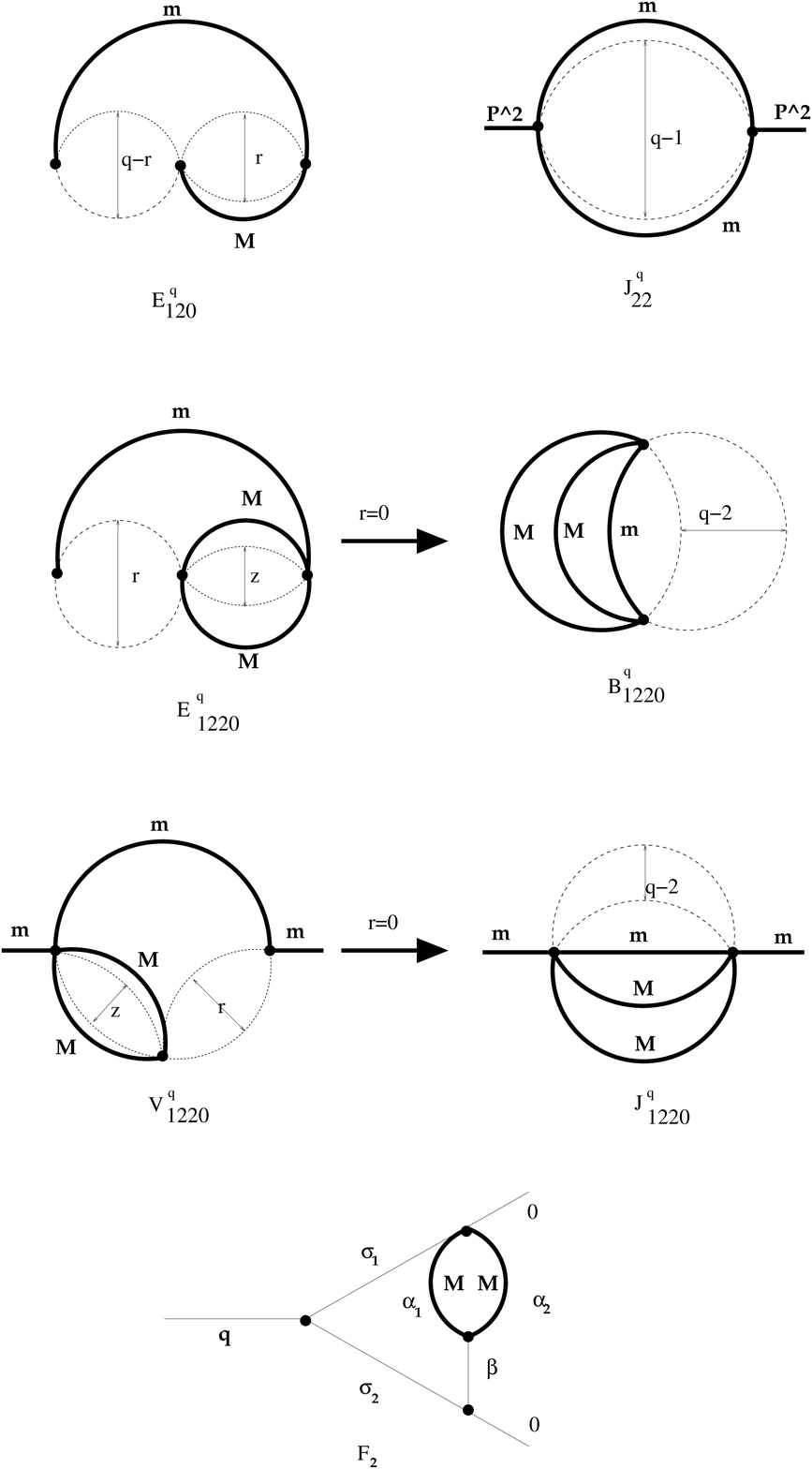}}}
\caption{
\label{diagrams}
Diagrams considered in the paper.
Bold and thin lines correspond to massive and massless propagators,
respectively.}
\end{figure}

As an illustration of the differential-reduction algorithm, let us consider
the diagrams shown in Figs.~\ref{one-loop} and \ref{diagrams}.\footnote{%
A few examples of the differential-reduction algorithm applied to the
reduction of Feynman diagrams were presented in Refs.~\cite{MKL06,Bytev}.}
In general, Feynman diagrams suffer from irreducible numerators.
Using the Davydychev-Tarasov algorithm \cite{Davydychev-Tarasov}, any tensor
integral may be represented in terms of scalar integrals with shifted
space-time dimensions and arbitrary (positive) powers of propagators.
In our analysis of the structures of the coefficients of the $\ep$ expansions,
we distinguish between two cases corresponding to even value $n=2m-2\ep$ and
odd value $n=2m-1-2\ep$ of space-time dimension, where $m$ is an integer. 


Since all diagrams shown in Fig.~\ref{diagrams} contain massless subloops, we
present for completeness the result for the $q$-loop massless sunset-type
propagator with $q+1$ massless lines.
It is given by
\begin{eqnarray}
J_{0}(p^2,\sigma_1, \sigma_2, \cdots, \sigma_{q+1}) & = & 
\int \frac{d^n (k_1 k_2 \cdots k_q)}
{[k_1^2]^{\sigma_1} [k_2^2]^{\sigma_2} \cdots 
[k_q^2]^{\sigma_q} [(k_1 + k_2 + \cdots + k_q + p)^2]^{\sigma_{q+1}}}
\nonumber \\ \hspace{-5mm}
& = &
\left\{ \prod_{k=1}^{q+1} \frac{\Gamma(\frac{n}{2}-\sigma_k)}{\Gamma(\sigma_k)} \right\}
\frac{
\left[ i^{1-n} \pi^{n/2} \right]^q
\Gamma\left(\sigma-\frac{n}{2}q\right)
}{
\Gamma\left(\frac{n}{2}(q+1) - \sigma \right)} 
(p^2)^{\frac{n}{2}q-\sigma}
\; ,\quad
\label{massless}
\end{eqnarray}
where
\begin{equation}
\sigma = \sum_{k=1}^{s} \sigma_k,
\label{eq:sigma}
\end{equation}
with $s=q+1$.
Some of the massless lines can be dressed by insertions of massless chains.
In this case, the respective ${\sigma_k}$ can be represented as
$\sigma_k = r_k - R_k \frac{n}{2}$, where $r_k$ and $R_k$ and integers.

\subsection{One-loop vertex: particular cases}
\label{Published}

We wish to remind the reader that any one-loop vertex diagram with
arbitrary masses, external momenta, and powers of propagators can be reduced
by recurrence relations, derived with the help of the integration-by-parts
technique \cite{ibp}, to a vertex master integral plus propagator
master integrals and bubble integrals (with all powers of propagators being
equal to unity), or, in the case of zero Gram and/or Cayley determinants, to
a linear combination of propagator and bubble integrals.
In terms of hypergeometric functions, vanishing Gram and/or Cayley determinants
correspond to the situation where the hypergeometric function describing the
original vertex diagram can be reduced to a Gauss hypergeometric function with
the following sets of parameters \cite{DK01,BD,BDST}: 
\begin{equation}
{}_{2}F_1\left(\begin{array}{c|}
1, I_1  -  \tfrac{n}{2}, \\
I_2 + \tfrac{n}{2} \end{array} ~z \right) \;, 
\quad 
{}_{2}F_1\left(\begin{array}{c|}
1, I_1  -  \tfrac{n}{2}, \\
I_2 + \tfrac{3}{2} \end{array} ~y \right) \;, 
\end{equation}
where $\{I_a\}$ are arbitrary integers. 
In the case of non-zero Gram and/or Cayley determinants, the one-loop vertex
master integral is expressible as a linear combination of Gauss hypergeometric
functions and Appell functions $F_1$ \cite{1loop}.

For the diagrams shown in Fig.~\ref{one-loop}, both the Gram and Cayley
determinants are non-zero. 
The hypergeometric representation of the corresponding master integral was
published in Ref.~\cite{KKWY08a}.
\begin{itemize}
%
%
\item
Diagram $C_1$ with arbitrary powers of propagators is expressible in terms
of two hypergeometric functions ${}_3F_2$ as
(see also Eq.~(3.44) in Ref.~\cite{AGO2})
\begin{eqnarray}
&& 
\frac{C_1}{i^{1-n} \pi^{\tfrac{n}{2}}} = 
(-m^2)^{\tfrac{n}{2} - j_{123}}
\nonumber \\ && \hspace{5mm}
\Biggl\{
\frac{\Gamma\left( j_{123}  -  \frac{n}{2} \right) \Gamma\left( \frac{n}{2}  -  j_{12} \right)}
     { \Gamma\left( \frac{n}{2} \right) \Gamma(j_3)}
{}_{3}F_2\left(\begin{array}{c|}
j_{123}  -  \tfrac{n}{2}, j_1, j_2 \\
\tfrac{n}{2}, 1  +  j_{12}  -  \tfrac{n}{2}  \end{array} ~-\frac{Q^2}{m^2} \right) 
\nonumber \\ && \hspace{5mm}
+ \left( - \frac{Q^2}{m^2}\right)^{\tfrac{n}{2} - j_{12}}
\frac{\Gamma\left( \frac{n}{2}  -  j_1\right) 
      \Gamma\left( \frac{n}{2}  -  j_2\right) 
      \Gamma\left( j_{12}  -  \frac{n}{2} \right)}
     { \Gamma\left( n  -  j_{12} \right) \Gamma(j_1) \Gamma(j_2)}
\nonumber \\ && \hspace{10mm}
\times
{}_{3}F_2\left(\begin{array}{c|}
j_3, \tfrac{n}{2} - j_1, \tfrac{n}{2} - j_2 \\
n - j_{12}, \tfrac{n}{2}  -  j_{12}  +  1 \end{array} ~-\frac{Q^2}{m^2} \right) 
\Biggr\} \;.
\label{eq:c1}
\end{eqnarray}
Here and in the following, we use the short-hand notations
$j_{ab} = j_a+j_b$ and $j_{abc} = j_a + j_b + j_c$ for compactness.
In accordance with the differential-reduction algorithm, 
each of the two ${}_3F_2$ functions in Eq.~(\ref{eq:c1}) is expressible in
terms of  a ${}_{2}F_1$ function with one unit upper parameter,
namely 
\begin{equation}
{}_{2}F_1\left(\begin{array}{c|}
1, 1 \\
I_1 + \tfrac{n}{2} \end{array} ~z \right) \;, 
\quad 
{}_{2}F_1\left(\begin{array}{c|}
1, I_1  +  \tfrac{n}{2} \\
I_2 + n \end{array} ~z \right) \;,
\end{equation}
respectively, plus rational functions of $z$.
Standard approaches yield one vertex master integral plus massive bubble and
massless propagator integrals.
The latter two types of integrals are expressible in terms of products of
Gamma functions and correspond to the rational functions in our approach.
%
%
\item
For diagram $C_2$ with arbitrary powers of propagators, the result is 
\begin{eqnarray}
&& 
\frac{C_2}{i^{1-n} \pi^{\tfrac{n}{2}}} = 
(-m^2)^{\tfrac{n}{2} - j_{123}}
\nonumber \\ && \hspace{5mm}
\Biggl\{
\frac{\Gamma\left( j_{123}  \!-\!  \frac{n}{2} \right) 
      \Gamma\left( \frac{n}{2}  \!-\!  j_{12} \right)
      \Gamma\left( n  \!-\!  j_{13}  \!-\!  2 j_2\right)}
     {\Gamma\left( n  \!-\!  j_{123} \right)
      \Gamma\left( \frac{n}{2}  \!-\!  j_2 \right) \Gamma(j_3)}
{}_{3}F_2\left(\begin{array}{c|}
j_{123}  \!-\!  \tfrac{n}{2}, j_1, j_2 \\
1  \!+\!  j_{12}  \!-\!  \frac{n}{2},  
1  \!+\!  j_{13}  \!+\!  2 j_2 \!-\!  n  \end{array} ~\frac{Q^2}{m^2} \right) 
\nonumber \\ && \hspace{5mm}
+ \left( - \frac{Q^2}{m^2}\right)^{\tfrac{n}{2} - j_{12}}
\frac{\Gamma\left( \frac{n}{2}  -  j_1\right) 
      \Gamma\left( j_{12}  -  \frac{n}{2} \right)
     \Gamma\left( \frac{n}{2}  -  j_{23} \right)}
     { \Gamma\left( n  -  j_{123} \right) \Gamma(j_1) \Gamma(j_2)}
\nonumber \\ && \hspace{10mm}
\times
{}_{3}F_2\left(\begin{array}{c|}
j_3, \tfrac{n}{2} - j_2,  \tfrac{n}{2} - j_1 \\ 
\tfrac{n}{2}  -  j_{12}  +  1, 1  +  j_{23}  -  \tfrac{n}{2} 
\end{array} ~\frac{Q^2}{m^2} \right) 
\Biggr\} \;.
\label{eq:c2}
\end{eqnarray}
Similarly to the previous case, each of the two ${}_{3}F_2$ functions in
Eq.~(\ref{eq:c2}) is expressible in terms of a ${}_{2}F_1$ function with one
unit upper parameter, namely
\begin{equation}
{}_{2}F_1\left(\begin{array}{c|}
1, 1 \\
I_1 - n \end{array} ~z \right) \;, 
\quad 
{}_{2}F_1\left(\begin{array}{c|}
1, I_1  +  \tfrac{n}{2}, \\
I_2 - \tfrac{n}{2} \end{array} ~z \right) \;,
\end{equation}
respectively, plus rational functions.
Standard approaches yield one vertex master integral plus massive bubble and
massless propagator integrals.
%
%
\item
Diagram $C_3$ with arbitrary powers of propagators is expressible in terms 
of one $_4F_3$ function as 
\begin{eqnarray}
&& 
\frac{C_3}{i^{1-n} \pi^{\tfrac{n}{2}}} = 
(-m^2)^{\tfrac{n}{2} - j_{123}}
\frac{\Gamma\left( j_{123}  -  \frac{n}{2} \right) 
      \Gamma\left( \frac{n}{2}  -  j_{3} \right)}
     {\Gamma\left( j_{12} \right)
      \Gamma\left( \frac{n}{2} \right) }
{}_{4}F_3\left(\begin{array}{c|}
j_{123}  -  \tfrac{n}{2}, j_1, j_2, \tfrac{n}{2}  -  j_3 \\
\tfrac{n}{2},  \tfrac{j_{12}}{2}, \tfrac{j_{12} + 1}{2} \end{array} ~\frac{Q^2}{4m^2} \right) 
\;.
\nonumber \\ 
\end{eqnarray}
In accordance with the differential-reduction algorithm, this function may be
written in terms of a $_3F_2$ function with one unit upper parameter and its
first derivative, 
\begin{equation}
\{ 
1, \theta
\}
\times 
{}_{3}F_2\left(\begin{array}{c|}
1, I_1-\tfrac{n}{2}, \tfrac{n}{2} - 1 + I_2 \\
\tfrac{n}{2} + I_2, \tfrac{1}{2} + I_3 \end{array} ~z \right) \;,
\end{equation}
and a rational function.
Standard approaches yield one vertex and one propagator master integral plus
massive bubble integrals.
%
%
\item
Diagram $C_4$ with arbitrary powers of propagators is expressible in terms of
one $_3F_2$ function as
\begin{eqnarray}
&& 
\frac{C_4}{i^{1-n} \pi^{\tfrac{n}{2}}} = 
(-m^2)^{\tfrac{n}{2} - j_{123}}
\frac{\Gamma\left( j_{123}  -  \frac{n}{2} \right) 
      \Gamma\left( \frac{n}{2}  -  j_{1} \right)}
     {\Gamma\left( j_{23} \right)
      \Gamma\left( \frac{n}{2} \right) }
{}_{3}F_2\left(\begin{array}{c|}
j_{123}  -  \frac{n}{2}, j_1, j_2 \\
\frac{n}{2},  j_{23} \end{array} ~\frac{Q^2}{m^2} \right) 
\;.\qquad
\end{eqnarray}
In accordance with the differential-reduction algorithm, this function may be
written in terms of a $_3F_2$ function with one unit upper parameter and its
first derivative, 
\begin{equation}
\{ 
1, \theta
\}
\times 
{}_{3}F_2\left(\begin{array}{c|}
1, I_1, I_2-\tfrac{n}{2} \\
I_1+1, \tfrac{n}{2} + I_2 \end{array} ~z \right) \;,
\end{equation}
and a rational function.
Standard approaches yield one vertex and one propagator master integral plus
massive bubble integrals.
\end{itemize}
\boldmath
\subsection{$E^{q}_{120}$}
\unboldmath

Let us consider the $q$-loop bubble diagram $E^q_{120}$ depicted in
Fig.~\ref{diagrams} with two massive lines of different masses and $q-r$ plus
$r$ massless propagators,
\begin{eqnarray}
&& \hspace{-5mm}
E^q_{120}(m^2, M^2, \alpha, \beta, \sigma_1, \cdots, \sigma_{q-r}, \rho_1, \cdots, \rho_r) = 
\nonumber \\ && \hspace{-5mm}
\int \frac{d^n (k_1 \cdots k_q)} 
{[k_1^2]^{\sigma_1} \cdots [k_{q-r-1}^2]^{\sigma_{q-r-1}} 
[(k_1  +  k_2  +  k_{q-r-1}+k_{q})^2]^{\sigma_{q-r}} 
[k_{q-1}^2-M^2]^{\alpha} [k_{q}^2-m^2]^{\beta} }
\nonumber \\ && \hspace{0mm}
\times
\frac{1}
{
[k_{q-r}^2]^{\rho_1} \cdots [(k_{q-r} + \cdots k_{q-1}+k_q)^2)]^{\rho_{r}}
}\; .
\label{eq:e120q} 
\end{eqnarray}
The special case $q=r$ was analyzed in Ref.~\cite{MKL06}.
The first non-trivial diagram of this type corresponds to a three-loop bubble
($q=3$) with $r=1$.
The Mellin-Barnes representation of Eq.~(\ref{eq:e120q}) reads: 
\begin{eqnarray}
&& \hspace{-5mm}
E^q_{120}(m^2, M^2, \alpha, \beta, \sigma_1, \cdots, \sigma_{q-r}, \rho_1, \cdots, \rho_r) = 
\frac{
(-M^2)^{\tfrac{n}{2}r \!-\! \alpha  \!-\!  \rho}
(-m^2)^{\tfrac{n}{2}(q-r) \!-\! \sigma \!-\! \beta }
}{
\left[ i^{1-n} \pi^\frac{n}{2} \right]^{-q}
\Gamma(\alpha) \Gamma(\beta) \Gamma\left(\frac{n}{2} \right)
}
\nonumber \\ && \hspace{-5mm}
\left\{ \prod_{j=1}^{q-r} \frac{\Gamma(\tfrac{n}{2} - \sigma_j)}{\Gamma(\sigma_j)} \right\}
\left\{ \prod_{k=1}^{r} \frac{\Gamma(\tfrac{n}{2} - \rho_k)}{\Gamma(\rho_k)} \right\}
\frac{ \Gamma\left(\sigma-\tfrac{n}{2}(q - r - 1)\right)}
     { \Gamma\left(\tfrac{n}{2}(q - r) - \sigma\right)}
\int dt 
\left( \frac{m^2}{M^2}\right)^t 
\nonumber \\ && \hspace{-5mm}
\times
\frac{\Gamma(-t) 
      \Gamma\left(\rho \!-\! \tfrac{n}{2}(r \!-\! 1) \!+\! t\right) 
      \Gamma\left(\alpha \!+\! \rho \!-\! \tfrac{n}{2}r \!+\! t \right) 
      \Gamma\left( \tfrac{n}{2}(q \!-\! r)  \!-\!  \sigma  \!+\! t \right) 
      \Gamma\left( \sigma  \!+\!  \beta  \!-\!  \tfrac{n}{2}(q \!-\! r)  \!-\! t \right) 
      }
     {\Gamma\left(\tfrac{n}{2}+t\right)} \;,
\nonumber\\
&&\label{MB:E120}
\end{eqnarray}
where $\sigma$ is defined by Eq.~(\ref{eq:sigma}) with $s=q-r$ and
\begin{equation}
\rho =  \sum_{k=1}^{r} \rho_k \;.
\label{eq:rho}
\end{equation}
Closing the contour of integration in Eq.~(\ref{MB:E120}) on the left, we
obtain in the notation of Ref.~\cite{BD}:
\begin{eqnarray}
&& \hspace{-5mm}
E^q_{120}(m^2, M^2, \alpha, \beta, \sigma_1, \cdots, \sigma_{q-r}, \rho_1, \cdots, \rho_r) = 
\left[ i^{1-n} \pi^\frac{n}{2} \right]^q
\frac{
(-M^2)^{\tfrac{n}{2}q - \alpha  -  \beta  -  \sigma  -  \rho}
}{\Gamma(\alpha) \Gamma(\beta) \Gamma\left(\frac{n}{2} \right)}
\nonumber \\ && \hspace{-5mm}
\left\{ \prod_{j=1}^{q-r} \frac{\Gamma(\tfrac{n}{2}-\sigma_j)}{\Gamma(\sigma_j)} \right\}
\left\{ \prod_{k=1}^{r} \frac{\Gamma(\tfrac{n}{2}-\rho_k)}{\Gamma(\rho_k)} \right\}
\frac{ \Gamma\left(\sigma-\tfrac{n}{2}(q - r - 1)\right)}{ \Gamma\left(\tfrac{n}{2}(q - r) - \sigma \right)}
\nonumber \\ && \hspace{-5mm}
\times
%
%
\Biggl\{
\frac{
 \Gamma(\beta) 
\Gamma\left( \tfrac{n}{2}(q - r)  -  \sigma  -  \beta \right) 
\Gamma\left( \rho  +  \beta  +  \sigma  -  \tfrac{n}{2}(q - 1) \right) 
\Gamma\left( \alpha  +  \rho  +  \beta  +  \sigma  -  \tfrac{n}{2}q \right) 
}
{
\Gamma\left( \sigma  +  \beta  -  \tfrac{n}{2}(q - r - 1) \right) 
}
\nonumber \\ && 
\times
~_{3}F_2\left(\begin{array}{c|}
\beta, 
\beta + \sigma + \rho  -  \frac{n}{2}(q - 1), 
\alpha + \beta + \sigma + \rho  -  \frac{n}{2}q \\
\sigma + \beta - \frac{n}{2}(q - r - 1),
1 + \sigma + \beta - \frac{n}{2}(q - r)
\end{array} ~\frac{m^2}{M^2} \right)
\nonumber \\ && \hspace{-5mm}
%
%
+ 
\left( \frac{m^2}{M^2} \right)^{\tfrac{n}{2}(q-r)  \!-\!  \sigma  \!-\!  \beta}
\frac{
\Gamma\left( \tfrac{n}{2}(q\!-\!r)  \!-\!  \sigma \right) 
\Gamma\left( \sigma  \!+\!  \beta  \!-\!  \tfrac{n}{2}(q \!-\! r) \right) 
\Gamma\left( \alpha  \!+\!  \rho  \!-\!  \tfrac{n}{2}r \right) 
\Gamma\left(  \rho  \!-\!  \tfrac{n}{2}(r\!-\!1) \right) 
}
{
\Gamma\left( \tfrac{n}{2} \right) 
}
\nonumber \\ && 
\times
~_{3}F_2\left(\begin{array}{c|}
\rho  \!-\!  \tfrac{n}{2}(r \!-\! 1),
\alpha  \!+\!  \rho  \!-\!  \tfrac{n}{2}r,
\tfrac{n}{2}(q \!-\! r)  \!-\!  \sigma \\
\tfrac{n}{2}, 
1  +  \tfrac{n}{2}(q - r)  -  \sigma  -  \beta, 
\end{array} ~\frac{m^2}{M^2} \right)
\Biggr\} \;. 
\label{E_Q_120}
\end{eqnarray}
Let us discuss the differential reduction of the two hypergeometric functions
in Eq.~(\ref{E_Q_120}) assuming that $\beta$, $\sigma_j$ ($\sigma \geq 2$),
and $\rho_k$ are integers. 
We distinguish between two cases: $r=1$ ($q \geq 3$) and $r \geq 2$
($q \geq 4$). 
For $r=1$, both hypergeometric functions in Eq.~(\ref{E_Q_120}) are
reducible to ${}_2F_1$ functions with one unit upper parameter, namely 
\begin{equation}
{}_{2}F_1\left(\begin{array}{c|}
1, I_1 - \tfrac{n}{2}q \\
I_2 - \tfrac{n}{2}(q-2) \end{array} ~z \right) \;, 
\quad 
{}_{2}F_1\left(\begin{array}{c|}
1, I_1  -  \tfrac{n}{2}, \\
I_2 + \tfrac{n}{2} \end{array} ~z \right) \;,
\end{equation}
respectively.
Standard approaches yield one three-loop bubble master integral and
integrals expressible in terms of Gamma functions. 
For $r \geq 2$, the first hypergeometric function is expressible in terms of a
${}_3F_2$ function with one unit upper parameter, and the second one is
reducible to a ${}_2F_1$ function with both upper parameters containing
non-zero $\ep$ parts, namely
\begin{equation}
(1, \theta) \times
{}_{3}F_2\left(\begin{array}{c|}
1, I_1  \!-\!  \tfrac{n}{2}(q \!-\! 1), I_2  \!-\!  \tfrac{n}{2}q \\
I_3 \!-\! \tfrac{n}{2}(q \!-\! r \!-\! 1), I_4 \!-\! \tfrac{n}{2}(q \!-\! r) \end{array} ~z \right) \;,
\quad
(1, \theta) \times
{}_{2}F_1\left(\begin{array}{c|}
I_1 \!-\! \tfrac{n}{2}(r \!-\! 1), I_2 \!-\! \tfrac{n}{2}r \\
I_3 \!+\! \tfrac{n}{2} \end{array} ~z \right) \;, 
\end{equation}
respectively.
For $q=3$ and $r=1$, we present the explicit form of the master integral, with 
the powers of propagator being all equal to unity, in dimension $n=4-2\ep$.
It reads:
\begin{eqnarray}
&& \hspace{-5mm}
E^3_{120}(m^2, M^2, 1,1,1,1,1) = 
\left[ i \pi^{2-\ep} \right]^3
(M^2)^{1-3\ep}
\frac{
\Gamma^2(1-\ep)
\Gamma\left(1+2\ep \right)
\Gamma\left(1+\ep \right)
}{
{2\ep^3(1-2\ep)(1-\ep)^2}
} \times
\nonumber \\ && \hspace{-9mm}
%
%
\Biggl\{
\frac{(1\!-\!\ep)}{3(1\!-\!3\ep)}
\frac{\Gamma(1\!+\!3\ep)}{\Gamma(1\!+\!\ep)} 
~_{2}F_1\left(\begin{array}{c|}
1, -1 \!+\! 3\ep \\
1 \!+\! \ep
\end{array} ~\frac{m^2}{M^2} \right)
%
%
+ 
\left( \frac{m^2}{M^2} \right)^{1-2\ep}
\frac{\Gamma(1\!+\!\ep)}{\Gamma(1\!-\!\ep)} 
~_{2}F_1\left(\begin{array}{c|}
1, \ep \\
2  \!-\!  \ep
\end{array} ~\frac{m^2}{M^2} \right)
\Biggr\} \;. 
\nonumber \\
\label{eq:e1203}
\end{eqnarray}
The two Gauss hypergeometric functions in Eq.~(\ref{eq:e1203}) can be reduced
to the basis functions considered in Refs.~\cite{KWY07a,DK04,MKL06} by the
following relations: 
\begin{eqnarray}
&& 
(1-2\ep)
~_{2}F_1\left(\begin{array}{c|}
1, \ep \\
2  -  \ep
\end{array} ~z \right)
 = 1 - \ep 
- \ep (1-z)
~_{2}F_1\left(\begin{array}{c|}
1, 1+\ep \\
2  -  \ep
\end{array} ~z \right) \;,
\nonumber \\ && 
(1-2\ep)
~_{2}F_1\left(\begin{array}{c|}
1, -1  +  3\ep \\
1  +  \ep
\end{array} ~z \right)
= 1  -  2 \ep 
 -  z (1 - 3\ep) 
\nonumber \\ && 
\hspace{35mm}
+ 
\frac{3\ep(1-3\ep)z(1-z)}{1+\ep}
~_{2}F_1\left(\begin{array}{c|}
1, 1+3\ep \\
2  +  \ep
\end{array} ~z \right) \;.
\end{eqnarray}
For illustration, we present here the first few coefficients of the $\ep$
expansion of Eq.~(\ref{eq:e1203}):
\begin{eqnarray}
&& \hspace{-5mm}
\frac{E^3_{120}(m^2, M^2, 1,1,1,1,1)}{\left[ i \pi^{2-\ep} \right]^3 \Gamma^3(1 + \ep)} = 
(M^2)^{1-3\ep}
\Biggl( 
\frac{1+2z}{6\ep^3} 
+ 
\frac{3+z\left[ 5-3\ln z )\right]}{3\ep^2} 
\nonumber \\ && 
+ 
\frac{1}{\ep} 
\left\{ 
\frac{25+34z}{6}  
 -  
(1 - z) 
\left[ \Li{2}{z}  +  \ln(z)\ln(1 - z) \right]  
 +  z \ln z (\ln z -  5)   +  \zeta_2
\right\}
\nonumber \\ &&  
+ (1-z) \left[ 
2 \Snp{1,2}{z} 
+ 2 \left( \Li{2}{z}  -  \zeta_2 \right) \ln(1-z) 
+ \Li{2}{z} \ln z
\right]
\nonumber \\ &&  
+ (1 - z) \left\{ 
\ln(1 - z) \ln z \left[ \ln z  +  \ln(1 - z) \right]
- 5 \left[ \Li{2}{z}  +  \ln(z)\ln(1 - z) \right]  
\right\}
\nonumber \\ &&  
+ \frac{45+49z}{3}
- 17 z \ln z 
+ 5 z \ln^2 z 
+ 2 \left( 3  -   z \ln z \right) \zeta_2 
- \frac{5-2z}{3} \zeta_3 
- \frac{2}{3} z \ln^3 z 
\nonumber \\ &&  
+ \ep (1 -z) \Biggl\{ 
\Li{2}{z} 
\left[ 
2 \ln(1 - z)  
\left( 
5 
 -  \ln (1 - z) 
 -  \ln z 
\right) 
 -  \ln z (\ln z  -  5)
 -  4  \zeta_2 
 -  17 
\right]
\nonumber \\ &&  
\hspace{20mm}
+ \Li{3}{z} \ln z 
- \Li{4}{z}
+ 2 \Snp{1,2}{z} 
\left[
 5 
- 2 \ln(1 - z) 
- \ln z 
\right]
- 4\Snp{1,3}{z}
\nonumber \\ &&  
\hspace{20mm}
+ 2 \ln(1 - z) \zeta_2 
\left[ 
\ln (1 - z) 
 -  5
 -  \ln z 
\right]
+ 4 \ln(1 - z) \zeta_3 
\nonumber \\ &&  
\hspace{20mm}
+ \ln z \ln(1 - z)
\left[ \ln(1 - z)  +  \ln z \right]
\left[ 5  -  \frac{2}{3} ( \ln(1 - z)  +  \ln z ) \right]
\nonumber \\ &&  
\hspace{20mm}
+ \ln z \ln(1 - z)
\left[
\frac{1}{3} \ln z \ln(1 - z)
 -  17 
\right]
\Biggr\}
\nonumber \\ &&  
+ 
\ep 
\Biggl\{ 
z \ln z 
\left( 
\frac{1}{3} \ln^2 z ( \ln z  -  10 ) 
 +  17 \ln z
+ 2 \zeta_2 (\ln z  -  5) 
 +  2\zeta_3 
 -  49 
\right) 
\nonumber \\ &&  
\hspace{20mm}
+ \frac{301}{6}  +  43 z
 +  25 \zeta_2 
 -  \frac{10}{3} (3 - z) \zeta_3 
 +  \frac{1}{2} \zeta_4 (23 - 14z)
\Biggr\}
+ {\cal O}(\ep)
\Biggr) \;,
\label{eq:e1203e}
\end{eqnarray}
where
\begin{equation}
z=\frac{m^2}{M^2}.
\label{eq:z}
\end{equation}
To check Eq.~(\ref{eq:e1203e}), we evaluate the first few coefficients of the
$\ep$ expansion of the original diagram in the large-mass limit \cite{heavy}
using  the program packages developed in Refs.~\cite{avdeev,heavy:package}
to find agreement.
\boldmath
\subsection{$J^{q}_{22}$}
\unboldmath
\label{J22}

Let us consider the $q$-loop sunset diagram $J^q_{22}$ in Fig.~\ref{diagrams}
with two massive lines of the same mass $m$ and $q-2$ massless subloops, which
is one of the most frequently studied Feynman diagrams.
It is defined as
\begin{eqnarray}
&& \hspace{-5mm}
J^q_{22}(m^2, p^2, \alpha_1, \alpha_2, \sigma_1, \cdots, \sigma_{q-1}) = 
\nonumber \\ && \hspace{-5mm}
\int \frac{d^n (k_1 \cdots k_q)}{[k_1^2]^{\sigma_1} \cdots [k_{q-1}^2]^{\sigma_{q-1}} 
[k_q^2-m^2]^{\alpha_1} [(k_1  +  k_2  +  \cdots  +  k_q+p)^2-m^2]^{\alpha_2} } \; . 
\label{eq:J22q}
\end{eqnarray}
The Mellin-Barnes representation of Eq.~(\ref{eq:J22q})
reads:\footnote{\label{foot:mb}%
In the one-loop case ($q=1$), the factor
$\left\{\prod_{k=1}^{q-1}\frac{\Gamma(\frac{n}{2}-\sigma_k)}{\Gamma(\sigma_k)}
\right\}$
is equal to unity, and the representation of Eq.~(\ref{MB:J22}) agrees with the
one in Ref.~\cite{BD}.}
\begin{eqnarray}
&& \hspace{-5mm}
J^q_{22}(m^2, p^2, \alpha_1, \alpha_2, \sigma_1, \cdots, \sigma_{q-1}) = 
\frac{
(-m^2)^{\tfrac{n}{2} - \alpha_{1,2}}
(p^2)^{\tfrac{n}{2}(q-1) - \sigma}
}{
\left[ i^{1-n} \pi^{n/2} \right]^{-q} 
\Gamma(\alpha_1) \Gamma(\alpha_2)}
\left\{ \prod_{k=1}^{q-1} \frac{\Gamma(\frac{n}{2}-\sigma_k)}{\Gamma(\sigma_k)} \right\}
\nonumber \\ && \hspace{-5mm}
\times
\int dt 
\left( -\frac{p^2}{m^2}\right)^t 
\frac{\Gamma(\alpha_1  +  t)
      \Gamma(\alpha_2  +  t)
      \Gamma(\alpha_{1,2}  -  \tfrac{n}{2}  +  t)
      \Gamma(\tfrac{n}{2}  +  t)
      \Gamma(\sigma  -  \tfrac{n}{2}(q-1) - t) 
      }
     {\Gamma(\alpha_{1,2} + 2t)
      \Gamma(\tfrac{n}{2}q - \sigma  + t) 
     } \;,\qquad
\label{MB:J22}
\end{eqnarray}
where $\sigma$ is defined by Eq.~(\ref{eq:sigma}) with $s=q-1$ and
\begin{equation}
\alpha_{1,2} = \alpha_1 + \alpha_2\;.
\label{eq:alpha12}
\end{equation}
Closing the contour of integration in Eq.~(\ref{MB:J22}) on the left, we
obtain in the notation of Ref.~\cite{BD}:
\begin{eqnarray}
\lefteqn{
J^q_{22}(m^2, p^2, \alpha_1, \alpha_2, \sigma_1, \cdots, \sigma_{q-1}) = 
\left[ i^{1-n} \pi^{n/2} \right]^q 
\frac{
(-m^2)^{\tfrac{n}{2}q - \alpha_{1,2} - \sigma}
}{\Gamma(\alpha_1) \Gamma(\alpha_2) }
\left\{ \prod_{k=1}^{q-1} \frac{\Gamma(\frac{n}{2}-\sigma_k)}{\Gamma(\sigma_k)} \right\}}
\nonumber \\ 
&&{}\times
\frac{
\Gamma\left( \alpha_1  +  \sigma  -  \tfrac{n}{2}(q-1) \right) 
\Gamma\left( \alpha_2  +  \sigma  -  \tfrac{n}{2}(q-1) \right) 
\Gamma\left( \sigma  -  \tfrac{n}{2} (q-2) \right) 
\Gamma\left( \alpha_{1,2}  +  \sigma  -  \tfrac{n}{2} q \right) 
}{
\Gamma \left( \alpha_{1,2}  +  2 \sigma  -  n(q-1) \right)
\Gamma\left( \tfrac{n}{2} \right) 
}
\nonumber\\
&&{}\times
~_{4}F_3\left(\begin{array}{c|}
\alpha_1  +  \sigma  -  \frac{n}{2}(q-1),
\alpha_2  +  \sigma  -  \frac{n}{2}(q-1),
\sigma  -  \frac{n}{2}(q - 2),
\alpha_{1,2}  +  \sigma  -  \frac{n}{2}q \\
\frac{n}{2}, 
\frac{1}{2} ( \alpha_{1,2}  -  n(q-1) )  +  \sigma, 
\frac{1}{2} ( 1  +  \alpha_{1,2}  -  n(q-1) )  +  \sigma, 
\end{array} ~\frac{p^2}{4m^2} \right) \;.
\nonumber\\
&&\label{eq:j22q}
\end{eqnarray}
For $q=1$ (see Footnote~\ref{foot:mb}), 
the ${}_4F_3$ function in Eq.~(\ref{eq:j22q}) is reduced to a ${}_3F_2$
function, in agreement with Ref.~\cite{BD}.
In the two-loop case ($q=2$), the hypergeometric representation of this
diagram was derived in Refs.~\cite{D91,BFT}.

Let us analyze the reduction of the hypergeometric function in
Eq.~(\ref{eq:j22q}) assuming that all parameters,
$\alpha_1$, $\alpha_2$, and $\sigma_k$, are integer. 
For  $q=1$ ($\sigma_k=0$), we obtain a ${}_3F_2$ function with integer
differences between upper and lower parameters, which, according to
{\it Criterion~1}, is reducible to a ${}_2F_1$ function with one integer
upper parameter, so that we have a basis hypergeometric function plus a
rational function. 
(For the one-loop propagator, there are one master integral and bubble
integrals.) 
For $q=2$, we get a ${}_4F_3$ function with integer parameter differences and
one integer upper parameter, so that it is reducible to a ${}_3F_2$ function
with one integer upper parameter and its first derivative,
\begin{equation}
(1,\theta) \times
~_{3}F_2\left(\begin{array}{c|}
1, I_1-\tfrac{n}{2}, I_2-n \\
I_3 + \tfrac{n}{2}, I_4+\tfrac{1}{2}-\tfrac{n}{2}
\end{array} ~z \right) \;,
\end{equation}
plus a rational function.
In this case, there are two nontrivial master integrals of the same topology
and bubble integrals, in accordance with the results of Ref.~\cite{2loop}.
For $q \geq 3$, we have a ${}_4F_3$ function with integer differences of
parameters, which, according to {\it Criterion~1}, is reducible to a ${}_3F_2$
function and its first two derivatives,
\begin{equation}
(1,\theta,\theta^2) \times
~_{3}F_2\left(\begin{array}{c|}
I_1-\tfrac{n}{2}(q-1), I_2-\tfrac{n}{2}(q-2), I_3-\tfrac{n}{2}q \\
\tfrac{n}{2}, I_4+\tfrac{1}{2}-\tfrac{n}{2}(q-1) 
\end{array} ~z \right) \;.
\end{equation}

\boldmath
\subsection{$E^{q}_{1220}$ and $B^{q}_{1220}$}
\label{B1220}
\unboldmath

Let us consider the $q$-loop bubble diagram $E^q_{1220}$ in
Fig.~\ref{diagrams} with two massive lines of different masses and $s$ plus
$r$ massless propagators,
\begin{eqnarray}
&& \hspace{-5mm}
E^q_{1220}(m^2, M^2, \alpha_1, \alpha_2, \beta, \sigma_1, \cdots, \sigma_{s}, \rho_1, \cdots, \rho_r) = 
\nonumber \\ && \hspace{-5mm}
\int \frac{d^n (k_1 \cdots k_q)} 
{[k_1^2]^{\sigma_1} \cdots [k_{s}^2]^{\sigma_{s}} 
[k_{s+1}^2-M^2]^{\alpha_1} [(k_1  +  k_2  +  \cdots  +  k_{s+1}+k_{q})^2-M^2]^{\alpha_2} }
\nonumber \\ && \hspace{0mm}
\times
\frac{1}{
[k_{s+2}^2]^{\rho_1} \cdots [k_{q-1}^2]^{\rho_{r-1}} [(k_{s+2} + \cdots k_{q-1}+k_q)^2)]^{\rho_{r}} 
(k_q^2-m^2)^\beta} 
\; ,
\label{eq:e1220q}
\end{eqnarray}
where, by construction, $s \geq 0$ and $r \geq 2$ and, as a consequence,
$q=s+r+1 \geq 3$.
The Mellin-Barnes representation of Eq.~(\ref{eq:e1220q}) reads:
\begin{eqnarray}
\lefteqn{
E^q_{1220}(m^2, M^2, \alpha_1, \alpha_2, \beta, \sigma_1, \cdots, \sigma_{s}, \rho_1, \cdots, \rho_r) = 
\frac{
(-M^2)^{\tfrac{n}{2} - \alpha_{1,2}}
(-m^2)^{\tfrac{n}{2}(q-1) - \sigma - \beta  -  \rho}
}{\Gamma(\alpha_1) \Gamma(\alpha_2) \Gamma(\beta) \Gamma\left(\tfrac{n}{2}\right)}}
\nonumber \\ &&{}\times
\left[ i^{1-n} \pi^\myfrac{n}{2} \right]^q
\left\{ \prod_{j=1}^{s} \frac{\Gamma(\tfrac{n}{2}-\sigma_j)}{\Gamma(\sigma_j)} \right\}
\left\{ \prod_{k=1}^{r} \frac{\Gamma(\tfrac{n}{2}-\rho_k)}{\Gamma(\rho_k)} \right\}
\frac{ \Gamma\left(\rho-\tfrac{n}{2}(r-1)\right)}{ \Gamma\left(\tfrac{n}{2}r-\rho\right)}
\nonumber \\
&&{}\times
\int dt
\left( \frac{m^2}{M^2}\right)^t 
\frac{\Gamma(\alpha_1 + t) \Gamma(\alpha_2 + t) 
      \Gamma(\alpha_{1,2} - \tfrac{n}{2} + t) \Gamma(\tfrac{n}{2}+t)
      }
     {\Gamma(\alpha_{1,2} + 2t) }
\nonumber \\ &&
{}\times
\frac{\Gamma\left( \sigma  -  \tfrac{n}{2}s  - t \right) 
      \Gamma\left( \sigma  +  \beta  +  \rho -  \tfrac{n}{2}(q-1)  - t \right) 
      \Gamma\left( \tfrac{n}{2}(q-1)  -  \sigma  -  \rho + t \right) 
      }
     {\Gamma\left(\tfrac{n}{2}(s+1) - \sigma + t \right) 
     } \;,
\label{MB:E1220}
\end{eqnarray}
where $\sigma$, $\rho$, and $\alpha_{1,2}$ are defined in
Eqs.~(\ref{eq:sigma}), (\ref{eq:rho}), and (\ref{eq:alpha12}), respectively.

Diagram $B_{1220}^q$ in Fig.~\ref{diagrams} emerges from
$E^q_{1220}$ with $r=1$ and $s=q-2$ in the smooth limit
$\rho=\rho_1 \to 0$.
Then, Eq.~(\ref{MB:E1220}) simplifies due to
$
\frac{\Gamma\left( \tfrac{n}{2}(q-1)  -  \sigma  -  \rho + s \right)}
     {\Gamma\left(\tfrac{n}{2}(s+1) - \sigma + s \right)} = 1 \;.
$

Closing the contour of integration in Eq.~(\ref{MB:E1220}) on the left, we
obtain in the notation of Ref.~\cite{BD}:
\begin{eqnarray}
&& \hspace{-5mm}
E^q_{1220}(m^2, M^2, \alpha_1, \alpha_2, \beta, \sigma_1, \cdots, \sigma_{z}, \rho_1, \cdots, \rho_r) = 
\left[ i^{1-n} \pi^\myfrac{n}{2} \right]^q
\frac{
(-M^2)^{\tfrac{n}{2}q - \alpha_{1,2} - \sigma - \beta  -  \rho}
}{\Gamma(\alpha_1) \Gamma(\alpha_2) \Gamma(\beta) \Gamma\left(\tfrac{n}{2}\right)}
\nonumber \\ && \hspace{-5mm}
\left\{ \prod_{j=1}^{s} \frac{\Gamma(\tfrac{n}{2}-\sigma_j)}{\Gamma(\sigma_j)} \right\}
\left\{ \prod_{k=1}^{r} \frac{\Gamma(\tfrac{n}{2}-\rho_k)}{\Gamma(\rho_k)} \right\}
\frac{ \Gamma\left(\rho-\tfrac{n}{2}(r-1)\right)}{ \Gamma\left(\tfrac{n}{2}r-\rho\right)}
\nonumber \\ && \hspace{-5mm}
\times
%
%
\Biggl\{
\left( \frac{m^2}{M^2} \right)^{\tfrac{n}{2}r  -  \beta  -  \rho}
\frac{
\Gamma\left( \sigma  -  \tfrac{n}{2}(s - 1) \right) 
\Gamma\left( \tfrac{n}{2}r  -  \rho \right) 
\Gamma\left( \rho  +  \beta  -  \tfrac{n}{2}r \right) 
}
{
\Gamma\left( \tfrac{n}{2} \right) 
\Gamma\left( \alpha_{1,2}  +  2 \sigma  -  ns \right) 
}
\nonumber \\ && 
\times
\Gamma\left(a_{1,\sigma}  -  \tfrac{n}{2}s \right) 
\Gamma\left(a_{2,\sigma}  -  \tfrac{n}{2}s \right) 
\Gamma\left(a_{1,2,\sigma}  -  \tfrac{n}{2}(s+1) \right) 
\nonumber \\ && \hspace{-5mm}
~_{5}F_4\left(\begin{array}{c|}
a_{1, \sigma}  -  \frac{n}{2}s,
a_{2, \sigma}  -  \frac{n}{2}s,
a_{1,2,\sigma}  -  \frac{n}{2}(s + 1),
\sigma  -  \frac{n}{2}(s - 1), 
\frac{n}{2}r  - \rho \\
\frac{n}{2}, 
\frac{\alpha_{1,2}  -  ns}{2}  +  \sigma, 
\frac{\alpha_{1,2}  +  1  -  ns}{2}  +  \sigma, 
\frac{n}{2}r  -  \rho  +   1  -  \beta 
\end{array} ~\frac{m^2}{4M^2} \right)
\nonumber \\ && \hspace{-5mm}
%
%
+ 
\frac{
\Gamma(\beta)
\Gamma\left( \tfrac{n}{2}r  -  a_{\beta,\rho} \right) 
\Gamma\left( a_{\sigma,\beta,\rho}  -  \tfrac{n}{2}(q - 2) \right) 
}
{
\Gamma\left( \alpha_{1,2}  +  2\sigma  +  2 \beta  +  2 \rho  -  n(q - 1) \right) 
\Gamma\left( a_{\beta,\rho}  - \tfrac{n}{2}(r - 1)  \right) 
}
\nonumber \\ && 
\times
\Gamma\left(a_{1,\sigma,\beta,\rho}  -  \tfrac{n}{2}(q-1) \right) 
\Gamma\left(a_{2,\sigma,\beta,\rho}  -  \tfrac{n}{2}(q-1) \right) 
\Gamma\left(a_{1,2,\sigma,\beta,\rho}  -  \tfrac{n}{2}q \right) 
\nonumber \\ && \hspace{-5mm}
~_{5}F_4\left(\begin{array}{c|}
a_{1,\sigma,\beta,\rho}  \!-\!  \tfrac{n}{2}(q \!-\! 1),
a_{2,\sigma,\beta,\rho}  \!-\!  \tfrac{n}{2}(q \!-\! 1),
a_{1,2,\sigma,\beta,\rho}  \!-\!  \tfrac{n}{2}q,
a_{\sigma,\beta,\rho}  \!-\!  \tfrac{n}{2}(q \!-\! 2), 
\beta \\
1  \!+\!  a_{\beta,\rho}  \!-\!  \tfrac{n}{2}r , 
a_{\sigma,\beta,\rho}  \!+\!  \tfrac{\alpha_{1,2}  \!-\!  n(q  \!-\!  1)}{2}, 
a_{\sigma,\beta,\rho}  \!+\!  \tfrac{\alpha_{1,2}  \!+\!  1  \!-\!  n(q  \!-\!  1)}{2}, 
a_{\beta,\rho}  \!-\!  \tfrac{n}{2}(r  \!-\! 1)
\end{array} ~\frac{m^2}{4M^2} \right)
\Biggr\} \;,
\nonumber\\
&&
\label{E_Q_1220}
\end{eqnarray}
where we have introduced the short-hand notations 
\begin{eqnarray}
a_{i, \sigma} &=& \alpha_1  +  \sigma \;, \qquad 
a_{\beta,\rho}  = \beta + \rho \;,
\nonumber\\
a_{1,2,\sigma}&=& \alpha_1  +  \alpha_2  +  \sigma \;, \qquad
a_{\sigma,\beta,\rho} = \sigma  +  \beta  +  \rho \;,
\nonumber \\
a_{1,2,\sigma,\beta} &=& \alpha_1  +  \alpha_2  +  \sigma  +  \beta \;,\qquad
a_{i,\sigma,\beta,\rho} = \alpha_i  +  \sigma  +  \beta  +  \rho \;,
\nonumber\\
a_{1,2,\sigma,\beta,\rho} &=&
\alpha_1  +  \alpha_2  +  \sigma  +  \beta  +  \rho \;,\qquad
i=1,2\;.
\label{coefficients}
\end{eqnarray}
Let us analyze the result of the differential reduction of the two
hypergeometric functions in Eq.~(\ref{E_Q_1220}) assuming that all parameters,
$\alpha_1,\alpha_2,\beta,\sigma_k,\rho_j$, are integer. 
We distinguish between the three cases: $s=0$, $s=1$, and $s \geq 2$. 
For $s=0$ ($\sigma_k=0$, $r=q-1$), both hypergeometric functions in
Eq.~(\ref{E_Q_1220}) are reducible to ${}_2F_1$ functions with one unit upper
parameter,
\begin{eqnarray}
{}_{2}F_1\left(\begin{array}{c|}
1, I_1-\tfrac{n}{2} \\
I_2 + \tfrac{1}{2} \end{array} ~z \right) \;, 
\qquad 
{}_{2}F_1\left(\begin{array}{c|}
1, I_1  -  \tfrac{n}{2}q, \\
\tfrac{1}{2} + I_2 - \tfrac{n}{2}(q - 1) \end{array} ~z \right) \;.
\label{E1220:i}
\end{eqnarray}
Standard approaches yield one master integral.
For $s=1$ $(r=q-2)$, each of the two hypergeometric functions is reducible to a
${}_3F_2$ function with one unit upper parameter,
\begin{equation}
(1,\theta) \times
{}_{3}F_2\left(\begin{array}{c|}
1, I_1 \!-\! n, I_2 \!-\! \tfrac{n}{2} \\
I_3 \!+\! \tfrac{n}{2}, I_4 \!+\! \tfrac{1}{2} \!-\! \tfrac{n}{2} \end{array} ~z \right) \;, 
\qquad 
(1,\theta) \times
{}_{3}F_2\left(\begin{array}{c|}
1, I_1 \!-\! \tfrac{n}{2}(q\!-\!1), I_2 \!-\! \tfrac{n}{2}q \\
I_3 \!-\! \tfrac{n}{2}(q \!-\! 3), I_4 \!+\! \tfrac{1}{2} \!-\! \tfrac{n}{2}(q\!-\!1) \end{array} ~z \right) \;.
\label{E1220:ii}
\end{equation}
For $s\geq 2$, the first hypergeometric function is reducible to a ${}_3F_2$
function and its first two derivatives, while the second one is expressible in
terms of a ${}_4F_3$ function and its first two derivatives, namely
\begin{eqnarray}
&& 
(1,\theta,\theta^2) \times
{}_{3}F_2\left(\begin{array}{c|}
I_1 - \tfrac{n}{2}s, I_2 - \tfrac{n}{2}(s + 1), I_3 - \tfrac{n}{2}(s - 1) \\
I_4 + \tfrac{n}{2}, I_5 + \tfrac{1}{2} - \tfrac{n}{2}s  
\end{array} ~z \right) \;, 
\nonumber \\ && 
(1,\theta,\theta^2) \times
{}_{4}F_3\left(\begin{array}{c|}
1, I_1  -  \tfrac{n}{2}q, I_2  -  \tfrac{n}{2}(q - 1), I_3  -  \tfrac{n}{2}(q - 2) \\
\tfrac{1}{2} + I_4 - \tfrac{n}{2}(q - 1), 
I_5 - \tfrac{n}{2}r, I_6 - \tfrac{n}{2}(r - 1) \end{array} ~z \right) \;,
\label{E1220:iii}
\end{eqnarray}
respectively.
Notice that, according to {\it Criterion 4}, a third derivative does not appear
because one of the upper parameters is integer.

We now present an explicit result for the three-loop case in which there is no
massless propagator inside the massive loop, $q=3$, $s=0$,
$\sigma_k=\sigma=0$, $r=2$, and $n=4-2\ep$.
In this case, there is one master integral with
$\alpha_1=\alpha_2=\beta=\rho_1=\rho_2=1$ ($\alpha_{1,2}=\rho=2$), 
\begin{eqnarray}
&& \hspace{-5mm}
E^3_{1220}(m^2, M^2,1,1,1,1,1) = 
\left[ i \pi^{2 - \ep} \right]^3
(M^2)^{1-3\ep}
\nonumber \\ && \hspace{-5mm}
\frac{\Gamma(1-\ep) \Gamma^2(1+\ep) \Gamma(1+2\ep)}
{2\ep^3 (1-\ep) (1-2\ep)}
%
%
\Biggl\{
\left( \frac{m^2}{M^2} \right)^{1-2\ep}
~_{2}F_1\left(\begin{array}{c|}
1, \ep \\
\frac{3}{2}
\end{array} ~\frac{m^2}{4M^2} \right)
\nonumber \\ && \hspace{-5mm}
%
%
+ 
\frac{2}{3(1-3\ep)}
\frac{
\Gamma(1 + 2\ep) \Gamma(1 + 3\ep)
}
{
\Gamma(1 +  4\ep)
\Gamma\left(1  +  \ep \right) 
}
~_{2}F_1\left(\begin{array}{c|}
1, -1 + 3\ep \\
\tfrac{1}{2}  +  2 \ep
\end{array} ~\frac{m^2}{4M^2} \right)
\Biggr\} \;.
\label{eq:e12203x} 
\end{eqnarray}
To reduce the Gauss hypergeometric functions in Eq.~(\ref{eq:e12203x}) to sets
of functions studied in Refs.~\cite{KWY07a,DK04,MKL06}, we apply the following
relations:
\begin{eqnarray}
(1-2\ep)
~_{2}F_1\left(\begin{array}{c|}
1, \ep  \\
\tfrac{3}{2}
\end{array} ~z\right) &=& 
1 - 2 \ep (1-z)
~_{2}F_1\left(\begin{array}{c|}
1, 1 +  \ep  \\
\tfrac{3}{2}
\end{array} ~z\right) \;,
\nonumber\\
~_{2}F_1\left(\begin{array}{c|}
1, -1 + 3\ep \\
\tfrac{1}{2}  +  2 \ep
\end{array} ~z\right)
&=& 
1 - 2z \frac{(1-3\ep)}{1-2\ep}
+ \frac{12\ep(1-3\ep) z(1-z)}{(1-2\ep)(1+4\ep)}
~_{2}F_1\left(\begin{array}{c|}
1, 1 + 3\ep \\
\tfrac{3}{2}  +  2 \ep
\end{array} ~z\right) \;.\qquad
\end{eqnarray}
The first few coefficients of the $\ep$ expansion of Eq.~(\ref{eq:e12203x})
are given by
\begin{eqnarray}
&& 
\frac{E^3_{1220}(m^2, M^2,1,1,1,1,1)}{\Gamma^3(1+\ep)\left[ i \pi^{2 - \ep} \right]^3}
= (M^2)^{1-3\ep}
\Biggl(
\frac{(1+z)}{3\ep^3}
+ \frac{6+z(5-3 \ln z)}{3\ep^2}
\nonumber \\ && 
+ \frac{1}{3\ep}
\left\{ 
25 + z(17  -  15 \ln z  +  3 \ln^2 z  +  3 \zeta_2)
\right\}
\nonumber \\ && \hspace{5mm}
+ 2\frac{(4-z)}{\ep } \frac{(1-y)}{(1+y)}
\left[
\Li{2}{1 - y}   +  \frac{1}{4} \ln^2 y  +  \frac{1}{2} \ln z \ln y  
\right]
\nonumber \\ && 
+ \frac{1}{3} 
( 90  +  49 z )
+ \frac{1}{3} \zeta_3 (8  -  7 z)
- z \left[ 17 \ln z  -  5 \ln^2 z  +  \frac{2}{3} \ln^3 z \right]
 +  \zeta_2 z (5 - 2 \ln z)
\nonumber \\ && 
+ \frac{1-y}{1+y}(4 - z)
\Biggl\{
4 \Snp{1,2}{y}
 +  2 \Snp{1,2}{y^2}
 -  4 \Snp{1,2}{-y}
 -  6 \Li{3}{y}
 -  2 \Li{3}{-y}
\nonumber \\ && \hspace{5mm}
+ \Li{2}{-y} \left[ 4 \ln(1 - y)  -  2 \ln z\right]
 +  \Li{2}{1-y} \left[ 10  -  8 \ln(1 - y)  -  4 \ln (1 + y) \right]
\nonumber \\ && \hspace{5mm}
+ \ln(1 - y)
\left[
2 \zeta_2 
 -  4 \ln y \ln(1 - y)
 -  3 \ln^2 y 
\right]
+ \ln^2 y \left[
\frac{5}{2} 
 -  \ln(1 + y) 
 +  \frac{2}{3} \ln y 
\right]
\nonumber \\ && \hspace{5mm}
 -  \zeta_2 (\ln z  -  4 \ln y)
 -  \zeta_3 
+ \ln z \ln y 
\left[ 
5  +  \frac{1}{2} \ln y  -  2  \ln(1 + y)  -  \ln z 
\right]
\Biggr\}
+ {\cal O}(\ep)
\Biggr) 
\;,\qquad
\label{eq:e12203e}
\end{eqnarray}
where $z$ is defined in Eq.~(\ref{eq:z}) and the variable $y$ is defined as
\begin{eqnarray}
y = \frac{1-\sqrt{\frac{z}{z-4}}}{1+\sqrt{\frac{z}{z-4}}} \;, \qquad 
z = -\frac{(1-y)^2}{y} \;, \qquad 
4  -  z = \frac{(1+y)^2}{y} \;.
\label{y}
\end{eqnarray}
The ${\cal O}(\ep)$ term is too long to be presented here, but is available
from Ref.~\cite{MKL:hyper}.
We mention here only that, in accordance with the expansion of the Gauss
hypergeometric function constructed in Refs.~\cite{DK04,MKL06}, this term is
expressible just in terms of Nielsen polylogarithms. 
To check Eq.~(\ref{eq:e12203e}), we evaluate the first few coefficients of the
$\ep$ expansion of the original diagram in the large-mass limit \cite{heavy}
using  the program packages developed in Refs.~\cite{avdeev,heavy:package}
to find agreement.
In order to construct the series expansion about $1-y$, which serves as the
small parameter of the large-mass expansion, we employ the trick described in
Ref.~\cite{JKV}.
In fact, the variable $1-y$ can be written as a series in the small parameter
$z$ as
\begin{equation}
1 - y = 
i \sqrt{z} 
\left[ 
\sqrt{1-\frac{z}{4}}  -  i \sqrt{\frac{z}{4}}
\right]
= 
i \sqrt{z} 
\left[ 1  -  i \frac{\sqrt{z}}{2}  -  \frac{z}{8} 
\sum_{k=0}^\infty \frac{(2k)!}{(k!)^2} \frac{1}{k+1}\left( \frac{z}{16}\right)^k 
\right]
 \;.
\end{equation}

For completeness, we also present the explicit result for diagram
$B^{q}_{1220}$ in Fig.~\ref{diagrams}:\footnote{%
For the four-loop case ($q=4$) with equal masses $m=M$, the Mellin-Barnes
representation of this diagram was published in Ref.~\cite{B1220:MB}. 
In order to recover the result of Ref.~\cite{B1220:MB} from 
Eq.~(\ref{MB:E1220}), it is sufficient to redefine the variable of integration
$t$ to be $t=N/2-\alpha_{12}-s$.}
\begin{eqnarray}
&& \hspace{-5mm}
B^q_{1220}(m^2, M^2, \alpha_1, \alpha_2, \beta, \sigma_1, \cdots, \sigma_{q-2}) 
\nonumber \\ && \hspace{-5mm}
= 
\frac{
(-M^2)^{\tfrac{n}{2}q  - \alpha_{1,2}  - \sigma  -  \beta}
\Gamma\left( \tfrac{n}{2}  -  \beta \right) 
}
{\Gamma(\alpha_1) \Gamma(\alpha_2) \Gamma(\beta) 
\Gamma\left( \tfrac{n}{2} \right) 
\left[ i^{1 - n} \pi^\myfrac{n}{2} \right]^{-q}
}
\left\{ \prod_{k=1}^{q-2} \frac{\Gamma(\frac{n}{2}-\sigma_k)}{\Gamma(\sigma_k)} \right\}
\nonumber \\ && \hspace{-5mm}
\times
\Biggl \{
\frac{
\Gamma\left( a_{1,\sigma,\beta}  -  \tfrac{n}{2}(q-1) \right) 
\Gamma\left( a_{2,\sigma,\beta}  -  \tfrac{n}{2}(q-1) \right) 
\Gamma\left( a_{1,2,\beta,\sigma}  -  \tfrac{n}{2} q \right) 
\Gamma\left( a_{\sigma,\beta}  -  \tfrac{n}{2} (q-2) \right) 
}{
\Gamma \left( \alpha_{1,2}  +  2 \beta  +  2 \sigma  -  n(q-1) \right)
}
\nonumber \\ && 
~_{4}F_3\left(\begin{array}{c|}
a_{1,\beta,\sigma}  -  \tfrac{n}{2}(q-1),
a_{2,\beta,\sigma}  -  \tfrac{n}{2}(q-1),
\alpha_{1,2}  +  \beta  +  \sigma  -  \tfrac{n}{2}q, 
a_{\sigma, \beta}  -  \tfrac{n}{2}(q - 2) \\
\frac{1}{2} (\alpha_{1,2}  -  n(q - 1) )  +  \sigma  +  \beta, 
\frac{1}{2} (\alpha_{1,2}  +  1  -  n(q - 1) )  +  \sigma  +  \beta, 
1  +  \beta  -  \tfrac{n}{2}
\end{array} ~\frac{m^2}{4M^2} \right) 
%
%
\nonumber \\ && \hspace{5mm}
+ 
\left(
\frac{m^2}{M^2}
\right)^{\tfrac{n}{2}-\beta}
\frac{
\Gamma\left( a_{1,\sigma}  -  \tfrac{n}{2}(q-2) \right) 
\Gamma\left( a_{2,\sigma}  -  \tfrac{n}{2}(q-2) \right) 
\Gamma\left( a_{1,2,\sigma}  -  \tfrac{n}{2} (q-1) \right) 
}{
\Gamma\left( \tfrac{n}{2}  -  \beta \right) 
\Gamma \left( \alpha_{1,2}  +  2 \sigma  -  n(q-2) \right)
}
\nonumber \\ && \hspace{5mm}
\times
\Gamma\left( \sigma  -  \tfrac{n}{2} (q-3) \right) 
\Gamma\left( \beta  -  \tfrac{n}{2} \right) 
\nonumber \\ && 
~_{4}F_3\left(\begin{array}{c|}
a_{1,\sigma}  -  \tfrac{n}{2}(q - 2),
a_{2,\sigma}  -  \tfrac{n}{2}(q - 2),
a_{1,2,\sigma}  -  \tfrac{n}{2}(q - 1), 
\sigma  -  \tfrac{n}{2}(q - 3) \\
\frac{1}{2} (\alpha_{1,2}  -  n(q - 2) )  +  \sigma, 
\frac{1}{2} (\alpha_{1,2}  +  1  -  n(q - 2) )  +  \sigma, 
1  -  \beta  +  \tfrac{n}{2}
\end{array} ~\frac{m^2}{4M^2} \right) 
\Biggr \} \;,
\label{B_Q_1220}
\end{eqnarray}
where $q \geq 2$.
For $q=2$, the hypergeometric representation was derived in
Ref.~\cite{DT}.\footnote{%
In this case, we have $\sigma=0$ and
$\left\{ \prod_{k=1}^{q-2} \frac{\Gamma(\frac{n}{2}-\sigma_k)}{\Gamma(\sigma_k)} \right\}=1$.}
The result of the differential reduction of the hypergeometric functions in
Eq.~(\ref{B_Q_1220}) assuming that all parameters,
$\alpha_1$, $\alpha_2$, $\beta$, and $\sigma_k$, are integer may be derived for
$q=2$, $q=3$, and $q\geq 4$ by directly substituting $r=1$ and $s=q-2$ in
Eqs.~(\ref{E1220:i})--(\ref{E1220:iii}), respectively.
We only point out here that, for $q=3$, 
the second hypergeometric function in
Eq.~(\ref{E1220:ii}) is reducible to a ${}_2F_1$ function and its first
derivative, so that the differential reduction yields
\begin{equation}
(1,\theta) \times 
~_{2}F_1\left(\begin{array}{c|}
I_1 - n, I_2 - \tfrac{3n}{2}  \\
\tfrac{1}{2} + I_3 - n
\end{array} ~z\right) \;,  
\quad 
(1,\theta) \times 
~_{3}F_2\left(\begin{array}{c|}
1, I_1 - n, I_2  -  \tfrac{n}{2}  \\
\tfrac{1}{2} + I_3  -  \tfrac{n}{2},  I_4  +  \tfrac{n}{2} \end{array} ~z\right)  \;,
\end{equation}
and that, for $q \geq 4$, the second hypergeometric function in
Eq.~(\ref{E1220:iii}) is reducible to a ${}_3F_2$ function and its first
two derivatives, so that the differential reduction yields
\begin{eqnarray}
&& 
(1,\theta,\theta^2) \times
{}_{3}F_2\left(\begin{array}{c|}
I_1  -  \tfrac{n}{2}q, I_2  -  \tfrac{n}{2}(q - 1), I_3  -  \tfrac{n}{2}(q - 2) \\
\tfrac{1}{2} + I_4 - \tfrac{n}{2}(q - 1), I_5 - \tfrac{n}{2} \end{array} ~z \right) \;,
\nonumber \\ && 
(1,\theta,\theta^2) \times
{}_{3}F_2\left(\begin{array}{c|}
I_1 - \tfrac{n}{2}(q-2), I_2 - \tfrac{n}{2}(q - 1), I_3 - \tfrac{n}{2}(q - 3) \\
I_4 + \tfrac{n}{2}, I_5 + \tfrac{1}{2} - \tfrac{n}{2}(q-2)   
\end{array} ~z \right) \;. 
\end{eqnarray}
It is interesting to note that, in the single-scale case $m=M$ \cite{ibs},
there is only one master integral \cite{avdeev}. 
This is because the hypergeometric functions ${}_3F_2$ and ${}_2F_1$ with
special values of parameters and argument $z=1/4$ are expressible as products
of Gamma functions. 
For details, see Eqs.~(4.36) and (4.42) in Ref.~\cite{DK01}.
A diagrammatic interpretation of similar identities was presented in
Ref.~\cite{Czakon}.
On the other hand, this is a manifestation of the existence of functional
relations between Feynman diagrams \cite{Tarasov:2008hw}.

We now present the explicit result for the three-loop case $q=3$,
$n=4-2\ep$. 
In this case, the first master integral corresponds to
$\alpha_1=\alpha_2=\beta=\sigma_1=1$, while we choose
$\alpha_1=\beta=\sigma_1=1$ and $\alpha_2=2$ for the second one,
\begin{eqnarray}
&& \hspace{-5mm}
B^3_{1220}(m^2, M^2, 1,1,1,1) 
= 
(M^2)^{2-3\ep}
\frac{\Gamma^{3}(1 + \ep) (i\pi^{2-\ep})^3}{\ep^3(1-\ep)^2(1-2\ep)}
\nonumber \\ && \hspace{-5mm}
\times
\Biggl\{
\frac{2(1 - \ep)(1 - 4\ep)
\Gamma^2\left( 1  +  2 \ep \right) 
\Gamma\left( 1  +  3\ep \right) 
\Gamma(1-\ep)
}{
3(1-3\ep)(2-3\ep)
\Gamma(1 + 4\ep)
\Gamma^2(1 + \ep)
}
~_{2}F_1\left(\begin{array}{c|}
-1  +  2\ep,
-2  +  3\ep \\
-\frac{1}{2}  +  2 \ep
\end{array} ~\frac{m^2}{4M^2} \right) 
\nonumber \\ && \hspace{5mm}
+ 
\left( 
\frac{m^2}{M^2}
\right)^{1 - \ep}
~_{3}F_2\left(\begin{array}{c|}
1, \ep, - 1  +  2\ep\\
2 - \ep, \frac{1}{2}  +  \ep 
\end{array} ~\frac{m^2}{4M^2} \right) 
\Biggr \}
\;,
\nonumber
\\
&& \hspace{-5mm}
%
%
%
%
B^3_{1220}(m^2, M^2,1,2,1,1) 
= 
(M^2)^{1-3\ep}
\frac{\Gamma^{3}(1 + \ep) (i\pi^{2-\ep})^3}{2\ep^3(1-\ep)^2}
\nonumber \\ && \hspace{-5mm}
\times
\Biggl\{
\frac{2(1-\ep)(1-4\ep)
\Gamma^2\left( 1  +  2 \ep \right) 
\Gamma\left( 1  +  3\ep \right) 
\Gamma(1-\ep)
}{
3
(1-2\ep)
(1-3\ep)
\Gamma(1 + 4\ep)
\Gamma^2(1 + \ep)
}
~_{2}F_1\left(\begin{array}{c|}
-1 + 2\ep,  -1  +  3\ep \\
- \frac{1}{2}  +  2 \ep
\end{array} ~\frac{m^2}{4M^2} \right) 
\nonumber \\ && \hspace{5mm}
+ 
\left( 
\frac{m^2}{M^2}
\right)^{1-\ep}
~_{3}F_2\left(\begin{array}{c|}
1, \ep, 2\ep\\
2 - \ep, \frac{1}{2}  +  \ep 
\end{array} ~\frac{m^2}{4M^2} \right) 
\Biggr \} \;.
\label{B3_1220_3}
\end{eqnarray}
In order to express all hypergeometric functions entering
Eq.~(\ref{B3_1220_3}) in terms of functions which were
analyzed in Ref.~\cite{DK04} (see Appendix~\ref{appendix} for details), we
apply the following set of relations:
\begin{eqnarray}
&& 
%
%
(1 - 4\ep)
~_{2}F_1\left(\begin{array}{c|}
-1  +  2\ep, -2  +  3 \ep \\
-\frac{1}{2}  +  2 \ep  \end{array} ~z\right) 
= 
-
\frac{24 z (1 - z) (1 + 2z) \ep^2}{(1+4\ep)}
~_{2}F_1\left(\begin{array}{c|}
1 +  2\ep, 1  +  3 \ep \\
\frac{3}{2}  +  2 \ep  \end{array} ~z\right) 
\nonumber \\ && \hspace{5mm}
+ \left[ 1 - 4\ep  -  4 z  +  8 \ep z^2  +  14 \ep z \right]
~_{2}F_1\left(\begin{array}{c|}
2\ep, 3\ep \\
\frac{1}{2}  +  2\ep  \end{array} ~z\right) \;, 
\nonumber\\ 
&& 
%
%
(1 - 4\ep)
~_{2}F_1\left(\begin{array}{c|}
-1  +  2\ep, -1  +  3\ep \\
-\frac{1}{2}  +  2 \ep  \end{array} ~z\right) 
= 
- \frac{24 z (1 - z) \ep^2}{1  +  4\ep}
~_{2}F_1\left(\begin{array}{c|}
1 + 2\ep, 1 + 3\ep \\
\frac{3}{2} + 2\ep  \end{array} ~z\right) 
\nonumber \\ && \hspace{5mm}
+ \left[ 1 - 4 \ep  -  2 z  +  10 z \ep) \right]
~_{2}F_1\left(\begin{array}{c|}
2 \ep, 3 \ep \\
\frac{1}{2}  +  f \ep  \end{array} ~z\right) \;,
\nonumber\\
&& 
\frac{2z(1 - 2\ep)(1 - 3\ep)}{1-\ep}
~_{3}F_2\left(\begin{array}{c|}
1, \ep, 2\ep \\
2 - \ep, \frac{1}{2} + \ep  \end{array} ~z\right) 
= 
1 - 2\ep
\nonumber \\ && \hspace{3mm}
+ \left[ 2 \ep (1 - 4z)  -  (1 - 2z) \right] 
~_{3}F_2\left(\begin{array}{c|}
1, \ep, 2\ep \\
1 - \ep, \frac{1}{2} + \ep  \end{array} ~z\right) 
+ \frac{8z(1 - z) \ep^2}{(1 - \ep)(1 + 2\ep)}
~_{3}F_2\left(\begin{array}{c|}
2, 1 + \ep, 1 + 2\ep \\
2 - \ep, \frac{3}{2} + \ep \end{array} ~z\right) \;, 
\nonumber \\ && 
z(1 - 3\ep)(2 - 3\ep)
~_{3}F_2\left(\begin{array}{c|}
1, \ep, -1 + 2\ep \\
2 - \ep, \frac{1}{2} + \ep  \end{array} ~z\right) 
= 
\frac{1}{2} (1 - \ep) (1  -  2 \ep  +  2 z \ep) 
\nonumber \\ && \hspace{10mm}
+ 
\frac{4 \ep^2 z (1 - z)(1 + 2z)}{(1 + 2\ep)}
~_{3}F_2\left(\begin{array}{c|}
2, 1 + \ep, 1 + 2\ep \\
2 - \ep, \frac{3}{2} + \ep \end{array} ~z\right) 
\nonumber \\ && \hspace{10mm}
- \frac{(1 - \ep)}{2} 
\left[ 1 - 4 z  -  2 \ep (1 -  8z  -  2z^2)\right]
~_{3}F_2\left(\begin{array}{c|}
1, \ep, 2\ep \\
1 - \ep, \frac{1}{2} + \ep \end{array} ~z\right) \;. 
\end{eqnarray}
Here, the following relation was used:
\begin{eqnarray}
&& 
z
(b - a_2) (b - a_3) (1 + b - a_3) (f - a_3)
~_{3}F_2\left(\begin{array}{c|}
1, a_2, a_3-1 \\
b+1, f  \end{array} ~z\right) 
\nonumber \\ && 
= 
- 
b
(1-a_3)(1-z) 
\left[ 
a_3-f+(a_2-b)z
\right]
\theta
~_{3}F_2\left(\begin{array}{c|}
1, a_2, a_3 \\
b, f  \end{array} ~z\right) 
\nonumber \\ && 
+ 
b (1-a_3)
\Biggl\{
a_2 (a_2 - b)z^2 
+ (a_3-f)(1-f) 
- b (1+a_2+2a_3-b) z
\nonumber \\ && \hspace{5mm}
- 
\left[ 
f (2a_2+a_3-2b) 
+  a_2(a_2-1) 
- (a_2+a_3)^2 
\right] z
\Biggr\} 
~_{3}F_2\left(\begin{array}{c|}
1, a_2, a_3 \\
b, f  \end{array} ~z\right) 
\nonumber \\ && 
- b(f-1) \left[(1-a_3)(f-a_3)-(b-1)(b-a_2)z \right] 
\;. 
\end{eqnarray}
The first coefficients of the $\ep$ expansions of Eq.~(\ref{B3_1220_3}) read: 
\begin{eqnarray}
&& 
\frac{B^3_{1220}(m^2,M^2,1,1,1,1)}{\Gamma^3(1 + \ep)(i\pi^{2 - \ep})^3}
= \left( M^{2} \right)^{2-3\ep}
\Biggl(
\frac{ 1 + 2z}{3\ep^3}  
+ \frac{1}{\ep^2}\
\left\{ \frac{7}{6} +  \frac{8}{3} z  -  \frac{1}{12} z^2   -   z \ln z \right\}
\nonumber \\ && 
+ \frac{1}{\ep}
\left\{ 
\frac{25}{12}  +  \frac{20}{3} z  -  \frac{5}{8} z^2 
+ \frac{1}{4} z \ln z \left[  z  +  2 \ln z  -  16 \right]
\right\}
\nonumber \\ && 
+ \frac{8}{3} \zeta_3 (1 - z)
 -  \frac{5}{24}  +  \frac{35}{3} z  -  \frac{145}{48} z^2
 -  \frac{1}{6} z \ln^3 z 
 +  \frac{(16 - z)}{8} z\ln^2 z 
 -  \frac{(80 - 15z)}{8} z \ln z 
\nonumber \\ && 
- 4(1 - z) \left[ \Snp{1,2}{1 - y}  +  \ln y \Li{2}{1 - y} \right]
- (1 - z) \ln^2 y \left[ \ln z  +  \frac{2}{3} \ln y   \right]
\nonumber \\ && 
-  (8  +  2 z  -  z^2) \frac{(1 - y)}{(1 + y)} 
\left[ \Li{2}{1-y}  +  \frac{1}{2} \ln z \ln y + \frac{1}{4} \ln^2 y \right] 
+ {\cal O}(\ep)
\Biggr) \;, 
\nonumber\\ && 
\frac{B^3_{1220}(m^2,M^2,1,2,1,1)}{\Gamma^3(1 + \ep)(i\pi^{2 - \ep})^3}
= \left( M^{2} \right)^{1-3\ep}
\Biggl(
\frac{1 + z}{3\ep^3} 
+ 
\frac{1}{6\ep^2} 
\left\{ 
4  +  5 z   -  3 z \ln z 
\right\}
\nonumber \\ && 
+ 
\frac{1}{3\ep} 
\left\{ 
1  +  4 z  
\right\}
- 
\frac{1}{4 \ep} z \ln z 
\left\{ 
4  -  \ln z 
\right\}
- \frac{10}{3} + \frac{5}{6}z 
- \frac{1}{2} z \ln z \left[ 
1  -  \ln z  +  \frac{1}{6} \ln^2 z 
\right]
\nonumber \\ && 
+ (4-z) \frac{(1-y)}{(1+y)}
\left[ 
2 \Li{2}{1-y} 
+ 
\frac{1}{2} \ln^2 y 
 +  
\ln z \ln y 
\right]
\nonumber \\ && 
+ 2(z-2) \left[ 
\Snp{1,2}{1-y}  +  \ln y \Li{2}{1-y}
+ \frac{1}{6} \ln^3 y 
+ \frac{1}{4} \ln z \ln^2 y  
- \frac{2}{3} \zeta_3 
\right]
+ {\cal O}(\ep)
\Biggr) \;,\qquad 
\label{eq:b12203e}
\end{eqnarray}
where $y$ and $z$ are defined in Eqs.~(\ref{y}) and (\ref{eq:z}), respectively.
The ${\cal O}(\ep)$ term is too long to be presented here, but is available
from Ref.~\cite{MKL:hyper}.
To check Eq.~(\ref{eq:b12203e}), we evaluate the first few coefficients of the
$\ep$ expansion of the original diagram in the large-mass limit \cite{heavy}
using  the program packages developed in Refs.~\cite{avdeev,heavy:package}
to find agreement.

\boldmath
\subsection{$V^{q}_{1220}$ and $J^{q}_{1220}$}
\unboldmath

Let us consider the $q$-loop on-shell propagator diagram $V^q_{1220}$ in
Fig.~\ref{diagrams} with $s$ plus $r$ massless propagators,
\begin{eqnarray}
\lefteqn{
V^q_{1220}(m^2, M^2, \alpha_1, \alpha_2, \beta, \sigma_1, \cdots, \sigma_{s}, \rho_1, \cdots, \rho_r)} 
\nonumber \\ &=&
\int \frac{d^n (k_1 \cdots k_q)} 
{[k_1^2]^{\sigma_1} \cdots [k_{s}^2]^{\sigma_{s}} 
[k_{s+1}^2-M^2]^{\alpha_1} [(k_1  +  k_2  +  \cdots  +  k_{s+1}+k_{q})^2-M^2]^{\alpha_2} }
\nonumber \\ &&{}\times
\left. 
\frac{1}{
[k_{s+2}^2]^{\rho_1} \cdots [k_{q-1}^2]^{\rho_{r-1}} [(k_{s+2}  +  \cdots  +  k_{q-1}+k_q)^2)]^{\rho_{r}} 
((k_q-p)^2-m^2)^\beta} 
\right|_{p^2=m^2}
\; , \qquad
\label{eq:v1220q}
\end{eqnarray}
where $q=s+r+1$.
The Mellin-Barnes representation of Eq.~(\ref{eq:v1220q}) reads:
\begin{eqnarray}
\lefteqn{
V^q_{1220}(m^2, M^2, \alpha_1, \alpha_2, \beta, \sigma_1, \cdots, \sigma_{s}, \rho_1, \cdots, \rho_r) = 
\frac{
(-M^2)^{\tfrac{n}{2} - \alpha_{1,2}}
(-m^2)^{\tfrac{n}{2}(q-1) - \sigma - \beta  -  \rho}
}{\Gamma(\alpha_1) \Gamma(\alpha_2) \Gamma(\beta)}}
\nonumber \\ &&{}
\times
\left[ i^{1-n} \pi^\myfrac{n}{2} \right]^q
\left\{ \prod_{j=1}^{s} \frac{\Gamma(\tfrac{n}{2}-\sigma_j)}{\Gamma(\sigma_j)} \right\}
\left\{ \prod_{k=1}^{r} \frac{\Gamma(\tfrac{n}{2}-\rho_k)}{\Gamma(\rho_k)} \right\}
\frac{ \Gamma\left(\rho-\tfrac{n}{2}(r-1)\right)}{ \Gamma\left(\tfrac{n}{2}r-\rho\right)}
\nonumber \\ &&{}\times
\int dt 
\left( \frac{m^2}{M^2}\right)^t 
\frac{\Gamma(\alpha_1 + t) \Gamma(\alpha_2 + t) 
      \Gamma(\alpha_{1,2} - \tfrac{n}{2} + t) \Gamma(\tfrac{n}{2}+t)
      }
     {\Gamma(\alpha_{1,2} + 2t) }
\nonumber \\ &&{}\times
\frac{\Gamma\left( \sigma  -  \tfrac{n}{2}s  - t \right) 
      \Gamma\left( \sigma  +  \beta  +  \rho -  \tfrac{n}{2}(q-1)  - t \right) 
      \Gamma\left( n(q-1)  -  2\sigma  -  \beta  -  2 \rho + 2t \right) 
      }
     {\Gamma\left(\tfrac{n}{2}(s+1) - \sigma + t \right) 
      \Gamma\left(\tfrac{n}{2}q - \sigma  -  \beta  -  \rho + t \right) 
     } \;,\qquad
\label{MB:V1220}
\end{eqnarray}
where $\sigma$, $\rho$, and $\alpha_{1,2}$ are defined in
Eqs.~(\ref{eq:sigma}), (\ref{eq:rho}), and (\ref{eq:alpha12}), respectively.

Diagram $J_{1220}^q$ in Fig.~\ref{diagrams} emerges from
$V^q_{1220}$ with $r=1$ and $s=q-2$ in the smooth limit
$\rho=\rho_1 \to 0$.
For the special case $q=3$, the Mellin-Barnes representation has recently been
presented in Ref.~\cite{grozin}.

Closing the contour of integration in Eq.~(\ref{MB:V1220}) on the left, we
obtain:
\begin{eqnarray}
\lefteqn{
V^q_{1220}(m^2, M^2, \alpha_1, \alpha_2, \beta, \sigma_1, \cdots, \sigma_{s}, \rho_1, \cdots, \rho_r) 
= 
\frac{
\left( i^{1-n} \pi^\myfrac{n}{2} \right)^q
(-M^2)^{\tfrac{n}{2}q - \alpha_{1,2,\sigma,\beta,\rho}}
}
{\Gamma(\alpha_1) \Gamma(\alpha_2) \Gamma(\beta)
\Gamma\left( \tfrac{n}{2} \right) 
}}
\nonumber \\
&&{}\times
\left\{ \prod_{j=1}^{s} \frac{\Gamma(\tfrac{n}{2}-\sigma_j)}{\Gamma(\sigma_j)} \right\}
\left\{ \prod_{k=1}^{r} \frac{\Gamma(\tfrac{n}{2}-\rho_k)}{\Gamma(\rho_k)} \right\}
\frac{ \Gamma\left(\rho-\tfrac{n}{2}(r-1)\right)}{ \Gamma\left(\tfrac{n}{2}r-\rho\right)}
\nonumber \\
&&{}\times
%
%
\Biggl\{
\left( \frac{m^2}{M^2} \right)^{\tfrac{n}{2}r - a_{\beta,\rho}}
\frac{
\Gamma\left( \sigma  -  \tfrac{n}{2}(s - 1) \right) 
\Gamma\left( nr  - 2 \rho  -  \beta \right) 
\Gamma\left( \rho  +  \beta  -  \tfrac{n}{2}r \right) 
}
{
\Gamma\left( \alpha_{1,2}  +  2 \sigma  -  ns \right) 
\Gamma\left( \tfrac{n}{2}(r + 1)  -  \rho  -  \beta \right) 
}
\nonumber \\
&&{}\times
\Gamma\left(a_{1,\sigma}  -  \tfrac{n}{2}s \right) 
\Gamma\left(a_{2,\sigma}  -  \tfrac{n}{2}s \right) 
\Gamma\left(a_{1,2,\sigma}  -  \tfrac{n}{2}(s+1) \right) 
\nonumber \\
&&{}\times
~_{6}F_5\left(\begin{array}{c|}
a_{1, \sigma}  \!-\!  \frac{n}{2}s,
a_{2, \sigma}  \!-\!  \frac{n}{2}s,
a_{1,2,\sigma}  \!-\!  \frac{n}{2}(s \!+\! 1),
\sigma  \!-\!  \frac{n}{2}(s \!-\! 1), 
\frac{nr  \!-\!  \beta}{2}  \!-\! \rho,
\frac{nr  \!-\!  \beta  \!+\!  1}{2}  \!-\! \rho \\
\frac{n}{2}, 
\frac{\alpha_{1,2}  \!-\!  ns}{2}  \!+\!  \sigma, 
\frac{\alpha_{1,2}  \!+\!  1  \!-\!  ns}{2}  \!+\!  \sigma, 
\frac{n}{2}(r \!+\! 1)  \!-\!  \beta  \!-\!  \rho, 
\frac{n}{2}r  \!-\!  \beta  \!-\!  \rho  \!+\!  1
\end{array} ~\frac{m^2}{M^2} \right)
\nonumber \\
&&{}+
%
%
\frac{
\Gamma\left( \tfrac{n}{2}r  -  a_{\beta,\rho} \right) 
\Gamma\left( a_{\sigma,\beta,\rho}  -  \tfrac{n}{2}(q - 2) \right) 
\Gamma(\beta)
}
{
\Gamma\left( \alpha_{1,2}  +  2\sigma  +  2 \beta  +  2 \rho  -  n(q - 1) \right) 
\Gamma\left( a_{\beta,\rho}  - \tfrac{n}{2}(r - 1)  \right) 
}
\nonumber \\
&&{}\times
\Gamma\left(a_{1,\sigma,\beta,\rho}  -  \tfrac{n}{2}(q-1) \right) 
\Gamma\left(a_{2,\sigma,\beta,\rho}  -  \tfrac{n}{2}(q-1) \right) 
\Gamma\left(a_{1,2,\sigma,\beta,\rho}  -  \tfrac{n}{2}q \right) 
\nonumber \\
&&{}\times
~_{6}F_5\left(\begin{array}{c|}
a_{1,\sigma,\beta,\rho}  \!-\!  \tfrac{n}{2}(q \!-\! 1),
a_{2,\sigma,\beta,\rho}  \!-\!  \tfrac{n}{2}(q \!-\! 1),
a_{1,2,\sigma,\beta,\rho}  \!-\!  \tfrac{n}{2}q,
a_{\sigma,\beta,\rho}  \!-\!  \tfrac{n}{2}(q \!-\! 2), 
\tfrac{\beta}{2},
\tfrac{\beta \!+\! 1}{2} \\
\tfrac{n}{2}, 
1  \!+\!  a_{\beta,\rho}  \!-\!  \tfrac{n}{2}r, 
\tfrac{\alpha_{1,2}  \!-\!  n(q  \!-\!  1)}{2}  \!+\!  a_{\sigma,\beta,\rho},
\tfrac{\alpha_{1,2}  \!+\!  1  \!-\!  n(q  \!-\!  1)}{2}  \!+\!  a_{\sigma,\beta,\rho}, 
a_{\beta,\rho}  \!-\!  \tfrac{n}{2}(r  \!-\! 1)
\end{array} ~\frac{m^2}{M^2} \right)
\Biggr\} \;, 
\nonumber \\
&& \label{V_Q_1220}
\end{eqnarray}
where 
$a_{1, \sigma}$,
$a_{2, \sigma}$,
$a_{\beta,\rho}$,
$a_{1,2,\sigma}$,
$a_{\sigma,\beta,\rho}$,
$a_{1,\sigma,\beta,\rho}$, and
$a_{2,\sigma,\beta,\rho}$
are defined in Eq.~(\ref{coefficients}).
In the special case in which there is no massless propagator inside the
massive loop, $s=0$ ($r=q-1$, $\sigma=0$,
$\prod_{k=1}^{q-2} \frac{\Gamma(\frac{n}{2}-\sigma_k)}{\Gamma(\sigma_k)}=1$),
the ${}_6F_5$ functions in Eq.~(\ref{V_Q_1220}) are reduced to ${}_5F_4$
functions. 
For $q=2$, $s=0$, and $r=1$, the hypergeometric representation was given in
Ref.~\cite{DG}. 

Let us analyze the result of the differential reduction of the hypergeometric
functions in Eq.~(\ref{V_Q_1220}) assuming that all parameters,
$\alpha_1$, $\alpha_2$, $\beta$, $\sigma_k$, and $\rho_k$, are integer. 
For $r \neq 0$, we distinguish between the three cases $s=0$, $s=1$, and
$s \geq 2$. 
For $s=0$ ($r=q-1$, $\sigma=0$), each of the two hypergeometric functions in
Eq.~(\ref{V_Q_1220}) is reducible to a ${}_3F_2$ function with one unit upper
parameter and its first derivative,
\begin{eqnarray}
&& 
(1,\theta ) \times 
~_{3}F_2\left(\begin{array}{c|}
1, I_1 - \tfrac{n}{2},  I_2 + \tfrac{1}{2} + \tfrac{n}{2}(q - 1) \\ 
I_3 + \tfrac{1}{2}, I_4 + \tfrac{n}{2}q 
\end{array} ~z\right) \;,  
\nonumber \\ && 
(1,\theta) \times 
~_{3}F_2\left(\begin{array}{c|}
1, I_1 + \tfrac{1}{2},  I_2 - \tfrac{n}{2}q \\
I_3 + \tfrac{n}{2}, \tfrac{1}{2}  +  I_4 - \tfrac{n}{2}(q - 1)
\end{array} ~z\right) \;.  
\label{V1220:i} 
\end{eqnarray}
For $s=1$ ($r=q-2$), each of the two hypergeometric functions in
Eq.~(\ref{V_Q_1220}) is reducible to a ${}_4F_3$ function with one unit upper
parameter and its first two derivatives,
\begin{eqnarray}
&& 
(1,\theta, \theta^2 ) \times 
~_{4}F_3\left(\begin{array}{c|}
1, I_1 - n, I_2 - \tfrac{n}{2},  I_3 + \tfrac{1}{2} + \tfrac{n}{2}(q - 2) \\ 
I_4 + \tfrac{n}{2}, I_5 + \tfrac{1}{2} - \tfrac{n}{2}, I_6 + \tfrac{n}{2}(q - 1) 
\end{array} ~z\right) \;,  
\quad 
\nonumber \\ && 
(1,\theta, \theta^2) \times 
~_{4}F_3\left(\begin{array}{c|}
1, I_1 + \tfrac{1}{2},  I_2 - \tfrac{n}{2}q, I_3 - \tfrac{n}{2}(q - 1) \\
I_4 + \tfrac{n}{2}, \tfrac{1}{2}  +  I_5 - \tfrac{n}{2}(q - 1), I_6 - \tfrac{n}{2}(q - 3)
\end{array} ~z\right) \;.  
\label{V1220:ii} 
\end{eqnarray}
For $s \geq 2$, one of the hypergeometric functions in Eq.~(\ref{V_Q_1220}) is
reducible to a ${}_4F_3$ function with all upper parameters having
non-zero $\ep$ parts, whereas the second one is expressible in terms of a
${}_5F_4$ function with one unit upper parameter,
\begin{eqnarray}
&& 
(1,\theta, \theta^2, \theta^3 ) \times 
~_{4}F_3\left(\begin{array}{c|}
I_1 - \tfrac{n}{2}(s-1), I_2 - \tfrac{n}{2}s,   I_3 - \tfrac{n}{2}(s + 1), I_4 + \tfrac{1}{2} + \tfrac{n}{2}r \\ 
I_5 + \tfrac{n}{2}, I_6 + \tfrac{1}{2} - \tfrac{n}{2}s, I_7 + \tfrac{n}{2}(r + 1) 
\end{array} ~z\right) \;,  
\quad 
\nonumber \\ && 
(1,\theta, \theta^2, \theta^3 ) \times 
~_{5}F_4\left(\begin{array}{c|}
1, I_1 + \tfrac{1}{2},  I_2 - \tfrac{n}{2}q, I_3 - \tfrac{n}{2}(q - 1), I_4 - \tfrac{n}{2}(q - 2) \\
I_5 + \tfrac{n}{2}, \tfrac{1}{2}  +  I_6 - \tfrac{n}{2}(q - 1), I_7 - \tfrac{n}{2}r, I_8 - \tfrac{n}{2}(r - 1)
\end{array} ~z\right) \;.  
\label{V1220:iii} 
\end{eqnarray}

For completeness, we also present an explicit result for diagram
$J^{q}_{1220}$ in Fig.~\ref{diagrams}:
\begin{eqnarray}
\lefteqn{
J^q_{1220}(m^2, M^2, \alpha_1, \alpha_2, \beta, \sigma_1, \cdots,
 \sigma_{q-2})}
\nonumber \\ 
&=&
\frac{
\left[ i^{1-n} \pi^\myfrac{n}{2} \right]^q
\Gamma\left( \beta  -  \tfrac{n}{2}\right) 
(-M^2)^{\frac{n}{2}q - \alpha_{1,2,\sigma,\beta}}
}{
\Gamma(\alpha_1) \Gamma(\alpha_2) \Gamma(\beta)
\Gamma\left( \tfrac{n}{2} \right) 
}
\left\{ \prod_{k=1}^{q-2} \frac{\Gamma(\frac{n}{2}-\sigma_k)}{\Gamma(\sigma_k)} \right\}
\nonumber \\
&&{}\times
\Biggl\{
\left( \frac{m^2}{M^2} \right)^{\tfrac{n}{2}  \!-\!  \beta}
\frac{
\Gamma\left( \sigma        \!-\!  \tfrac{n}{2}(q  \!-\! 3) \right) 
\Gamma\left( a_{1,\sigma}   \!-\!  \tfrac{n}{2}(q  \!-\! 2) \right) 
\Gamma\left( a_{2,\sigma}   \!-\!  \tfrac{n}{2}(q  \!-\! 2) \right) 
\Gamma\left( a_{1,2,\sigma} \!-\!  \tfrac{n}{2}(q  \!-\! 1) \right) 
}
{
\Gamma\left( \alpha_{1,2}  +  2 \sigma  -  n(q  -  2) \right) 
}
\nonumber \\
&&{}\times
~_{6}F_5\left(\begin{array}{c|}
a_{1,\sigma}  \!-\!  \tfrac{n}{2}(q \!-\! 2),
a_{2,\sigma}  \!-\!  \tfrac{n}{2}(q \!-\! 2),
a_{1,2,\sigma}  \!-\!  \tfrac{n}{2}(q \!-\! 1),
\sigma  \!-\!  \tfrac{n}{2}(q \!-\! 3), 
\tfrac{n \!-\! \beta}{2},
\frac{n \!-\! \beta \!+\! 1}{2} \\
\tfrac{n}{2}, 
n \!-\! \beta, 
\tfrac{n}{2} \!+\!  1  \!-\!  \beta, 
\tfrac{\alpha_{1,2}  \!-\!  n(q  \!-\!  2)}{2}  \!+\!  \sigma, 
\tfrac{\alpha_{1,2}  \!+\!  1  \!-\!  n(q  \!-\!  2)}{2}  \!+\!  \sigma
\end{array} ~\frac{m^2}{M^2} \right)
\nonumber \\
%
%
&&{}+
\frac{
\Gamma\left( \tfrac{n}{2}       \!-\!  \beta \right) 
\Gamma\left( a_{\sigma,\beta}    \!-\!  \tfrac{n}{2}(q \!-\! 2) \right) 
\Gamma\left( a_{1,\sigma,\beta}  \!-\!  \tfrac{n}{2}(q \!-\! 1) \right) 
\Gamma\left( a_{2,\sigma,\beta}  \!-\!  \tfrac{n}{2}(q \!-\! 1) \right) 
\Gamma\left( \alpha_{1,2,\sigma,\beta}  \!-\!  \tfrac{n}{2}q \right) 
}
{
\Gamma\left( \beta  \!-\!  \tfrac{n}{2}\right) 
\Gamma\left( \alpha_{1,2}  \!+\!  2\sigma  \!+\!  2 \beta  \!-\!  n(q \!-\! 1) \right) 
}
\nonumber \\
&&{}\times
~_{6}F_5\left(\begin{array}{c|}
a_{1,\sigma, \beta}  \!-\!  \tfrac{n}{2}(q \!-\! 1),
a_{2,\sigma, \beta}  \!-\!  \tfrac{n}{2}(q \!-\! 1),
a_{1,2,\sigma,\beta} \!-\!  \tfrac{n}{2}q,
a_{\sigma,\beta}  \!-\!  \tfrac{n}{2}(q \!-\! 2), 
\tfrac{\beta}{2},
\tfrac{\beta + 1}{2} \\
\tfrac{n}{2}, 
\beta, 
1  \!+\!  \beta  \!-\!  \tfrac{n}{2} , 
\tfrac{\alpha_{1,2}  \!-\!  n(q  \!-\!  1)}{2}  \!+\!  \sigma  \!+\!  \beta, 
\tfrac{\alpha_{1,2}  \!+\!  1  \!-\!  n(q  \!-\!  1)}{2}  \!+\!  \sigma  \!+\!  \beta
\end{array} ~\frac{m^2}{M^2} \right)
\Biggr\} \;.
\nonumber\\
&&\label{J_Q_1220}
\end{eqnarray}
In this case, the result of the differential reduction assuming that all
parameters, $\alpha_1$, $\alpha_2$, $\beta$, and $\sigma_k$, are integer can be
derived for $q=2$, $q=3$, and $q\geq 4$ by directly substituting $r=1$ and
$s=q-2$ in Eqs.~(\ref{V1220:i})--(\ref{V1220:iii}), respectively.
We only point out here that, for $q=3$, 
the second hypergeometric function in
Eq.~(\ref{V1220:ii}) is reducible to a ${}_3F_2$ function and its first
two derivatives, so that the differential reduction yields
\begin{eqnarray}
&& 
(1,\theta, \theta^2 ) \times 
~_{4}F_3\left(\begin{array}{c|}
1, I_1 - n, I_2 - \tfrac{n}{2},  I_3 + \tfrac{1}{2} + \tfrac{n}{2} \\ 
I_4 + \tfrac{n}{2}, I_5 + \tfrac{1}{2} - \tfrac{n}{2}, I_6 + n
\end{array} ~z\right) \;,  
\quad 
\nonumber \\ && 
(1,\theta, \theta^2) \times 
~_{3}F_2\left(\begin{array}{c|}
I_1 + \tfrac{1}{2},  I_2 - n, I_3 - \tfrac{3n}{2} \\
I_4 + \tfrac{n}{2}, \tfrac{1}{2}  +  I_5 - n 
\end{array} ~z\right) \;,  
\end{eqnarray}
and that, for $q \geq 4$, the ${}_5F_4$  hypergeometric functions in Eq.~(\ref{V1220:iii}) is
reducible to a ${}_4F_3$ function with all upper parameters having
non-zero $\ep$ parts:
\begin{eqnarray}
&& 
(1,\theta, \theta^2, \theta^3 ) \times 
~_{4}F_3\left(\begin{array}{c|}
I_1 - \tfrac{n}{2}(q-3), I_2 - \tfrac{n}{2}(q-2),   I_3 - \tfrac{n}{2}(q - 1), I_4 + \tfrac{1}{2} + \tfrac{n}{2} \\ 
I_5 + \tfrac{n}{2}, I_6 + \tfrac{1}{2} - \tfrac{n}{2}(q-2), I_7 + n 
\end{array} ~z\right) \;,  
\quad 
\nonumber \\ && 
(1,\theta, \theta^2, \theta^3 ) \times 
~_{4}F_3\left(\begin{array}{c|}
I_1 + \tfrac{1}{2},  I_2 - \tfrac{n}{2}q, I_3 - \tfrac{n}{2}(q - 1), I_4 - \tfrac{n}{2}(q - 2) \\
I_5 + \tfrac{n}{2}, \tfrac{1}{2}  +  I_6 - \tfrac{n}{2}(q - 1), I_7 - \tfrac{n}{2}
\end{array} ~z\right) \;.  
\end{eqnarray}
In particular, for $q=2$, in accordance with Ref.~\cite{sunset2}, there are
two nontrivial master integrals plus bubble integrals.
For $q=3$, in accordance with Ref.~\cite{grozin}, there are three nontrivial 
master integrals plus bubble integrals.

\subsection{Two-loop vertex}

Let us consider the diagram $F_2$ shown in Fig.~\ref{diagrams} 
\begin{eqnarray}
&& 
F_2(M^2, p^2, \alpha_1, \alpha_2, \sigma_1, \sigma_2, ) = 
\nonumber \\ && \hspace{-5mm}
\left. 
\int \frac{d^n (k_1 k_2)} 
{[(k_1-p_1)^2]^{\sigma_1} [(k_1-p_2)^2]^{\sigma_2} [k_1^2]^\beta [k_2^2-M^2]^{\alpha_1} [(k_1-k_2)^2-M^2]^{\alpha_2}}
\right|_{p_1^2=p_2^2=0}
\nonumber \\ && 
= 
\left[ i^{1-n} \pi^\frac{n}{2} \right]^2
\frac{
(-M^2)^{\tfrac{n}{2}\!-\!\alpha_{12}}
(p^2)^{\tfrac{n}{2}\!-\!\sigma_{12}\!-\!\beta }
}{\Gamma(\alpha_1) \Gamma(\alpha_2) \Gamma(\sigma_1) \Gamma(\sigma_2)}
\nonumber \\ && \hspace{5mm}
\times 
\int dt 
\left( -\frac{p^2}{M^2}\right)^t 
\frac{\Gamma(-t) 
      \Gamma(\alpha_1\!+\!t) 
      \Gamma(\alpha_2\!+\!t) 
      \Gamma(\alpha_{12}\!-\!\tfrac{n}{2}\!+\!t) 
      }
     {
\Gamma\left(\alpha_{12}+2t\right)
} 
\nonumber \\ && \hspace{10mm}
\frac{
      \Gamma\left( \tfrac{n}{2} \!-\! \sigma_1 \!-\! \beta \!+\!t \right) 
      \Gamma\left( \tfrac{n}{2} \!-\! \sigma_2 \!-\! \beta \!+\!t \right) 
      \Gamma\left( \sigma_{12} \!+\! \beta \!-\! \tfrac{n}{2}\!-\!t \right) 
}
{
\Gamma(n-\sigma_{12}-\beta+t)
} \;.
\label{F2}
\end{eqnarray}
where $\sigma$ is defined in Eq.~(\ref{eq:sigma}) with $s=2$ and $\alpha_{12}$
in Eq.~(\ref{eq:alpha12}).
Closing the contour of integration on the left we obtain in the notation of
Ref.~\cite{BD}:
\begin{eqnarray}
&& 
F_2(M^2, p^2, \alpha_1, \alpha_2, \sigma_1, \sigma_2, ) = 
\left[ i^{1-n} \pi^\frac{n}{2} \right]^2
\frac{
(-M^2)^{\tfrac{n}{2}\!-\!\alpha_{12} }
(p^2)^{\tfrac{n}{2}\!-\!\sigma_{12}\!-\!\beta } 
\Gamma\left( \tfrac{n}{2} \!-\! \sigma_{12} \!-\! \beta \right) 
}{\Gamma(\alpha_1) \Gamma(\alpha_2)} 
\nonumber \\ && \hspace{-5mm}
\times
%
%
\Biggl\{
\frac{
\Gamma(\alpha_1) 
\Gamma(\alpha_2) 
\Gamma\left( \tfrac{n}{2} \!-\! \sigma_1 \!-\! \beta \right) 
\Gamma\left( \tfrac{n}{2} \!-\! \sigma_2 \!-\! \beta \right) 
\Gamma\left(\sigma_{12} \!+\! \beta \!-\! \tfrac{n}{2} \right) 
\Gamma\left( \alpha_{12} \!-\! \tfrac{n}{2} \right) 
}
{
\Gamma(\sigma_1) \Gamma(\sigma_2)
\Gamma(\alpha_{12}) \Gamma\left( n\!-\!\sigma_{12} \!-\! \beta \right) 
\Gamma\left( \tfrac{n}{2} \!-\! \sigma_{12} \!-\! \beta \right) 
}
\nonumber \\ && 
\times
~_{5}F_4\left(\begin{array}{c|}
\alpha_1,  \alpha_2, 
\alpha_{12} \!-\! \frac{n}{2}, 
\frac{n}{2} \!-\! \sigma_1 \!-\! \beta, 
\frac{n}{2} \!-\! \sigma_2 \!-\! \beta 
\\
n\!-\!\sigma_{12} \!-\! \beta, 
1\!+\! \frac{n}{2} \!-\! \sigma_{12} \!-\! \beta,  
\frac{\alpha_{12}}{2},  \frac{\alpha_{12}+1}{2}
\end{array} ~-\frac{p^2}{4M^2} \right)
\nonumber \\ && \hspace{-5mm}
%
%
+ 
\left( -\frac{p^2}{M^2} \right)^{\beta \!+\! \sigma_{12} \!-\! \tfrac{n}{2}}
\frac{
\Gamma\left( \alpha_1 \!+\! \sigma_{12} \!+\! \beta \!-\! \tfrac{n}{2} \right) 
\Gamma\left( \alpha_2 \!+\! \sigma_{12} \!+\! \beta \!-\! \tfrac{n}{2} \right) 
\Gamma\left( \sigma_{12} \!+\! \alpha_{12} \!+\! \beta \!-\! n \right) 
}
{
\Gamma(\alpha_{12} \!+\! 2\sigma_{12} \!+\! 2 \beta \!-\! n)
\Gamma\left( \tfrac{n}{2} \right) 
}
\nonumber \\ && 
\times
~_{5}F_4\left(\begin{array}{c|}
\sigma_1, \sigma_2, 
\alpha_1 \!+\! \sigma_{12} \!+\! \beta \!-\! \tfrac{n}{2},
\alpha_2 \!+\! \sigma_{12} \!+\! \beta \!-\! \tfrac{n}{2},
\alpha_{12} \!+\! \sigma_{12} \!+\! \beta \!-\! n \\
\frac{n}{2}, 1 \!+\! \sigma_{12} \!+\! \beta \!-\! \frac{n}{2}, 
\sigma_{12} \!+\! \beta \!+\! \frac{\alpha_{12}-n}{2}, 
\sigma_{12} \!+\! \beta \!+\! \frac{\alpha_{12}+1-n}{2} 
\end{array} ~-\frac{p^2}{4M^2} \right)
\Biggr\} \;.\quad 
\label{F2_hyper}
\end{eqnarray}
By differential reduction, the two hypergeometric functions in
Eq.~(\ref{F2_hyper}) may be reduced to
\begin{eqnarray}
(1,\theta)  \times 
~_{3}F_2\left(\begin{array}{c|}
1, 
I_1 \!-\! \tfrac{n}{2},
\tfrac{n}{2} \!+\! I_2 \\
n \!+\! I_3, \tfrac{1}{2} \!+\! I_4 
\end{array} ~-\frac{p^2}{4M^2} \right)
\;, 
\quad 
(1,\theta)  \times 
~_{3}F_2\left(\begin{array}{c|}
1,1, I_1 \!-\! n \\ 
I_2 \!+\! \tfrac{n}{2}, I_3 \!+\! \tfrac{1}{2} \!-\! \tfrac{n}{2} 
\end{array} ~-\frac{p^2}{4M^2} \right)
\;,\quad 
\end{eqnarray}
respectively, where $\{I_a\}$ is a set of integers. 
In accordance with Ref.~\cite{DK04}, the first few coefficients of the $\ep$
expansion of this diagram are expressible in terms of Remiddi-Vermaseren
functions, i.e.\ multiple polylogarithms of the square roots of unity, of
argument
\begin{equation}
\tilde y = \frac{1-\sqrt{\frac{z}{z+4}}}{1+\sqrt{\frac{z}{z+4}}},
\end{equation}
where $z=p^2/M^2$ (see also the discussion in Ref.~\cite{Passarino}).
The on-shell value $z=1$ corresponds to Remiddi-Vermaseren functions of
argument $(3-\sqrt{5})/2=[(\sqrt{5}-1)/2]^2$, in agreement with results of
Ref.~\cite{FKKR}.
It is interesting to note that these constants are generated not only in
Higgs-boson decay \cite{FKKR}, but also in orthopositronium decay \cite{KKV}.

\section{Master integrals in differential reduction and IBP technique}
\label{camparison}

In this section, we compare the differential-reduction and IPB techniques with
respect to the numbers of master integrals which they produce.
For this purpose, we return to the Feynman diagrams depicted in Figs.~\ref{one-loop}
and \ref{diagrams}.
They all have the following hypergeometric structure: 
\begin{equation}
\Phi(n,\vec{j};z) = \sum_{a=1}^k z^{l_a} C_{l_a}(n,\vec{j})
{}_{p+1}F_{p}(\vec{A}_a;\vec{B}_a;\kappa z) \;, 
\label{eq}
\end{equation}
where $\vec{j}$ is the set of the powers of the propagators of the Feynman
diagram, 
$n$ is dimension of space-time, 
$k$ is the number of hypergeometric functions,  
$\{l_a,A_a,B_a \}$ are linear combinations of $\vec{j}$ and $n$ with rational
coefficients, 
$\kappa$ is a rational number (being $\kappa=1,1/4$ in the
considered cases), and 
$C_{l_a}$ are products of $\Gamma$ functions with
arguments only depending on $n$ and $\vec{j}$. 
Being a sum of holonomic functions, $\Phi(\vec{j};\vec{z})$ is also holonomic.
Thus, the number of basis elements on the r.h.s.\ of Eq.~(\ref{eq}) is equal to
the number of master-integrals $\{\Phi_k(\vec{z})\}$ that may be derived from
the l.h.s.\ of Eq.~(\ref{eq}) by applying the integration-by-parts technique,
which may be written symbolically as
\begin{equation}
\Phi(n,\vec{j};z) = \sum_{k=1}^h B_k(n,\vec{j};z) \Phi_k(n;z) \;.  
\label{eq2}
\end{equation}
Here, it is understood that diagrams that are expressible in terms of Gamma
functions \cite{vladimirov} are not counted.
The number of basis elements in the framework of differential reduction is
defined to be the highest power of the differential operator $\theta$ in
\begin{equation}
{}_{p+1}F_{p}(\vec{A};\vec{B};z) = \sum_{l=0}^v P_l(z) \theta^{l} {}_{s+1}F_{s}(\vec{A}-\vec{I_1};\vec{B}-\vec{I_2};z) \;,
\end{equation}
where $\vec{I_1},\vec{I_2}$ are sets of integers, $P_l(z)$ are rational
functions, and the differential-reduction algorithm for exceptional values of
parameters of hypergeometric function, constructed in Sections~(2.3) and (2.4)
is employed, in which case one has $v \leq  p$ in general. 
Our analysis demonstrates that there is a very simple relation between the
number $h$ of nontrivial master integrals follows from IBP, 
which are not expressible in terms of Gamma functions, 
and the maximal power $v$ of $\theta$ generated by the
differential reduction in Eq.~(\ref{eq3}), namely
\begin{equation}
h=v+1 \;.
\label{eq3}
\end{equation}
This relation does not depend on the number $k$ of hypergeometric functions
entering Eq.~(\ref{eq}).
In Table~\ref{tab:k}, these highest powers are collected for the Feynman
diagrams shown in Fig.~\ref{diagrams}.
\begin{table}
\caption{\label{tab:k}
Highest powers of the differential operator $\theta$ generated by the
differential reduction of the Feynman diagrams shown in Fig.~\ref{diagrams}.}
\begin{displaymath}
\begin{array}{cc|ccc|cccc}
h & E^q_{120} & s & E^q_{1220} & V^q_{1220} & q & J^q_{22} & B^q_{1220} &
J^q_{1220}\\
\hline
1 & 1 & 0 & 1 & 2 & 1 & 1 & - & - \\
{}\ge2 & 2 & 1 & 2 & 3 & 2 & 2 & 1 & 2 \\
 & & {}\ge2 & 3 & 4 & 3 & 3 & 2 & 3 \\
 & & & & & {}\ge4 & 3 & 3 & 4
\end{array}
\end{displaymath}
\end{table}

\section{Discussion and conclusion }
\label{conclusion}

The idea that any Feynman diagram can be associated with a generalized
Horn-type hypergeometric function or a linear combination of such functions was
born at the end of the 60s \cite{first} and was applied during the last twenty
years \cite{BD,BFT,davydychev}.
The Mellin-Barnes representation (\ref{MB}) of a Feynman diagram is universal.
Under certain conditions, it can be converted to a linear combination of
Horn-type hypergeometric functions (\ref{Horn}). 
The latter possess very interesting  properties.
In fact, the systems of differential equations which they satisfy are
sufficient for 
(i) the differential reduction of the original functions to restricted
sets of basis functions, whose number follows directly from the system of
differential equations;
(ii) the construction of the all-order $\ep$ expansions of the basis
hypergeometric functions in form of iterated integrals
\cite{KWY07a,KWY07b,KWY07c,KK08a}.  
To our knowledge, the first property has never been discussed in the context of
a reduction procedure of Feynman diagrams. 

The aim of the present paper was to show how the differential-reduction
algorithm can be applied to reduce Feynman diagrams without having to exploit
the integration-by-parts technique \cite{ibp} or its dimensional generalization
\cite{Davydychev-Tarasov}. 
In our analysis, we considered a phenomenologically interesting class of
Feynman diagrams, namely those shown in Figs.~\ref{one-loop} and
\ref{diagrams}, which are expressible in terms of generalized hypergeometric
functions ${}_{p+1}F_{p}$ of one variable and cannot be treated with currently
available program packages.

As first steps in our analysis, Takayama's differential-reduction
algorithm~\cite{takayama} was extended to the case of {\it exceptional} values
of parameters in Section~\ref{exceptional}, and the basis of differential
reduction was written explicitly in Section~\ref{criteria}. 
In particular, this algorithm allowed us to complete the proofs of the
theorems regarding the construction of all-order $\ep$ expansions of
hypergeometric functions presented in Refs.~\cite{KWY07a,KWY07b,KWY07c,KK08a}.
For non-exceptional values of parameters, this algorithm is implemented in the
Mathematica based program {\tt HYPERDIRE} to be described in a in a separate
communication \cite{preparation}.

The differential-reduction formalism via the construction of step-up and
step-down operators can be applied to the reduction of any Feynman diagram.
Its advantages reside in the simplicity and universality of the construction of
this type of operators, the full control over the analytical structure of the
Feynman diagrams via the hypergeometric representations, the simplicity of the
criterion of reducibility of hypergeometric functions to simpler functions, and
the existence of a few, recently developed algorithms
\cite{KWY07a,KWY07b,KWY07c,KK08a,DK04,nested} for analytically
constructing the coefficients of the $\ep$ expansions of hypergeometric
functions.

The main criteria of reducibility of hypergeometric functions, namely
integer difference between upper and lower parameters and integer value of one
of the upper parameters, are also valid for Horn-type hypergeometric functions
of more than one variables.
Moreover, the first type of reduction can be applied directly to Mellin-Barnes
representations of Feynman diagrams. 


This approach becomes incomplete when some of the variables of a Horn-type 
hypergeometric function belong to its surface of singularities;
for hypergeometric functions of one variable, this corresponds to $z=1$. 
In this case, the extended Ore algebra technique \cite{ore}, the telescoping
approach \cite{a=b}, or the Laporta algorithm \cite{laporta} can be employed to
obtain exact analytical results.
In addition to these algorithms, there is another approach based on the
analytical structure of the coefficients of the $\ep$ expansions of
hypergeometric functions: instead of looking for exact recurrence relations,
the limit $z\to1$ can be taken for the coefficients of the Laurent expansions.
In particular, this procedure can be applied to any hypergeometric function
whose $\ep$ expansion has coefficients that are expressible in terms of
multiple polylogarithms.
This algorithm may be applied to evaluate Feynman diagrams, if just the first
few coefficients of their $\ep$ expansions are required. 

As an illustration of the usefulness of our differential-reduction procedure,
we considered a few examples, namely the Feynman diagrams shown in
Figs.~\ref{one-loop} and \ref{diagrams} with arbitrary powers of propagators
and space-time dimension, and explained how to express them in terms of
hypergeometric functions and to construct their $\ep$ expansions to higher
orders.

An interesting observation we made here is that, up to products of one-loop
bubbles, the number of master integrals constructed through the
integration-by-parts technique is equal to the highest power of the
differential operator $\theta$ generated by the differential reduction of
hypergeometric functions plus one, independently of the number of
hypergeometric functions occuring in the expressions for the Feynman diagrams.


\vspace{5mm}
\noindent 
{\bf Acknowledgments.} \\
We are grateful to A.~Isaev, A.~Davydychev, V.~Spiridonov, O.~Tarasov,
O.~Veretin, B.~Ward, and S.~Yost for useful discussions, to  
G.~Sandukovskaya for carefully reading this manuscript, and to A.~Scheplyakov
for his contribution at an early stage of this work.
M.Y.K. is grateful to the members of Physics Department of Baylor University
for their hospitality when this project was started.
This research was supported in part by BMBF Grant No.\ 05~HT6GUA, by
DFG Grants No.\ KN~365/3--1 and KN~365/3--2, by HGF Grant No.\ HA~101, and by
Russian Academy of Science Grant No.\ MK--1607.2008.2.

\appendix
\section{Sets of basis functions for the Laurent expansions of hypergeometric
functions}
\label{appendix}

\subsection{All parameters are integer}

Using the algorithm described in Ref.~\cite{KWY07c}, the derivatives of a
basis hypergeometric function with integer parameters may be expressed in
terms of the $\rho_{p+k}^{(p-a)}(z)$ functions defined in that reference,
as\footnote{%
In Eq.~(2.11a) of Ref.~\cite{KWY07c}, the last parameter should have the value 
$p-2$ rather than $p-1$.}
\begin{eqnarray}
\theta^j
{}_{p}F_{p-1}(\vec{a}\ep;1 + \vec{b}\ep; z)
&=& \delta_{j0}
 +  \ep^p \left( \prod_{r=1}^p a_r \right) 
\sum_{k=0}^\infty \ep^k \rho_{p + k}^{(p - 1 - j)}(z) \;,
\qquad 
j = 0, \cdots, p - 1 \;.
\nonumber\\
&&\end{eqnarray}
The iterative solution of the $\rho_{p+k}^{(p-1-j)}(z)$ functions is given by
Eqs.~(2.13a)--(2.13b) in Ref.~\cite{KWY07c}, and their explicit forms in terms
of generalized polylogarithms are collected in Section~3 therein.
\subsection{One lower parameter is half-integer}

In Refs.~\cite{DK04,DK01}, the $\ep$ expansions of the functions 
\begin{eqnarray}
~_{P+1}F_{P}\left(\begin{array}{c|}
\{1+a_i \ep\}^{K}, \{2+d_i\ep \}^{P-K+1}   \\
\frac{3}{2} + f\ep, \{1+ e_j \ep\}^R, \{ 2+c_i\ep\}^{P-R-1} \end{array} ~ z \right) \;,
\label{dk04}
\end{eqnarray}
with $K-R \geq 2$, were constructed up to functions of weight 4 and, in the
case of $z=1/4$, up to weight 5.\footnote{%
The series representations of these hypergeometric functions are proportional
to $\sum_{j=0}^\infty\frac{z^j}{j^{K-R-1} }$.} 
For these parameters, after some trivial factorization, the coefficients of
the $\ep$ expansions contain functions of only one definite weight.
In this section, we present explicit relations between the basis functions of
the differential reduction, defined by Eq.~(\ref{decomposition}), and the
functions of Eq.~(\ref{dk04}).
One of the auxiliary relations intensively used in the following reads: 
\begin{eqnarray}
~_{p}F_{p-1}\left(\begin{array}{c|}
a_1, a_2, \cdots, a_{p}   \\
b_1, b_2, \cdots, b_{p-1} \end{array} ~ z \right) 
= 
1 
+ 
z \frac{\prod_{j=1}^p a_j }{\prod_{k=1}^{p-1} b_k}
~_{p+1}F_{p}\left(\begin{array}{c|}
1, 1 + a_1, 1 + a_2, \cdots, 1 + a_{p}   \\
2, 1 + b_1, 1 + b_2, \cdots, 1 + b_{p-1} \end{array} ~ z \right) \;.
\qquad
\label{shift}
\end{eqnarray}
%
\boldmath
\subsubsection{${}_2F_1$}
\unboldmath

A detailed discussion of the $_2F_1$ function was  presented in
Refs.~\cite{KWY07a,KK08a,MKL06}.

\boldmath
\subsubsection{${}_3F_2$}
\unboldmath

For the hypergeometric function
\begin{eqnarray}
~_{3}F_{2}\left(\begin{array}{c|}
I_1+a_1 \ep,  I_2+a_2 \ep, I_3+a_3 \ep \\
I_4+\frac{1}{2}+f\ep, I_5+c_1\ep \end{array} ~ z \right) \;,
\end{eqnarray}
where $\{I_k\}$ are arbitrary integers and all $a_i \neq 0$, the basis of the
differential reduction is
\begin{eqnarray}
&& 
\{ 1, \theta, \theta^2 \} \times
~_{3}F_2\left(\begin{array}{c|}
a_1 \ep, a_2 \ep, a_3 \ep  \\
\frac{1}{2}  +  f \ep, 1  +  c_1 \ep \end{array} ~ z \right) \;.
\label{3f2:basis:HYPERDIRE}
\end{eqnarray}
For $P=2$, there are four functions of the type of Eq.~(\ref{dk04}): 
\begin{eqnarray}
~_{4}F_3\left(\begin{array}{c|}
1 + a_1 \ep, 1 + a_2 \ep, 1 + a_3 \ep, 1  \\
\frac{3}{2}  +  f \ep, 2  +  c_1 \ep, 2 \end{array} ~ z \right) \;,
& \quad & 
~_{3}F_2\left(\begin{array}{c|}
1  +  a_1 \ep, 1  +  a_2 \ep, 1  +  a_3 \ep  \\
\frac{3}{2}  +  f \ep, 2  +  c_1 \ep \end{array} ~ z \right) \; ,
\nonumber \\  
~_{3}F_2\left(\begin{array}{c|}
1  +  a_1 \ep, 1  +  a_2\ep, 2  +  a_3\ep \\
\frac{3}{2}  +  f\ep,  2  +  c_1 \ep \end{array} ~ z \right) \;,
& \quad & 
~_{3}F_2\left(\begin{array}{c|}
1  +  a_1 \ep, 1  +  a_2 \ep, 1  +  a_3 \ep  \\
\frac{3}{2}  +  f \ep, 1  +  c_1 \ep \end{array} ~ z \right) \; .
\qquad\label{3f2:basis:dk04}
\end{eqnarray}
The functions of Eqs.~(\ref{3f2:basis:HYPERDIRE}) and (\ref{3f2:basis:dk04})
are related as
\begin{eqnarray}
&& \hspace{-10mm}
~_{3}F_2\left(\begin{array}{c|}
a_1 \ep, a_2 \ep, a_3 \ep  \\
\frac{1}{2}  +  f \ep, 1  +  c_1 \ep \end{array} ~ z \right) 
= 1  +  
\frac{2 a_1 a_2 a_3 \ep^3}{(1+2f\ep)(1+c_1\ep)} 
~_{4}F_{3}\left(\begin{array}{c|}
1 + a_1 \ep, 1 + a_2 \ep, 1 + a_{3} \ep, 1   \\
\frac{3}{2} + f \ep, 2 + c_1 \ep, 2 \end{array} ~ z \right) \;,
\nonumber \\ && \hspace{-10mm}
\theta
~_{3}F_2\left(\begin{array}{c|}
a_1 \ep, a_2 \ep, a_3 \ep  \\
\frac{1}{2}  +  f \ep, 1  +  c_1 \ep \end{array} ~ z \right) 
= 
2 z \frac{a_1 a_2 a_3 \ep^3}{(1 + 2f\ep)(1 + c_1\ep)} 
~_{3}F_2\left(\begin{array}{c|}
1  +  a_1 \ep, 1  +  a_2 \ep, 1  +  a_3 \ep  \\
\frac{3}{2}  +  f \ep, 2  +  c_1 \ep \end{array} ~ z \right) \; .
\nonumber\\
&&
\end{eqnarray}
The second derivative in Eq.~(\ref{3f2:basis:HYPERDIRE}) may be obtained from
any of the following relations: 
\begin{eqnarray}
&&
~_{3}F_2\left(\begin{array}{c|}
1 + a_1, 1 + a_2, 1 + a_3 \\
1  +  f, c_1 \end{array} ~ z \right) 
= 
\frac{1}{z}
\frac{f \theta (\theta+c_1-1)}{a_1a_2a_3}
~_{3}F_2\left(\begin{array}{c|}
a_1, a_2, a_3 \\
f, c_1 \end{array} ~ z \right) \;,
\nonumber \\ && 
~_{3}F_2\left(\begin{array}{c|}
1 + a_1, 1 + a_2, 2 + a_3 \\
1 + f, 1  +  c_1 \end{array} ~ z \right) 
= 
\frac{1}{z}
\frac{c_1 f \theta (\theta + a_3)}{a_1a_2a_3(1+a_3)}
~_{3}F_2\left(\begin{array}{c|}
a_1, a_2, a_3 \\
f, c_1 \end{array} ~ z \right) \;,\quad
\label{3f2:2}
\end{eqnarray}
where
\begin{equation}
a_i\to a_i \ep,\qquad
f\to\frac{1}{2}+f\ep,\qquad
c_1\to1+c_1\ep,
\label{eq:sub}
\end{equation}
are to be substituted.
In particular, it can be expressed as a linear combination of the second,
third, and fourth functions in Eq.~(\ref{3f2:basis:dk04}) as
\begin{eqnarray}
&& 
\frac{(\gamma_1 + \gamma_2)}{z}
\frac{f c_1}{a_1 a_2 a_3 }
\theta^2
~_{3}F_2\left(\begin{array}{c|}
a_1, a_2, a_3 \\
c_1, f  \end{array} ~ z \right) 
\nonumber \\ && 
= 
\gamma_1 c_1 
~_{3}F_2\left(\begin{array}{c|}
1  +  a_1, 1  +  a_2, 1 +  a_3 \\
c_1,  1 +  f \end{array} ~ z \right) 
+ 
\gamma_2 (1 + a_3)
~_{3}F_2\left(\begin{array}{c|}
1  +  a_1, 1  +  a_2, 2  +  a_3 \\
1  +  c_1,  1  +  f \end{array} ~ z \right) 
\nonumber \\ && 
- 
\left[ 
\gamma_1 (c_1 - 1)
 +  
\gamma_2 a_3
\right]
~_{3}F_2\left(\begin{array}{c|}
1  +  a_1, 1  +  a_2, 1  +  a_3 \\
1  +  c_1,  1  +  f \end{array} ~ z \right) 
\;,
\label{eq:t23f2}
\end{eqnarray}
where $\gamma_1$ and $\gamma_2$ are arbitrary numbers. 
Putting $\gamma_2 = - \gamma_1$, we obtain a linear relation between the
hypergeometric functions on the r.h.s.\ of Eq.~(\ref{eq:t23f2}).
We checked that the $\ep$ expansions of the hypergeometric functions 
constructed in Ref.~\cite{DK04} and collected in Ref.~\cite{MKL:hyper} satisfy
this relation.

%
In the case $a_1=1$, the basis of the differential reduction is 
\begin{eqnarray}
&& 
\{ 1, \theta \} \times
~_{3}F_2\left(\begin{array}{c|}
1, 1 + a_2 \ep, 1 + a_3 \ep  \\
\frac{3}{2}  +  f \ep, 2  +  c_1 \ep \end{array} ~ z \right) \;.
\label{3f2:basis:HYPERDIRE:integer}
\end{eqnarray}
The first derivative in Eq.~(\ref{3f2:basis:HYPERDIRE:integer}) can be
obtained from any of the two functions 
\begin{eqnarray}
~_{3}F_2\left(\begin{array}{c|}
1, 1  +  a_2 \ep, 1  +  a_3 \ep  \\
\frac{3}{2}  +  f \ep, 1  +  c_1 \ep \end{array} ~ z \right) \; ,
& \quad &  
~_{3}F_2\left(\begin{array}{c|}
1, 1  +  a_2\ep, 2  +  a_3\ep \\
\frac{3}{2}  +  f\ep,  2  +  c_1 \ep \end{array} ~ z \right) \;,
\label{3f2:basis:dk04:integer}
\end{eqnarray}
by the single application of the step-up and step-down operators defined in
Eq.~(\ref{universal:b}).
The result is
\begin{eqnarray}
&& 
(\gamma_1  +  \gamma_2 ) \theta
~_{3}F_2\left(\begin{array}{c|}
1, 1 + a_2 \ep, 1 + a_3 \ep  \\
\frac{3}{2}  +  f \ep, 2  +  c_1 \ep \end{array} ~ z \right) 
= 
\nonumber \\ && 
\gamma_1 (1 + a_3 \ep)
~_{3}F_2\left(\begin{array}{c|}
1, 1 + a_2 \ep, 2 + a_3 \ep  \\
\frac{3}{2}  +  f \ep, 2  +  c_1 \ep \end{array} ~ z \right) 
+ 
\gamma_2 
(1 + c_1\ep)
~_{3}F_2\left(\begin{array}{c|}
1, 1 + a_2 \ep, 1 + a_3 \ep  \\
\frac{3}{2}  +  f \ep, 1  +  c_1 \ep \end{array} ~ z \right) 
\nonumber \\ && 
- 
\left[
\gamma_1 (1 + a_3 \ep)
+ 
\gamma_2 (1 + c_1\ep)
\right]
~_{3}F_2\left(\begin{array}{c|}
1, 1 + a_2 \ep, 1 + a_3 \ep  \\
\frac{3}{2}  +  f \ep, 2  +  c_1 \ep \end{array} ~ z \right) 
\;,
\label{3f2:basis:dk04:1}
\end{eqnarray}
where $\gamma_1$ and $\gamma_2$ are  arbitrary numbers.


\boldmath
\subsubsection{${}_4F_3$}
\unboldmath

Similarly, for the hypergeometric function
\begin{eqnarray}
~_{4}F_{3}\left(\begin{array}{c|}
I_1+a_1 \ep,  I_2+a_2 \ep, I_3+a_3 \ep, I_4+a_4 \ep \\
I_5+\frac{1}{2}+f\ep, I_6+c_1\ep, I_7+c_2\ep \end{array} ~ z \right) \;,
\end{eqnarray}
where $\{I_k\}$ are arbitrary integers and all $a_i \neq 0$, the basis of the
differential reduction is
\begin{eqnarray}
&& 
\{ 1, \theta, \theta^2, \theta^3 \} \times
~_{4}F_3\left(\begin{array}{c|}
a_1 \ep, a_2 \ep, a_3 \ep, a_4 \ep  \\
\frac{1}{2}  +  f \ep, 1  +  c_1 \ep, 1 + c_2 \ep \end{array} ~ z \right) \;.
\label{4f3:HYPERDIRE}
\end{eqnarray}
There are seven functions of the type of Eq.~(\ref{dk04}), namely
\begin{eqnarray}
~_{5}F_4\left(\begin{array}{c|}
1 + a_1 \ep, 1 + a_2 \ep, 1 + a_3 \ep, 1 + a_4 \ep, 1  \\
\frac{3}{2}  +  f \ep, 2  +  c_1 \ep, 2  +  c_2\ep,  2  \end{array} ~ z \right) \; .
& \quad & 
~_{4}F_3\left(\begin{array}{c|}
1 + a_1 \ep, 1 + a_2 \ep, 1 + a_3 \ep, 1 + a_4 \ep  \\
\frac{3}{2}  +  f \ep, 2  +  c_1 \ep, 2  +  c_2 \ep \end{array} ~ z \right) \; ,
\nonumber \\ 
~_{4}F_3\left(\begin{array}{c|}
1 + a_1 \ep, 1 + a_2 \ep, 1 + a_3 \ep, 2 + a_4 \ep  \\
\frac{3}{2}  +  f \ep, 2  +  c_1 \ep, 2  +  c_2 \ep \end{array} ~ z \right) \; ,
& \quad & 
~_{4}F_3\left(\begin{array}{c|}
1 + a_1 \ep, 1 + a_2 \ep, 1 + a_3 \ep, 1 + a_4 \ep  \\
\frac{3}{2}  +  f \ep, 2  +  c_1 \ep, 1  +  c_2 \ep \end{array} ~ z \right) \; ,
\nonumber \\ 
~_{4}F_3\left(\begin{array}{c|}
1 + a_1\ep, 1 + a_2\ep, 2 + a_3\ep, 2 + a_4\ep  \\
\frac{3}{2}  +  f \ep, 2  +  c_1 \ep, 2  +  c_2 \ep \end{array} ~ z \right) \;, 
& \quad & 
~_{4}F_3\left(\begin{array}{c|}
1 + a_1 \ep, 1 + a_2 \ep, 1 + a_3 \ep, 2 + a_4\ep  \\
\frac{3}{2}  +  f \ep, 1  +  c_1 \ep, 1 + c_2\ep \end{array} ~ z \right) \; ,
\nonumber \\ 
~_{4}F_3\left(\begin{array}{c|}
1 + a_1 \ep, 1 + a_2 \ep, 1 + a_3 \ep, 1 + a_4 \ep  \\
\frac{3}{2}  +  f \ep, 1  +  c_1 \ep, 1  +  c_2 \ep \end{array} ~ z \right) \; .
&& 
\label{4f3:dk04}
\end{eqnarray}
The two bases defined by Eqs.~(\ref{4f3:HYPERDIRE}) and (\ref{4f3:dk04}) are
related as
\begin{eqnarray}
&& 
~_{4}F_3\left(\begin{array}{c|}
a_1 \ep, a_2 \ep, a_3 \ep, a_4 \ep  \\
\frac{1}{2}  +  f \ep, 1  +  c_1 \ep, 1+c_2 \ep \end{array} ~ z \right) 
= 
\nonumber \\ && 
1  +  
\frac{2 a_1 a_2 a_3 a_4 \ep^4}{(1+2f\ep) (1+c_1\ep) (1+c_2\ep)}
~_{5}F_4\left(\begin{array}{c|}
1 + a_1 \ep, 1 + a_2 \ep, 1 + a_3 \ep, 1 + a_4 \ep, 1  \\
\frac{3}{2}  +  f \ep, 2  +  c_1 \ep, 2+c_2 \ep, 2 \end{array} ~ z \right) \;,
\nonumber \\ && 
\theta
~_{4}F_3\left(\begin{array}{c|}
a_1 \ep, a_2 \ep, a_3 \ep, a_4 \ep  \\
\frac{1}{2}  +  f \ep, 1  +  c_1 \ep, 1  +  c_2 \ep \end{array} ~ z \right) 
\nonumber \\ && 
= 
2z \frac{a_1 a_2 a_3 a_4 \ep^4}{(1+2f\ep)(1+c_1\ep) (1+c_2\ep)}
~_{4}F_3\left(\begin{array}{c|}
1 + a_1 \ep, 1 + a_2 \ep, 1 + a_3 \ep, 1 + a_4 \ep  \\
\frac{3}{2}  +  f \ep, 2  +  c_1 \ep, 2  +  c_2 \ep \end{array} ~ z \right)
\; .\qquad
\end{eqnarray}
The second derivative in Eq.~(\ref{4f3:HYPERDIRE}) may be obtained from any
of the following relations:
\begin{eqnarray}
&& 
~_{4}F_3\left(\begin{array}{c|}
1 + a_1, 1 + a_2, 1 + a_3, 1 + a_4  \\
1  +  f, 1  +  c_1, c_2 \end{array} ~ z \right) 
= 
\frac{f}{z} 
\frac{c_1 \theta (\theta + c_2 - 1)}{a_1 a_2 a_3 a_4}
~_{4}F_3\left(\begin{array}{c|}
a_1, a_2, a_3,a_4 \\
f, c_1, c_2 \end{array} ~ z \right) 
\; ,
\nonumber \\ && 
~_{4}F_3\left(\begin{array}{c|}
1+a_1, 1+a_2, 1 + a_3, 2 + a_4  \\
1  +  f, 1  +  c_1, 1  +  c_2 \end{array} ~ z \right) 
= 
\frac{f}{z} 
\frac{c_1 c_2 \theta (\theta + a_4)}
     {a_1 a_2 a_3 a_4 (1 + a_4)}
~_{4}F_3\left(\begin{array}{c|}
a_1, a_2, a_3,a_4 \\
f, c_1, c_2 \end{array} ~ z \right) 
\;, 
\nonumber \\ 
\label{4f3:1}
\end{eqnarray}
with the substitutions of Eq.~(\ref{eq:sub}).
In particular, it can be expressed it as a linear combination of the second,
third, and fourth hypergeometric functions in Eq.~(\ref{4f3:dk04}) as
\begin{eqnarray}
&& 
\frac{(\gamma_1+\gamma_2)}{z}\frac{f c_1 c_2}{a_1 a_2 a_3 a_4}
\theta^2 
~_{4}F_3\left(\begin{array}{c|}
a_1, a_2, a_3,a_4 \\
f, c_1, c_2 \end{array} ~ z \right) 
= 
\gamma_1 c_2 
~_{4}F_3\left(\begin{array}{c|}
1 + a_1, 1 + a_2, 1 + a_3, 1 + a_4  \\
1  +  f, 1  +  c_1, c_2 \end{array} ~ z \right) 
\nonumber \\ && 
-\left[ a_4 \gamma_2  +  (c_2 - 1) \gamma_1 \right]
~_{4}F_3\left(\begin{array}{c|}
1 + a_1, 1 + a_2, 1 + a_3, 1 + a_4  \\
1  +  f, 1  +  c_1, 1 + c_2 \end{array} ~ z \right) 
\nonumber \\ && 
+ 
\gamma_2 (1+a_4)
~_{4}F_3\left(\begin{array}{c|}
1+a_1, 1+a_2, 1 + a_3, 2 + a_4  \\
1  +  f, 1  +  c_1, 1  +  c_2 \end{array} ~ z \right) 
\;, 
\label{4f3:dk04:2}
\end{eqnarray}
where $\gamma_1$ and $\gamma_2$ are arbitrary numbers.  
Putting $\gamma_2=-\gamma_1$, we obtain a linear relation between the
hypergeometric functions on the r.h.s.\ of Eq.~(\ref{4f3:dk04:2}).
We checked that the $\ep$ expansions of the hypergeometric functions 
constructed in Ref.~\cite{DK04} satisfy this relation.
The third derivative in Eq.~(\ref{4f3:HYPERDIRE}) may be obtained from any of
the following relations:
\begin{eqnarray}
~_{4}F_3\left(\begin{array}{c|}
1 + a_1, 1 + a_2, 1 + a_3, 1 + a_4  \\
1  +  f,  c_1, c_2  \end{array} ~ z \right) 
&=& 
\frac{f}{z} 
\frac{\theta (\theta + c_1 - 1)(\theta + c_2 - 1)}
     {a_1 a_2 a_3 a_4}
~_{4}F_3\left(\begin{array}{c|}
a_1, a_2, a_3,a_4 \\
f, c_1, c_2 \end{array} ~ z \right)  
\; ,
\nonumber \\
~_{4}F_3\left(\begin{array}{c|}
1 + a_1, 1 + a_2, 2 + a_3, 2 + a_4  \\
1  +  f, 1  +  c_1, 1  +  c_2 \end{array} ~ z \right) 
&=& 
\frac{f}{z} 
\frac{c_1 c_2 \theta (\theta + a_3) (\theta + a_4)}
     {a_1 a_2 a_3 a_4 (1 + a_3)(1 + a_4)}
~_{4}F_3\left(\begin{array}{c|}
a_1, a_2, a_3,a_4 \\
f, c_1, c_2 \end{array} ~ z \right)  \;,
\nonumber \\
~_{4}F_3\left(\begin{array}{c|}
1 + a_1, 1 + a_2, 1 + a_3, 2 + a_4  \\
1  +  f, 1  +  c_1, c_2 \end{array} ~ z \right) 
&=& 
\frac{f}{z} 
\frac{c_1 \theta (\theta + a_4) (\theta + c_2 - 1)}
     {a_1 a_2 a_3 a_4 (1 + a_4)}
~_{4}F_3\left(\begin{array}{c|}
a_1, a_2, a_3,a_4 \\
f, c_1, c_2 \end{array} ~ z \right)  
\;,
\nonumber\\
&&\label{4f3:2}
\end{eqnarray}
with the substitutions of Eq.~(\ref{eq:sub}).
As a consequence of Eq.~(\ref{4f3:2}), the following linear relations 
between hypergeometric functions are valid: 
\begin{eqnarray} 
%
%
&& 
c_1 
~_{4}F_3\left(\begin{array}{c|}
1+a_1, 1+a_2, 1 + a_3, 1 + a_4  \\
1  +  f, c_1, c_2 \end{array} ~ z \right) 
- 
(1 + a_4)
~_{4}F_3\left(\begin{array}{c|}
1+a_1, 1+a_2, 1 + a_3, 2 + a_4  \\
1  +  f, 1  +  c_1, c_2 \end{array} ~ z \right) 
\nonumber \\ && 
= 
(c_1 - a_4 - 1)
~_{4}F_3\left(\begin{array}{c|}
1+a_1, 1+a_2, 1 + a_3, 1 + a_4  \\
1  +  f, 1  +  c_1, c_2 \end{array} ~ z \right) \;,
%
%
\nonumber
\\ && 
c_2
~_{4}F_3\left(\begin{array}{c|}
1+a_1, 1+a_2, 1 + a_3, 2 + a_4  \\
1  +  f, 1 + c_1, c_2 \end{array} ~ z \right) 
- 
(1 + a_3)
~_{4}F_3\left(\begin{array}{c|}
1+a_1, 1+a_2, 2 + a_3, 2 + a_4  \\
1  +  f, 1  +  c_1, 1 + c_2 \end{array} ~ z \right) 
\nonumber \\ && 
= 
(c_2 - a_3 - 1)
~_{4}F_3\left(\begin{array}{c|}
1+a_1, 1+a_2, 1 + a_3, 2 + a_4  \\
1  +  f, 1  +  c_1, 1 + c_2 \end{array} ~ z \right) \;,
\nonumber\\
&& 
(\beta_1  +  \beta_2)
\Biggl\{ 
c_1 c_2
~_{4}F_3\left(\begin{array}{c|}
1 + a_1, 1 + a_2, 1 + a_3, 1 + a_4  \\
1  +  f, c_1, c_2 \end{array} ~ z \right) 
\nonumber \\ && 
- 
(1 + a_3)(1 + a_4)
~_{4}F_3\left(\begin{array}{c|}
1 + a_1, 1 + a_2, 2 + a_3, 2 + a_4  \\
1  +  f, 1  +  c_1, 1 + c_2 \end{array} ~ z \right) 
\Biggr\} 
\nonumber \\ && 
+ 
(c_2 - a_4 - 1)
\left[ 
\beta_1 (c_2  -  a_3  -  1) 
 -  
\beta_2 (c_1  -  a_4  -  1) 
\right]
~_{4}F_3\left(\begin{array}{c|}
1+a_1, 1+a_2, 1 + a_3, 1 + a_4  \\
1  +  f, 1  +  c_1, 1 + c_2 \end{array} ~ z \right) 
\nonumber \\ && 
- 
\beta_1
(c_1 + c_2 - 2 - a_3 - a_4)  c_2 
~_{4}F_3\left(\begin{array}{c|}
1+a_1, 1+a_2, 1 + a_3, 1 + a_4  \\
1  +  f, 1  +  c_1, c_2 \end{array} ~ z \right) 
\nonumber \\ && 
- 
\beta_2 
(c_1 + c_2 - 2 - a_3 - a_4) (1  +  a_4)
~_{4}F_3\left(\begin{array}{c|}
1+a_1, 1+a_2, 1 + a_3, 2 + a_4  \\
1  +  f, 1  +  c_1, 1 + c_2 \end{array} ~ z \right) 
= 0 \;, 
\end{eqnarray} 
where $\beta_1$ and $\beta_2$ are arbitrary numbers. 
We checked that the $\ep$ expansions of the hypergeometric functions ${}_4F_3$ 
constructed in Ref.~\cite{DK04} and collected in Ref.~\cite{MKL:hyper} satisfy
all these relations. 

In the case $a_1=1$, the basis of the differential reduction is 
\begin{eqnarray}
&& 
\{ 1, \theta, \theta^2 \} \times
~_{4}F_3\left(\begin{array}{c|}
1, 1 + a_2 \ep, 1 + a_3 \ep, 1 + a_4 \ep  \\
\frac{3}{2}  +  f \ep, 2  +  c_1 \ep, 2+c_2 \ep \end{array} ~ z \right) \;.
\label{4f3:HYPERDIRE:integer}
\end{eqnarray}
The first derivative in Eq.~(\ref{4f3:HYPERDIRE:integer}) can be related to
any of the two functions,
\begin{equation}
~_{4}F_3\left(\begin{array}{c|}
1, 1 + a_2 \ep, 1 + a_3 \ep, 2 + a_4 \ep  \\
\frac{3}{2}  +  f \ep, 2  +  c_1 \ep, 2  +  c_2 \ep \end{array} ~ z \right) \; ,
\qquad 
~_{4}F_3\left(\begin{array}{c|}
1, 1 + a_2 \ep, 1 + a_3 \ep, 1 + a_4 \ep  \\
\frac{3}{2}  +  f \ep, 2  +  c_1 \ep, 1  +  c_2 \ep \end{array} ~ z \right) \; ,
\end{equation}
by single application of the operators $B^{(+)}_{a_4}$ or $H^{(-)}_{c_2}$ given
in Eq.~(\ref{universal:b}).
The second derivative in Eq.~(\ref{4f3:HYPERDIRE:integer}) is related to any
of the following hypergeometric functions:
\begin{eqnarray}
&& 
~_{4}F_3\left(\begin{array}{c|}
1 + a_1 \ep, 1 + a_2 \ep, 1 + a_3 \ep, 1 + a_4 \ep  \\
\frac{3}{2}  +  f \ep, 1  +  c_1 \ep, 1  +  c_2 \ep \end{array} ~ z \right) \; ,
\quad 
~_{4}F_3\left(\begin{array}{c|}
1 + a_1 \ep, 1 + a_2 \ep, 2 + a_3 \ep, 2 + a_4 \ep  \\
\frac{3}{2}  +  f \ep, 2  +  c_1 \ep, 2  +  c_2 \ep \end{array} ~ z \right) \; ,
\nonumber \\ && 
~_{4}F_3\left(\begin{array}{c|}
1 + a_1 \ep, 1 + a_2 \ep, 1 + a_3 \ep, 2 + a_4 \ep  \\
\frac{3}{2}  +  f \ep, 2  +  c_1 \ep, 1  +  c_2 \ep \end{array} ~ z \right) \; ,
\end{eqnarray}
which are generated from the basis function of
Eq.~(\ref{4f3:HYPERDIRE:integer}) via application of 
$H^{(-)}_{b_2} H^{(-)}_{b_3}$, 
$B^{(+)}_{a_3} B^{(+)}_{a_4}$, or 
$H^{(-)}_{b_3} B^{(+)}_{a_4}$.

\boldmath
\subsubsection{${}_5F_4$}
\unboldmath

Similarly, for the hypergeometric function
\begin{eqnarray}
~_{5}F_{4}\left(\begin{array}{c|}
I_1 + a_1 \ep,  I_2 + a_2 \ep, I_3 + a_3 \ep, I_4 + a_4 \ep, I_5  +  a_5\ep \\
I_6 + \frac{1}{2} + f\ep, I_7 + c_1\ep, I_8 + c_2\ep, I_9  +  c_3 \ep \end{array} ~ z \right) \;,
\end{eqnarray}
where $\{I_k\}$ are arbitrary integers and all $a_i \neq 0$, the basis of the
differential reduction is 
\begin{eqnarray}
&& 
\{ 1, \theta, \theta^2, \theta^3, \theta^4 \} \times
~_{4}F_3\left(\begin{array}{c|}
a_1 \ep, a_2 \ep, a_3 \ep, a_4 \ep, a_5 \ep  \\
\frac{1}{2}  +  f \ep, 1  +  c_1 \ep, 1+c_2 \ep,  1 +  c_3 \ep\end{array} ~ z \right) \;.
\label{5f4:HYPERDIRE}
\end{eqnarray}
In this case, there are eleven functions of the type of Eq.~(\ref{dk04}).
The two bases defined by Eqs.~(\ref{5f4:HYPERDIRE}) and (\ref{dk04}) are
related as
\begin{eqnarray}
&& 
~_{5}F_4\left(\begin{array}{c|}
a_1 \ep, a_2 \ep, a_3 \ep, a_4 \ep, a_5 \ep  \\
\frac{1}{2}  +  f \ep, 1  +  c_1 \ep, 1 + c_2 \ep, 1 + c_3 \ep \end{array} ~ z \right) 
= 
1 + 
\nonumber \\ && 
\frac{2 \ep^5 \left( \prod_{i=1}^5 a_i \right) }{(1  +  2f\ep) \left[ \prod_{k=1}^3(1 + c_k\ep) \right]}
~_{6}F_5\left(\begin{array}{c|}
1 + a_1 \ep, 1 + a_2 \ep, 1 + a_3 \ep, 1 + a_4 \ep, 1  +  a_5 \ep, 1  \\
\frac{3}{2}  +  f \ep, 2  +  c_1 \ep, 2 + c_2 \ep, 2  +  c_3\ep, 2 \end{array} ~ z \right) \;,
\nonumber \\ && 
\theta
~_{5}F_4\left(\begin{array}{c|}
a_1 \ep, a_2 \ep, a_3 \ep, a_4 \ep, a_5 \ep  \\
\frac{1}{2}  +  f \ep, 1  +  c_1 \ep, 1 + c_2 \ep, 1 + c_3 \ep \end{array} ~ z \right) 
\nonumber \\ && 
= 
\frac{2 z \left( \prod_{i=1}^5 a_i \right) \ep^5}{(1  +  2f\ep) \left[ \prod_{k=1}^3(1 + c_k\ep) \right] }
~_{5}F_4\left(\begin{array}{c|}
1 + a_1 \ep, 1 + a_2 \ep, 1 + a_3 \ep, 1 + a_4 \ep, 1  +  a_5 \ep  \\
\frac{3}{2}  +  f \ep, 2  +  c_1 \ep, 2  +  c_2, 2  +  c_3 \ep \ep \end{array} ~ z \right) \; .\qquad
\label{5f4:dk04:1}
\end{eqnarray}
The second derivative in Eq.~(\ref{5f4:HYPERDIRE}) may be obtained as
\begin{eqnarray}
&& 
\frac{(\gamma_1+\gamma_2)}{z}
\frac{f \left( \prod_{k=1}^3 c_k \right) }{ \left( \prod_{j=1}^5 a_j \right)}
\theta^2 
~_{5}F_4\left(\begin{array}{c|}
a_1, a_2, a_3, a_4, a_5 \\
f, c_1, c_2, c_3 \end{array} ~ z \right) 
\nonumber \\ &=& 
\gamma_1 c_3 
~_{5}F_4\left(\begin{array}{c|}
1  + a_1, 1 +  a_2, 1 + a_3, 1 + a_4, 1  +  a_5   \\
1  +  f,  1  + c_1, 1 + c_2, c_3  \end{array} ~ z \right) 
\nonumber \\ &&{}+ \gamma_2  (1 + a_5)
~_{5}F_4\left(\begin{array}{c|}
1 + a_1, 1 + a_2, 1 + a_3, 1 + a_4, 2  +  a_5   \\
1  +  f \ep, 1  +  c_1,  1 + c_2,  1 + c_3  \end{array} ~ z \right) \
\nonumber \\ &&{}- \left[ 
\gamma_1 (c_3 - 1)  +  \gamma_2 a_5 
\right]
~_{5}F_4\left(\begin{array}{c|}
1 + a_1, 1 + a_2, 1 + a_3, 1 + a_4, 1  +  a_5   \\
1  +  f \ep, 1  +  c_1,  1 + c_2,  1 + c_3  \end{array} ~ z \right) \; ,
\label{5f4:dk04:2}
\end{eqnarray}
with the substitutions of Eq.~(\ref{eq:sub}),
where $\gamma_1$ and $\gamma_2$ are arbitrary numbers.
Putting $\gamma_2 = - \gamma_1$, we obtain a linear relation between the
hypergeometric functions on the r.h.s.\ of Eq.~(\ref{5f4:dk04:2}).
We checked that the $\ep$ expansions of the hypergeometric functions 
constructed in Ref.~\cite{DK04} satisfy this relation.
The third derivative in Eq.~(\ref{5f4:HYPERDIRE}) is related to one of the
following functions or their linear combination:
\begin{eqnarray}
~_{5}F_4\left(\begin{array}{c|}
1 + a_1, 1 + a_2, 1 + a_3, 2 + a_4, 2  +  a_5  \\
1  +  f, 1  +  c_1, 1  +  c_2,  1  +  c_3  \end{array} ~ z \right) 
&=&  
\frac{f}{z} 
\frac{\left( \prod_{k=1}^3 c_k \right) \theta (\theta  +  a_4) (\theta  +  a_5)}
     {\left(\prod_{j=1}^5 a_j \right)(1 + a_4)(1 + a_5)}
F(z)
\; ,
\nonumber \\
~_{5}F_4\left(\begin{array}{c|}
1 + a_1, 1 + a_2, 1 + a_3, 1 + a_4, 2  +  a_5  \\
1  +  f, 1  +  c_1, 1  +  c_2, c_3  \end{array} ~ z \right) 
&=&   
\frac{f}{z} 
\frac{\left( \prod_{k=1}^2 c_k \right) \theta (\theta  +  c_3  -  1)(\theta  +  a_5)}
     {\left(\prod_{j=1}^5 a_j \right)(1 + a_5)}
F(z)
\; ,
\nonumber \\
~_{5}F_4\left(\begin{array}{c|}
1 + a_1, 1 + a_2, 1 + a_3, 1 + a_4, 1  +  a_5   \\
1  +  f, 1  +  c_1, c_2,  c_3  \end{array} ~ z \right) 
&=&   
\frac{f}{z} 
\frac{c_1 \theta (\theta  +  c_2  -  1)(\theta  +  c_3  -  1)}
     {\left(\prod_{j=1}^5 a_j \right)}
F(z)
\; ,\qquad
\label{5f4:dk04:3}
\end{eqnarray}
where
\begin{equation}
F(z) \equiv 
~_{5}F_4\left(\begin{array}{c|}
a_1, a_2, a_3, a_4, a_5  \\
f, c_1, c_2,  c_3  \end{array} ~ z \right) \;,
\end{equation}
As a consequence of Eq.~(\ref{5f4:dk04:3}), the following linear relations
between hypergeometric functions hold:
\begin{eqnarray}
&& 
(1 + a_4  -  c_3)
~_{5}F_4\left(\begin{array}{c|}
1 + a_1, 1 + a_2, 1 + a_3, 1 + a_4, 2  +  a_5   \\
1  +  f, 1  +  c_1, 1 + c_2,  1 + c_3  \end{array} ~ z \right) 
\nonumber \\ &=&
(1 + a_4)
~_{5}F_4\left(\begin{array}{c|}
1 + a_1, 1 + a_2,   1 + a_3,  2 + a_4, 2  +  a_5   \\
1  +  f, 1  +  c_1, 1 + c_2,  1 + c_3  \end{array} ~ z \right) 
\nonumber \\ &&{}-  
c_3 
~_{5}F_4\left(\begin{array}{c|}
1 + a_1, 1 + a_2, 1 + a_3, 1 + a_4, 2  +  a_5   \\
1  +  f, 1  +  c_1, 1 + c_2,  c_3  \end{array} ~ z \right) \;, 
%
%
\nonumber \\
&& 
(1 + a_5  -  c_2)
~_{5}F_4\left(\begin{array}{c|}
1 + a_1, 1 + a_2, 1 + a_3, 1 + a_4, 1  +  a_5   \\
1  +  f, 1  +  c_1, 1 + c_2,  c_3  \end{array} ~ z \right) 
\nonumber \\ &=&
(1 + a_5)
~_{5}F_4\left(\begin{array}{c|}
1 + a_1, 1 + a_2,   1 + a_3,  1 + a_4, 2  +  a_5   \\
1  +  f, 1  +  c_1, 1 + c_2,  c_3  \end{array} ~ z \right) 
\nonumber \\ &&{}-  
c_2 
~_{5}F_4\left(\begin{array}{c|}
1 + a_1, 1 + a_2, 1 + a_3, 1 + a_4, 1  +  a_5   \\
1  +  f, 1  +  c_1, c_2,  c_3  \end{array} ~ z \right) \;, 
\nonumber\\
&& 
(a_5  -  c_3  +  1) 
\left[ 
\beta_1 \left(c_3  -  a_4  - 1 \right)
 -  
\beta_2 \left(c_2  -  a_5  -  1 \right)
\right]
~_{5}F_4\left(\begin{array}{c|}
1 + a_1, 1 +   a_2, 1 + a_3,  1 + a_4, 1  +  a_5   \\
1  +  f, 1  +  c_1, 1 + c_2,  1 + c_3  \end{array} ~ z \right) 
\nonumber \\
&=& 
(\beta_1 + \beta_2) c_2 c_3
~_{5}F_4\left(\begin{array}{c|}
1 + a_1, 1 + a_2, 1 + a_3, 1 + a_4, 1  +  a_5   \\
1  +  f, 1  +  c_1, c_2,  c_3  \end{array} ~ z \right) 
\nonumber \\
&&{}- 
(\beta_1 + \beta_2) 
(1 + a_4)(1 + a_5)
~_{5}F_4\left(\begin{array}{c|}
1 + a_1, 1 + a_2,   1 + a_3,  2 + a_4, 2  +  a_5   \\
1  +  f, 1  +  c_1, 1 + c_2,  1 + c_3  \end{array} ~ z \right) 
\nonumber \\
&&{}+  
\beta_1 c_3 \left( a_4 + a_5  -  c_2  -  c_3 + 2 \right) 
~_{5}F_4\left(\begin{array}{c|}
1 + a_1, 1 + a_2, 1 + a_3, 1 + a_4, 1  +  a_5   \\
1  +  f, 1  +  c_1, 1 + c_2,  c_3  \end{array} ~ z \right) 
\nonumber \\
&&{}+  
\beta_2 (1 + a_5) \left( a_4 + a_5  -  c_2  -  c_3 + 2 \right)
~_{5}F_4\left(\begin{array}{c|}
1 + a_1, 1 + a_2, 1 + a_3, 1 + a_4, 2  +  a_5   \\
1  +  f, 1  +  c_1, 1 + c_2,  1 + c_3  \end{array} ~ z \right) 
\;. 
\end{eqnarray}
The forth derivative in Eq.~(\ref{5f4:HYPERDIRE}) is related to any of the
following hypergeometric functions:
\begin{eqnarray}
~_{5}F_4\left(\begin{array}{c|}
1 + a_1, 1 + a_2, 2 + a_3, 2 + a_4, 2  +  a_5   \\
1  +  f, 1  +  c_1, 1  +  c_2,  1  +  c_3  \end{array} ~ z \right) 
&=&  
\frac{f}{z} 
\frac{c_1 c_2 c_3  \theta (\theta + a_3)(\theta  +  a_4) (\theta  +  a_5)}
     {\left(\prod_{j=1}^5 a_j \right)(1 + a_3)(1 + a_4)(1 + a_5)}
F(z)
\; ,
\nonumber \\
%
%
~_{5}F_4\left(\begin{array}{c|}
1 + a_1, 1 + a_2, 1 + a_3, 2 + a_4, 2  +  a_5   \\
1  +  f, 1  +  c_1, 1  +  c_2,  c_3  \end{array} ~ z \right) 
&=&  
\frac{f}{z} 
\frac{c_1 c_2 \theta (\theta + c_3 - 1)(\theta  +  a_4) (\theta  +  a_5)}
     {\left(\prod_{j=1}^5 a_j \right)(1 + a_4)(1 + a_5)}
F(z)
\; ,
\nonumber \\
%
%
~_{5}F_4\left(\begin{array}{c|}
1 + a_1, 1 + a_2, 1 + a_3, 1 + a_4, 2  +  a_5   \\
1  +  f, 1  +  c_1, c_2,  c_3  \end{array} ~ z \right) 
&=&  
\frac{f}{z} 
\frac{c_1 \theta (\theta + c_2 - 1)(\theta  +  c_3  -  1) (\theta  +  a_5)}
     {\left(\prod_{j=1}^5 a_j \right)(1 + a_5)}
F(z)
\; ,
\nonumber \\ 
%
%
~_{5}F_4\left(\begin{array}{c|}
1 + a_1, 1 + a_2, 1 + a_3, 1 + a_4, 1  +  a_5  \\
1  +  f, c_1, c_2,  c_3  \end{array} ~ z \right) 
&=&  
\frac{f}{z} 
\frac{\theta (\theta + c_1 - 1)(\theta  +  c_2  -  1) (\theta  +  c_3  -  1)}
     {\left(\prod_{j=1}^5 a_j \right)}
F(z)
\; ,
\nonumber \\ 
\label{5f4:dk04:4}
\end{eqnarray}
with the substitutions of Eq.~(\ref{eq:sub}).
The set of algebraic relations between the hypergeometric functions of
Eq.~(\ref{5f4:dk04:4}) is to long to be presented here. 
We checked that the $\ep$ expansions of the hypergeometric functions ${}_5F_4$ 
constructed in Ref.~\cite{DK04} satisfy all these linear relations. 


\boldmath
\subsubsection{${}_{p+1}F_p$, $p  \geq 6$}
\unboldmath

For the hypergeometric function
\begin{equation}
~_{p+1}F_{p}\left(\begin{array}{c|}
\{1\}_{p + 1}, \\
\frac{3}{2}, \{2\}_{p - 1} \end{array} ~ z \right) \;,
\end{equation}
the following representation was derived in Refs.~\cite{VK00}: 
\begin{eqnarray}
\frac{z}{2} 
~_{p+1}F_{p}\left(\begin{array}{c|}
\{1\}_{p + 1}, \\
\frac{3}{2}, \{2\}_{p - 1} \end{array} ~ z \right) 
& = &  
 \sum_{k=1}^\infty \frac{1}{\binom{2n}{n}} \frac{z^n}{n^p}
\nonumber \\ 
& = &  
-
\sum_{j=0}^{p-2} \frac{(-2)^j}{(a - 2 - j)!j!} (\ln z )^{a - 2 - j}
\LS{j+2}{1}{2\arcsin\frac{\sqrt{z}}{2}} \;, \qquad
\label{ibs}
\end{eqnarray}
where $0 \leq z \leq 4$, $p \geq 2$, and 
$\LS{a}{j}{z}$ denotes the generalized log-sine function\footnote{%
A computer program for the numerical evaluation of the generalized log-sine
functions with high precision was developed in Ref.~\cite{lsjk}.}
\cite{Lewin}.
Later, it was shown in Ref.~\cite{DK01} that any generalized log-sine function
of argument $\theta$ can be written in terms of Nielsen polylogarithms with 
argument\footnote{%
In terms of the variable $z$ in Eq.~(\ref{ibs}), we have
$$
y = 
\left. \cos \theta \pm i  \sin \theta  \right|_{\theta = 2 \arcsin \frac{\sqrt{z}}{2}}
= 
1  -  \frac{z}{2} \pm i \sqrt{z}{\sqrt{1 - \frac{z}{4}}}
\equiv 
\frac{1-\sqrt{\frac{z}{z-4}}}{1+\sqrt{\frac{z}{z-4}}} \;.
$$
}
$y = \exp(\pm i \theta)$
(see Eqs.~(2.18) and (A.21) in Ref.~\cite{DK01}).

\end{document}